\documentclass[a4paper]{article} 
\usepackage{amber_vltdoc}
\usepackage{supertabular}
\usepackage{url}
\usepackage[latin1]{inputenc}
\usepackage[OT1]{fontenc}
\usepackage{graphicx}

\makeatletter
\renewcommand{\ps@ambermain}{
  \renewcommand{\@oddhead}{\tabcolsep=10pt
    \renewcommand{\arraystretch}{1.}
    \begin{tabular}{|c|c|p{6cm}|}
      \hline
       \multirow{2}{2.8cm}{\bf \large OCA/UNSA}
      &\multirow{2}{7cm}{\bf \Large VLT / AMBER}
      &Doc. No \numerodoc\\
       \multirow{2}{2.8cm}{\bf \large LAOG}
      &
      &Issue : \numeroissue\\
      \cline{2-2}
       \multirow{2}{2.8cm}{\bf \large MPIfR/OAA}
      &\multirow{2}{7cm}{\bf \titredoc}
      &Date : \datedoc\\
      \cline{3-3}
      &
      &Page : \thepage~/~\pageref{LastPage}\\
      \hline
    \end{tabular}}
  \renewcommand{\@evenhead}{\@oddhead}
  \renewcommand{\@oddfoot}{}
  \renewcommand{\@evenfoot}{}
  }
\renewcommand{\ps@amberfront}{
  \renewcommand{\@oddhead}{\tabcolsep=10pt
    \renewcommand{\arraystretch}{1.}
    \begin{tabular}{|c|c|p{6cm}|}
      \hline
       \multirow{2}{2.8cm}{\bf \large OCA/UNSA}
      &\multirow{2}{7cm}{\bf \Large VLT / AMBER}
      &Doc. No \numerodoc\\
       \multirow{2}{2.8cm}{\bf \large LAOG}
      &
      &Issue : \numeroissue\\
      \cline{2-2}
       \multirow{2}{2.8cm}{\bf \large MPIfR/OAA}
      &\multirow{2}{7cm}{\bf \titredoc}
      &Date : \datedoc\\
      \cline{3-3}
      &
      &Page : \thepage\\
      \hline
    \end{tabular}}
  \renewcommand{\@evenhead}{\@oddhead}
  \renewcommand{\@oddfoot}{}
  \renewcommand{\@evenfoot}{}
  }
\makeatother

\newcounter{AD}
\newcounter{RD}
\newcommand{\ADlabel}[1]{\refstepcounter{AD}AD~\theAD\label{#1}}%
\newcommand{\RDlabel}[1]{\refstepcounter{RD}RD~\theRD\label{#1}}%

\renewcommand{\numerodoc}{VLT-TRE-AMB-15830-7120}        
\renewcommand{\numeroissue}{1.2}                   
\renewcommand{\datedoc}{16/04/2008}                      
\renewcommand{\titredoc}{February 2008 ATF run report$^1$}   
\renewcommand{\preparedfunction}{AMBER Task Force\footnote{With 
    the support of F.\ Rantakyrö}}                       
\renewcommand{\preparedname}{\parbox{4.5cm}{A. Chelli, 
    G. Duvert, P. Kern, F. Malbet (chair)}}              
\renewcommand{\approvedfunction}{Principal Investigator} 
\renewcommand{\approvedname}{Romain Petrov}              
\renewcommand{\releasedfunction}{Project Manager}        
\renewcommand{\releasedname}{Pierre Antonelli}           
\renewcommand{\bottomcomment}{$^1$ With the support of F.\ Rantakyrö (ESO)}

\newcounter{task}
\newcounter{taskmem}
\newcommand{\taskprefix}{}
                 {\list{\bf [Task~\taskprefix\arabic{task}]}%
                       {\usecounter{task}
                        \setcounter{task}{\thetaskmem}
                        \setlength{\rightmargin}{\leftmargin}}}%
                 {\setcounter{taskmem}{\thetask}
                  \endlist}

\begin{document}
\doctitlepage

\clearpage

\section*{Change record}

\tabcolsep=10pt
\renewcommand{\arraystretch}{2.}
\noindent
\begin{tabular}{|p{2cm}|p{2.6cm}|p{4cm}|p{6.5cm}|}
\hline
Issue&Date&Update sections&Reason / remarks\\
\hline\hline
0.1-0.5 &
\begin{tabular}{@{}l}
14/02/2008-\\*[-1em]
21/03/2008\\ 
\end{tabular}
 &all &drafts\\
1.0     &24/03/2008 &all &first released version\\
1.1     &14/04/2008 &~\ref{app:spectral-shifts}&Added new material related to spectral displacement\\
1.2     &16/04/2008 &~\ref{app:largep2vm} &Added recommendation, etc. on full-frame P2VM characterisation\\
\hline
\end{tabular}

\renewcommand{\arraystretch}{1.}
\clearpage
\tableofcontents
\addcontentsline{toc}{section}{CONTENTS}
\clearpage
\section*{Acronyms and abbreviations}
\addcontentsline{toc}{section}{ACRONYMS AND ABBREVIATIONS}
 \begin{supertabular}{ll}
  AD           &Applicable document\\
  ADC          &Atmospheric dispersion compensation\\
  AGN          &Active galactic Nucleus\\
  AIT          &Assembly, Integration and Tests\\
  AMBER        &Astronomical Multi-BEam Recombiner\\
  AO           &Adaptive optics\\
  AT           &Auxiliary telescope (1.8m)\\
  ATF          &AMBER task force\\
  BCD          &Beam commuting device\\
  COM          &Commissioning run\\
  DDL          &Differential delay line\\
  DIT          &Detector integration time\\
  DL           &Delay line\\
  EGP          &Extrasolar Giant Planet\\
  ETC          &Exposure Time Calculator\\
  FT           &Fringe tracking\\
  ITF          &Interferometry Task Force\\
  HR           &High resolution (12000)\\
  LR           &Low resolution (35)\\
  MR           &Medium resolution (1000)\\
  OB           &Observation block\\
  OPD          &Optical path difference\\
  P2VM         &Pixel to visibility matrix\\
  PDR          &Preliminary Design Review\\
  PPRS         &Paranal Problem Report System\\
  PRIMA        &Phase reference imaging and micro-arcsecond astrometry\\
  REF          &Reference documents\\
  SNR          &Signal-to-noise ratio\\
  SOB          &Sequence of Observation Blocks\\
  SOW          &Statement of Work\\
  SS           &Star separator\\
  TBC          &To be confirmed\\
  TBD          &To be defined\\
  TEC          &Technical run\\
  UT           &Unit telescope (8m)\\
  VIMA         &VLTI Main Array (array of 4 UTs)\\
  VISA         &VLTI Sub Array (array of ATs)\\
  VLT          &Very Large Telescope\\
  VLTI         &Very Large Telescope Interferometer\\
  YSO          & Young Stellar Object\\
  ZOPD         &Zero optical path difference\\
\end{supertabular}
\clearpage

\pagestyle{ambermain}
\pagenumbering{arabic}
\setcounter{page}{1}
\setlength{\parskip}{1ex}

\clearpage
\section{Introduction}
\label{sect:intro}

\subsection{Scope of the report}
\label{sec:scope-document}

The AMBER Task Force described in document [RD
\ref{rd:atfplan}] the objectives, its methodology and its plan to
investigate the various AMBER issues which prevent the instrument to
fulfill its initial specifications and therefore its original science
program.

The objectives of the February run was mainly to bring AMBER (see
AMB-ATF-001 memo) into contractual specifications the accuracy of the
absolute visibility, of the differential and of the closure phase
through a fundamental analysis of the instrument status and
limitations.

\subsection{Outline of the report}
\label{sec:outline-report}

Before the run, a new implementation of the AMBER software by A.~Chelli
and G.Duvert has been designed. This new and more accurate software
using the same philosophy as the \texttt{amdlib}
v2.1\footnote{\texttt{amdlib} v2.1 is the last public release of the
  AMBER data reduction software.} is described in Sect.\
\ref{sec:dataproc}.  The first days of the run were dedicated to the
alignment of AMBER and characterization of its behavior. Many issues
were tackled and the results are reported in Sect.\ \ref{sec:hardware}.
Then we focused our attention on the main objective of our run: the
performances limitations due to phase beating and to the lack of
absolute calibration reported in Sect.\ \ref{sec:performances}.  We end
our report with the AMBER Task Force recommendations given in Sect.\
\ref{sec:recommendations}.

\emph{Note: we tried to keep the main part of the report as concise as possible
  and we moved the details in annexes.}

\subsection{Documents}
\subsubsection{Applicable documents}
\noindent
\begin{tabular}{p{6ex}p{0.45\textwidth}p{0.37\textwidth}}
Code &Title &Number\\
\hline
\ADlabel{ad:techspec} &AMBER Technical Specifications
     &VLT-SPE-ESO-15830-2074 issue 1.0 \\
     &&dated 20.04.2000\\
\hline
\end{tabular}

\subsubsection{Reference documents}
\noindent
\begin{tabular}{p{6ex}p{0.45\textwidth}p{0.37\textwidth}}
  Code &Title &Number\\
  \hline
  \RDlabel{rd:atfplan} &AMBER Task Force: Objectives, Methodology, Plan
  &VLT-PLA-AMB-15830-7003 issue 1.0\\
  &&dated 13/12/2008\\
  \RDlabel{rd:amb-atf-001} &ATF detailed plan for Feb'08 run 
  &AMB-ATF-001 issue 1.5 dated 28/01/2008\\
  \RDlabel{rd:amb-igr-018} &AMBER data processing in the Image space
  &AMB-IGR-018 issue 1.0 dated 26/10/2000\\
  \RDlabel{rd:amb-det-007} &Report of the AMBER detector intervention
  from September 10 to 19, 2007 
  &AMB-DET-007 issue 1.0 dated 28/09/2007\\
  \RDlabel{rd:OPM-LLS} &OPM Warm Optics Design Report
  &VLT-TRE-AMB-15830-1001 issue  2.2\\
  &&dated 28/05/2001\\
  \RDlabel{rd:OPM-TR} &OPM Warm Optics Test Report
  &VLT-TRE-AMB-15830-1010 issue 1.0\\
  &&dated 10/10/2003\\
  \hline
\end{tabular}
\bigskip

\section{Data processing analysis}
\label{sec:dataproc}

In order to contribute to a better understanding of the AMBER/VLTI
system, we have developed a software to model the experimental P2VM,
and our own data reduction software to have a full control of each
step of the data reduction process which was not possible using the
actual \texttt{amdlib} package written in C language. The detailed
implementation is described in appendix \ref{app:dataproc}.

Although we have not been able to test all modes of AMBER with the ATF
software, we think that it brings significant improvements with a
visible impact on the final sensitivity limit. The P2VM modeling
allows to characterize fully the instrument performance and should be
used as a health check test. To achieve the best possible performance
with the P2VM scheme, we propose that the P2VM should be characterized
over all the camera, not on the 32 pixels wide strips of the
observation (see app.~\ref{app:largep2vm}).

\section{Hardware system analysis}
\label{sec:hardware}

The first days of the run were dedicated to the alignment of AMBER and
characterization of its behavior. We summarizes below the various issues
which we have been facing.

\subsection{Detector quality}
\label{sec:detector-quality}

The result of the work on the detector quality in September is found
excellent. No more detector fringes, very low number of bad pixels. The
reader is referred to the report on the AMBER detector intervention in
September 2007, by U.Beckmann (see [RD\ref{rd:amb-det-007}] for
details).

\subsection{Optical alignment}
\label{sec:optical-alignment}

\subsubsection{Focus and tip-tilt alignment of the detector and the spectrograph}

After the intervention on SPG, and the complete cooling of the whole
SPG/DET dewar, a proper alignment was performed between the
detector (DET) and the spectrograph (SPG) to provide a well focused
image of the input slit on the detector and to align it along a the
pixel line. The full alignment procedure is provided in appendix
\ref{app:optical-alignment} and will be inserted in the suitable AMBER
documentation.

\subsubsection{Alignment of the pupil and adjustment of the focalizer}

All pupils were properly adjusted in order to match the cold pupil mask
and the beam position launched by all the H and K dichroics and the J
mirrors. It corresponds to optimized values of the flux while inserting
the cold masks (\texttt{3T\_K} and \texttt{3T\_JHK}). After the
operation, all the translation stages of the mirror or dichroic support
were no longer at the limits of their possible strokes.  During the same
operation we properly aligned the last parabolic mirror (focalizer). The
warm optics adjustment improved significantly the image quality. All
0$^{\rm th}$ order images in J, H, and K are smaller than $25\times2.5$
pixels and the image angle is $180^\circ\,^{+0.5}_{- 2}$ (see appendix
\ref{app:optical-alignment} for details).

We recommend that this focalizing mirror must not be used for beam
adjustment on the spectrograph slit, but the periscope mirrors instead.

\subsubsection{Optical ghosts}
\label{sec:optical-ghosts}

Appendix \ref{app:char-ghosts} shows that in the data processing a major
problem came from a mismatch of the photometric calibration. We analyzed
the historical data acquired prior to the Feb 2008 ATF run as well as new data
taken during the run.

We found that there is a residual ghost at a 6\% level in the
interferometric channel.  The remaining ghost is most probably due to
internal reflexion inside the spectrograph dewar and it cannot be
removed by any alignment of the warm optics. The effect of the ghost can
be fully corrected by the software described in
Sect.~\ref{sec:dataproc} and Appendix~\ref{app:dataproc}.

We recommend to implement the compensation of the interferometric ghost
effects in the \texttt{amdlib} library and to identify some quality
checks for the P2VM calibration.

\subsection{Optical stability}
\label{sec:optical-stability}

The stability of the instrument has been an issue for a long
time. Therefore we tried to characterize this stability during the ATF run.

\subsubsection{Stability of output beams}
\label{sec:output-beam-stab}

We have taken daily the position of the beams on the detector with no
spectral resolution in order to detect possible drifts.  During this
survey, no significant drifts were observed within a time scale of a few
days.  During all this period, we did not perform any adjustment of the
optics located after the fiber outputs in the laboratory.  Day by day
survey of the beam positions and beam fluxes are reported in appendix
\ref{app:optical-stability}. Less than 10\% contrast losses requires
overlaping at the level of 5 pixels in X and of 0.6 pixels in Y. Figure
\ref{fig:optical-stab} in appendix \ref{app:optical-stability} shows
that during the ATF run the beams were always within specifications. Our
understanding is that the effort done by Paranal to achieve a better
stability of AMBER by cutting and strengthening the optomechanical
elements were fruitful.

I the same appendix \ref{app:optical-stability}, we have reported an
experiment to see if the output beams were subject to vibrations and the
definitive answer is negative.

We recommend to relax the alignment requirements, in order to avoid too
frequent interventions in the laboratory, and implement the strategy
described in Appendix~\ref{app:optical-stability},
section~\ref{sec:alignment-health-check}

\subsubsection{Optical path stability}
\label{sec:opd-stability}

The OPD is not stable especially after a major adjustment (several tens
of microns). Relaxation time is about 1 day also possibly due to the
large CAU mirror (see below). In Sect.~\ref{sec:p2vm-stability} of
appendix~\ref{app:optical-stability}, we show that 2 P2VMs taken
respectively at the beginning of the night and this end displays same basic
features except for the phase on one baseline.

\subsubsection{CAU stability}

It has been observed that the large parabolic mirror of the CAU present
significant instabilities. After tuning operations on the AMBER table,
fluctuations of the OPD are observed for hours, and stable situation is
attained again after one day. These instabilities may reach up to 100 nm
on 1-3 baseline. These fluctuations are directly observed under slight
knocking of the mirror support (see Fig.~\ref{fig:cau-bang}), and are
present each time a noise source happens on the AMBER table (BCD
movements, CAU motor movements certainly, large piezo movements...). We
suspect that photometric noise (due to changes in injection in the
fibers) is associated with each vibration of this mirror.

\subsubsection{Recommendations}

We recommend to avoid any too frequent realignment of the output optics
which appears not necessary at all. For the OPD, normally if the
translation stages of the input dichroics are not used, they should
reach their relaxation position. However, one could implement the P2VM
calibration which allow to not depend on the actual OPD. Finally, in
order to improve the calibration process, solutions to fix the large CAU
parabolic mirror must be found.

\subsection{Spectral dispersion}
\label{sec:spectral-dispersion}

\subsubsection{Grating position}
\label{sec:grating}

The grating motor seems to behave in pre-summer 2007 conditions. It
losses steps, shifting globally the images of the camera by a few
pixels (less than 20) . This does not preclude operation since the
camera sub-windows can in general be \emph{positioned according to this
shift}. 

However the software developed in Summer 2007 years to measure the shift
and recenter the images on predefined positions on the camera, results
in too many calls to the rotation of the grating.

We recommend to update the OS and templates in order to avoid moving the
grating more than once per configuration.

\emph{NOTE: a recent PPRS, No.~026605 reports that the MACCON board was
  defective and the change of the board improved a
  lot the reproducibility.}

\subsubsection{Spectral shifts}
\label{sec:spectral-shifts}

The spectrograph is well-aligned. The spectral shifts between
photometric channels and the interferometric ones is characterized and
should not be recomputed each time (see
appendix~\ref{app:spectral-resolution}) for details.).

The spectral shift is primarily due the combination of the position of
the slit (large, narrow, etc) and of a small angle between the optical
components that fold the 3 photometric beams on the camera (3 glued
total-reflexion prisms).  At the worst, this combination changes only by
a small amount every time the spectrograph is opened.

We recommend to stop using the spectral calibration procedure and use
the fixed shifts.

\subsubsection{Spectral calibration of the low resolution JHK mode}

We performed a Fourier Transform Spectrograph analysis of AMBER Low
Resolution JHK spectra by moving incrementally by steps of 0.01
microns the piezos on beams 1 and 3 for each band (see
appendix~\ref{app:spectral-resolution}). Each pixel of the 
Interferometric region on the camera of AMBER (32 by 60 pixels in this
LR mode) is thus modulated with a period equal to the wavelength it
sees.  The Dispersion law is found compatible with a linear dispersion
of coefficients, $\lambda=2.6917-3.2082\cdot10^{-2}\times\mathrm{i}$,
with i as the pixel number.

\subsection{Internal calibration source}
\label{sec:CAU}

For tests, the RAS (Remote Alignment Source) was replaced by a
laboratory source directly connected to the K band fiber of the CAU. In
this condition, while the K band dichroic of the RAS is bypassed, the
modulation of the flux in all photometric channels is significantly
reduced and is not detectable any more in the raw data in MR (see
appendix \ref{app:internal-sources}). LR image in JHK in this condition
allows a detection for all bands

The two CAU fibers suffer from incorrect superposition, since the JH fiber
seems to be moving and requires some readjustments. This misalignment
lead to non-zero phase closure measurement on the CAU, with an unstable
value (see below Sect.~\ref{sec:clos-phase-corr}).  

We recommend to remove the CAU dichroic and to use only one fiber as the
one for the CAU bypassing the 2 dichroics of the calibrations sources (RAS
and CAU).

\section{Performance limitations}
\label{sec:performances}

Our main goal for the AMBER Task Force run was to find the origin of the
performance limitations of AMBER. We have performed various experiences
until we found the principal origin of these limitations: the
polarizers.

\subsection{Closure phase corruption}
\label{sec:clos-phase-corr}

Previous data set showed that the closure phase
signal was corrupted. We realize that by shutting down the remote
alignment source (RAS) JH fiber, the K spectrum becomes very
different. Residual K band light the RAS JH fiber polluted the K
source. Therefore the instrument was seeing a binary made by the
strong K fiber and the residual K light going through the JH fiber. By
shutting down the RAS light leads to a closure of 0° all over the K
band (see Fig.~\ref{fig:RAS-CP} in appendix \ref{app:internal-sources}).

\subsection{Phase beating in differential phase}
\label{sec:phase-beating}

\begin{figure}[t]
  \centering
  \includegraphics[angle=270,width=0.49\hsize]{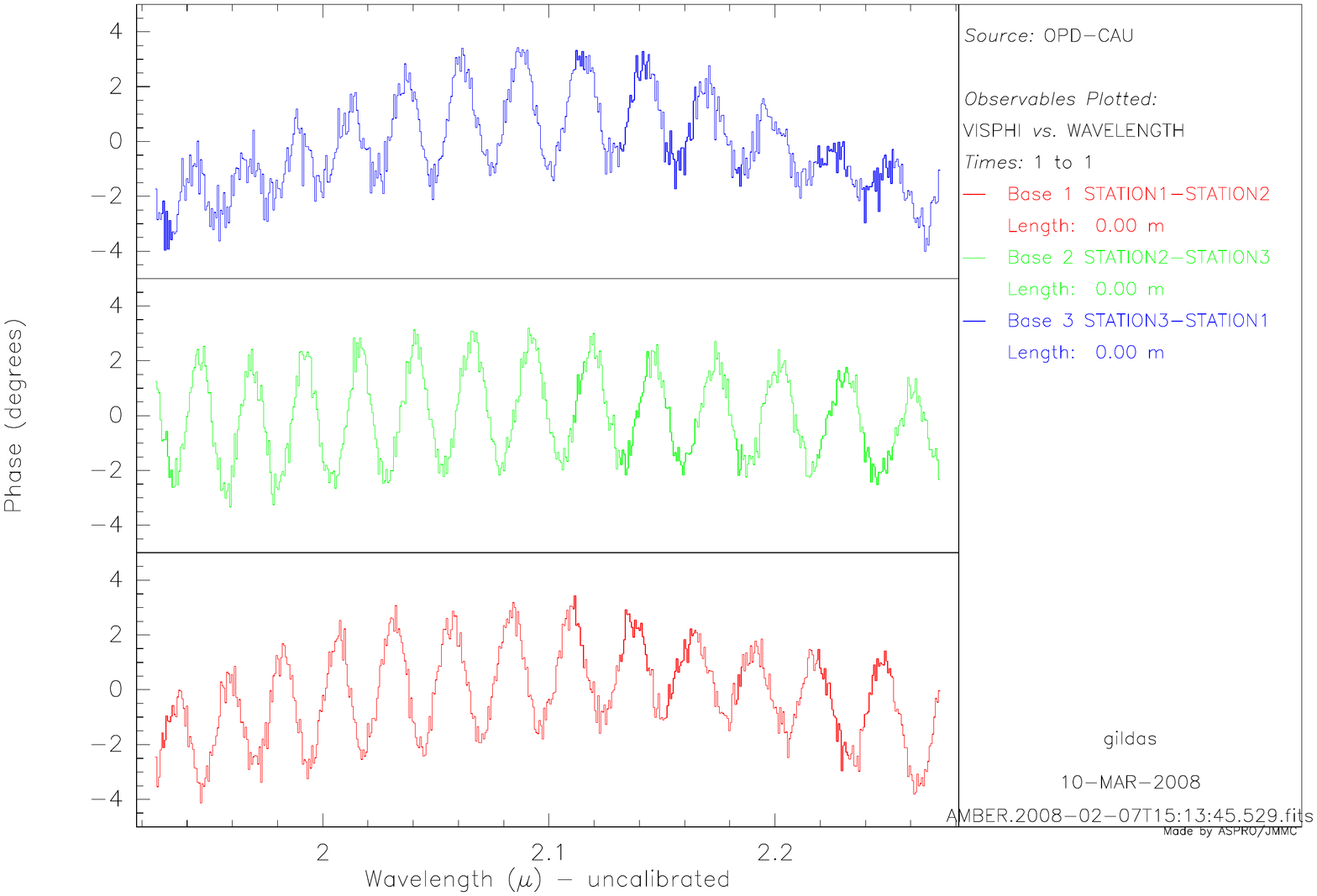}
  \hfill
  \includegraphics[angle=270,width=0.49\hsize]{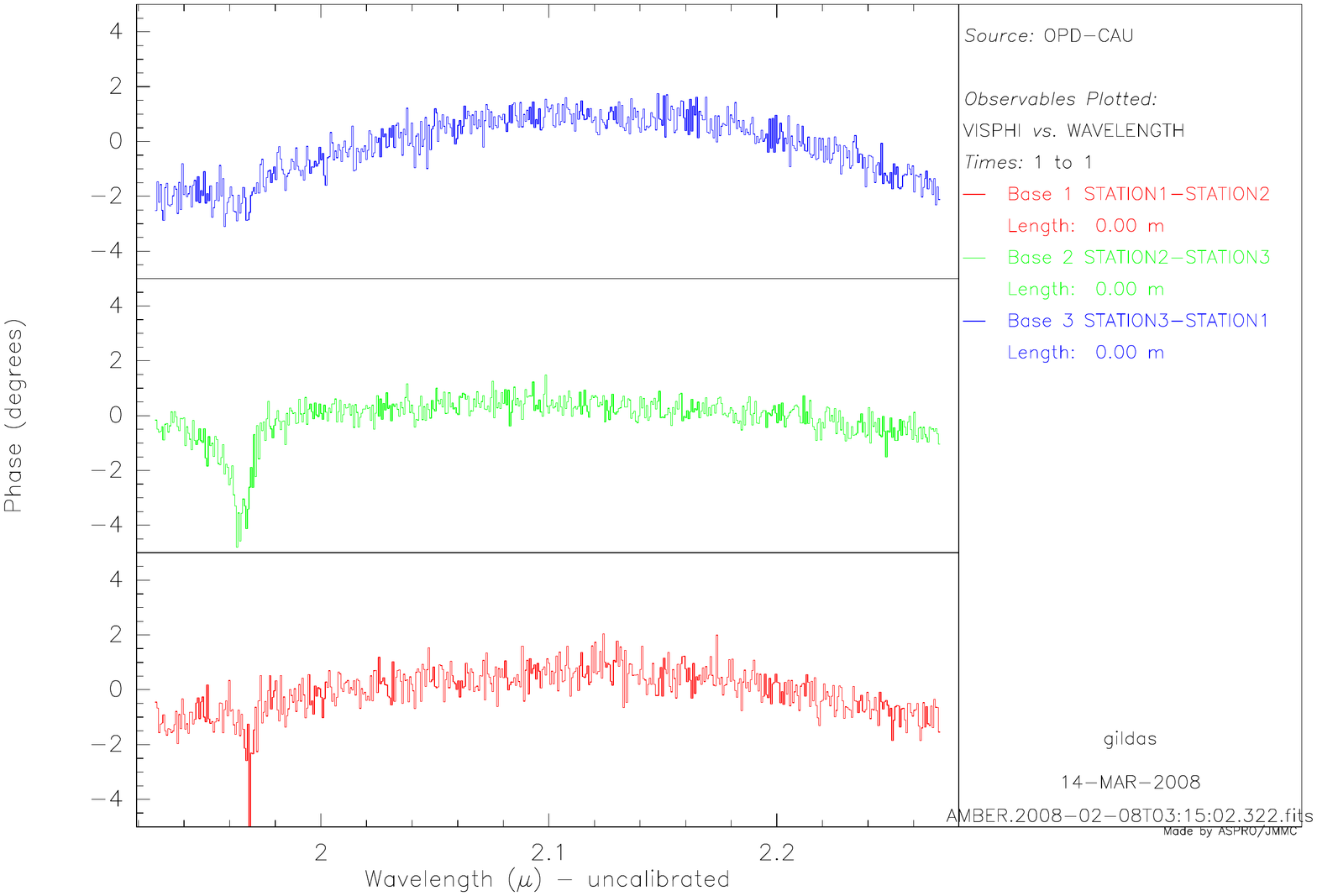}
  \caption{Differential phases measured on the internal source. Left
    figures were taken with the current set up using the polarizers
    whereas right figures were taken with the POL taken out. The feature
    at $1.97\mu$ is due to a strong absorption/polarisation of the RAS
    dichroic}
   \label{fig:phase-beating}
\end{figure}
Is called phase beating a periodic oscillation of the phase as
illustrated in the left part of Fig.~\ref{fig:phase-beating}. We did
intensive tests to find the origin of the phase beating:
\begin{itemize}
\item \textbf{Crossing the fibers.} No improvement was found. Crossing
  does not change injection significantly, but adds a 7.5\,mm
  piston. Although fringes are still visible in HR, there is no significant
  improvement on phase beating.
\item \textbf{Removing the polarizers.} The result was quite impressive:
  \begin{list}{--}{}
  \item no more fringe beating (see Fig.~\ref{fig:phase-beating});
  \item lower fringe contrast (see Fig.~\ref{fig:amber-without-pol}) by
    half which means that polarizers help 
    to increase the contrast; 
  \item the flux in K is much higher and the spectrum is very different
    with no longer stripes;
  \end{list}
\item \textbf{One of the polarizer inserted after the fiber
    outputs}. The polarizer (POL3) placed  at the output of the
  fibers close to the periscope in collimated beams, led to a much higher
  contrast: 0.9 0.9 0.85 for the different baselines.
\item \textbf{POL3 inserted just before the spectrograph entrance}. We
  put POL3 just before the spectrograph in a convergent beam. It was not
  successful because although there was a certain gain in contrast, it
  also generated a lot of unstable flux modulation.
\end{itemize}

The most probable explanation is a Fabry-Pérot effect (described in appendix
\ref{app:FP}). AMBER documentation provides a description
of the polarizers which is recalled in appendix \ref{app:POL}. The
polarizers are made of two prisms separated by an air blade whose
thickness is about 120 microns. These air blades can induce a
fringe beating in a more complex way than a single Fabry-Pérot effect,
because it is not the same in the different arms and leads to
additional beating fluctuations both in phase and in amplitude (see for
example Fig.~\ref{fig:amber-with-pol}).

This FP effect cannot be calibrated by the BCD because the fringe shape change too
rapidly probably due to the change of the Fabry-Pérot thickness.

We recommend therefore to use other types of polarizers with no parallel
optical faces and to use them in a common collimated beams after the
fibers. 
 
\subsection{Fluctuation of the contrast}
\label{sec:absol-visib}

The second effect that we investigated is the origin of the fluctuation
of the visibilities on sky. These fluctuations seems to be larger than
the ones expected because of the atmospheric turbulence. Therefore we
tried to identify the hardware on the instrument that could be at the
origin of these fluctuations (see log book in
appendix~\ref{app:logbook}). 

We did several experiments:
\begin{itemize}
\item \textbf{Piezos shut down}. To check that the piezos were not
  generating any vibrations, we shut them down. This was not easy, since
  when plugged off the fiber head are shifted by half stroke (about 40
  microns). The input dichroics had to be used to compensate for the
  differential OPDs. Figs.~\ref{fig:piezos-on} and \ref{fig:piezos-off}
  illustrate that the visibility fluctuations are still present and even
  reinforced due to the OPD instability (cf. Sect.~\ref{sec:CAU}): it
  could reach 10 to 15\%...
\item \textbf{Checking the beam overlap on the detector.} One possible
  explanation could be that due to vibrations the beam overlap is
  changing with time on the detector. A data set taken with no spectral
  resolution (see Fig.~\ref{fig:optical-fast}) showed that the beam
  positions in X and Y but also in flux do not change significantly with
  time or at a level much less than 1\%.
\item \textbf{Data recording with different DIT.} Using different
  Detector Integration Time (DIT), the fluctuations kept the same
  strength therefore rejecting the possibility to have fluctuations due
  to vibrations.
\item \textbf{Using the beam commuting device (BCD).} No special effect
  found and also the fluctuations are not reproductive.
\item \textbf{Removing the polarizers.} Data taken with the polarizers
  and without them (see Figs.~\ref{fig:amber-with-pol} and
  \ref{fig:amber-without-pol} in appendix \ref{app:POL}) show clearly
  that the phase beating can be very important and most of all variable!
  The remaining variability may come from an additional effect on the
  internal source (see Sect.~\ref{sec:opd-stability}).
\item \textbf{Injection into fibers.} The new piezo-controlled 'Iris
  Fast Guiding' (hereafter IFG) positioning of the AMBER entrance beams
  and the availability of the AT beacons were used to test whether the
  injection of the beam creates or increases the ``phase beating''
  effect. No effects are detectable when the polarizer was removed.  In
  the presence of the polarisers, however, amplitude effects are present
  and vary in shape with the injection. The small changes in the
  positioning of the beams demonstrates how sensitive the experiment is
  to the differential Fabry-P\'erot effect induced by the polarizers.
\item \textbf{CAU optical path stability.} One possibility for the
  fluctuations present even without the polarizers is an instability in
  the CAU subsystem. Monitoring the AMBER observables when touching
  various mechanical supports of the CAU showed that the large parabolic
  mirror of the CAU (see Fig.~\ref{fig:cau-parabolic} in Appendix
  \ref{app:optical-stability}) might be at the origin of some
  fluctuations. We stopped the investigation in lab and to check
  what would be the results on sky without using the CAU (see next section).
\end{itemize}

In conclusion, we believe that the origin of the visibility fluctuations
are the results of 2 effects: the Fabry-Perot differential effect
changing with time coupled to an instability of a large parabolic mirror
in the alignment unit.

\subsection{AMBER performance on sky}
\label{sec:expect-perf}

\begin{figure}[t]
  \centering
  \includegraphics[width=0.9\hsize]{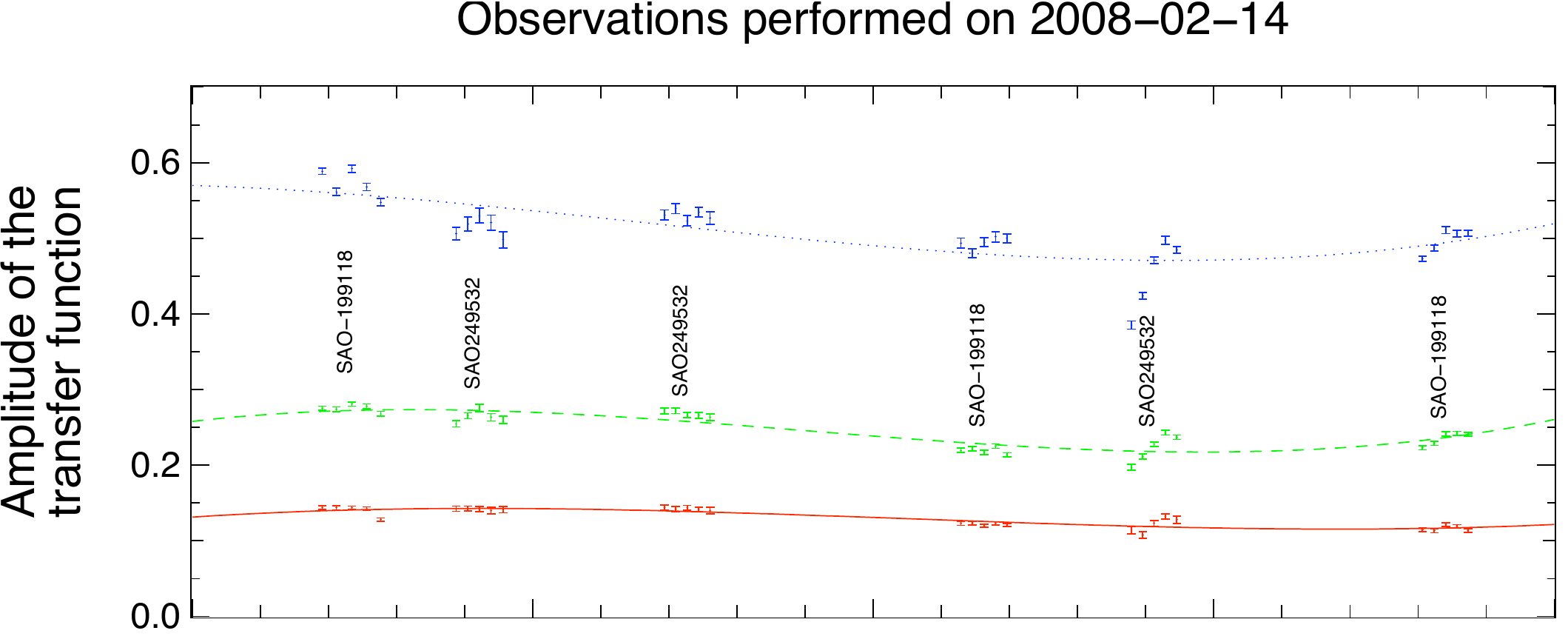}
  \caption{AMBER resulting transfer function without the polarizers.}
  \label{fig:atf-tf}
\end{figure}

It is too preliminary to conclude on the overall performances of the
instrument, but it seems that even with a reduced instrumental contrast,
AMBER limiting magnitude is similar and even better than previously
known and can measure small visibilities. The closure phase and
differential phases seems to be limited by fundamental (to be checked
with SNR calculations) noise and no longer with systematics from the
instruments.

The last night of the ATF run, we have observed repeatedly two
calibrator stars in medium resolution without the polarizers together
with another star of larger expected diameter. The resulting transfer
function is not fully stable, but can be fitted with a polynomial of order
3. The evolution is of the order of less than 5\% and can be accounted
for seeing (see appendix~\ref{app:observations}).

In addition differential phases on all baselines and closure phase are
stable all along the night with an accuracy of $5.10^{-3}$\,rad and
5\,deg respectively. 

Also AMBER has not been tested in all modes, we believe that the
polarizers and the CAU mirror were the main sources of visibility
fluctuations. We still have low resolution data with and with FINITO to
investigate in order to assess the actual performance. However the
only assessment which will be valid would be the one made with a new
polarizer to get higher contrast.



\section{Recommendations}
\label{sec:recommendations}

We recommend the AMBER consortium and the ESO responsibles:  
\begin{list}{}{}
\item (1) to take actions as soon as possible to replace the current POL unit 
with a single polarizer put in the common beam before the spectrograph 
without Fabry-Perot effect,
\item (2) to investigate the origin of the sensitivity to vibrations of the 
CAU parabolic mirror and fix it
\item (3) to simplify the calibration light sources (RAS)
\item (4) to fully characterize the instrument in its final state with a final 
commissioning run
\item (5) to correct the AMBER observing software to minimize spectral grating 
movements
\item (6) to update the observing software and data reduction pipeline with 
optimum visibility computation and quality controls.
\item (7) to take the P2VM with a much larger width than the
  observations ($\ge64$ pixels channel width).
\end{list}

\appendix
\clearpage
\
\vfill 
\begin{center}
  \textbf{\LARGE Annexes}
\end{center}
\vfill 
\

\clearpage
\section{Data processing analysis}
\label{app:dataproc}

In order to contribute to a better understanding of the AMBER/VLTI
system, we have developed an IDL software to model the experimental
P2VM, and our own data reduction software, also written in IDL
language, to have a full control of each step of the data reduction
process which was not possible using the actual \texttt{amdlib} package
written in C language.

\subsection{Characterization of the AMBER P2VM}
The P2VM is fully described by the $v_k$ parameters and by the
carrying waves $c_k$ and $d_k$ (see memo AMB-IGR-018 for the
definitions [RD~\ref{rd:amb-igr-018}], and also Tatulli et al.\ 2007, A\&A
464, 29). The $v_k$ parameters allow to compute the continuum in the
interferometric channel from the measurements of the photometric
channels. However, the presence of optical ghost (see Sect.~\ref{sec:optical-ghosts})
affects the $v_k$ calculation, and the computed continuum does not
strictly match anymore the true interferometric channel continuum.  To
solve this problem to the first order, we introduce a new way to
compute the $v_k$ parameters, which takes into account the flux in the
3 photometric channels.

The experimental P2VM is then fitted by a synthetic P2VM. The latter being
perfect by construction, the output of the P2VM modelling represents
the signature of the AMBER system from the calibration lamp to the
scientific detector. This signature can be expressed in terms of 4
observables for each spectral channel:
\begin{itemize}
\item the system visibility 
\item the spatial frequency
\item the phases of the carrying waves $c_k$ and $d_k$
\item a continuum regularization parameter
\end{itemize}
The continuum regularization parameter is introduced to take into
account an eventual residual continuum mismatch. If the $v_k$
parameters are correctly computed, then this parameter should be close
to 1, which is generally the case with the new procedure. 

\begin{figure}[p]
  \centering
  \includegraphics[width=0.85\hsize]{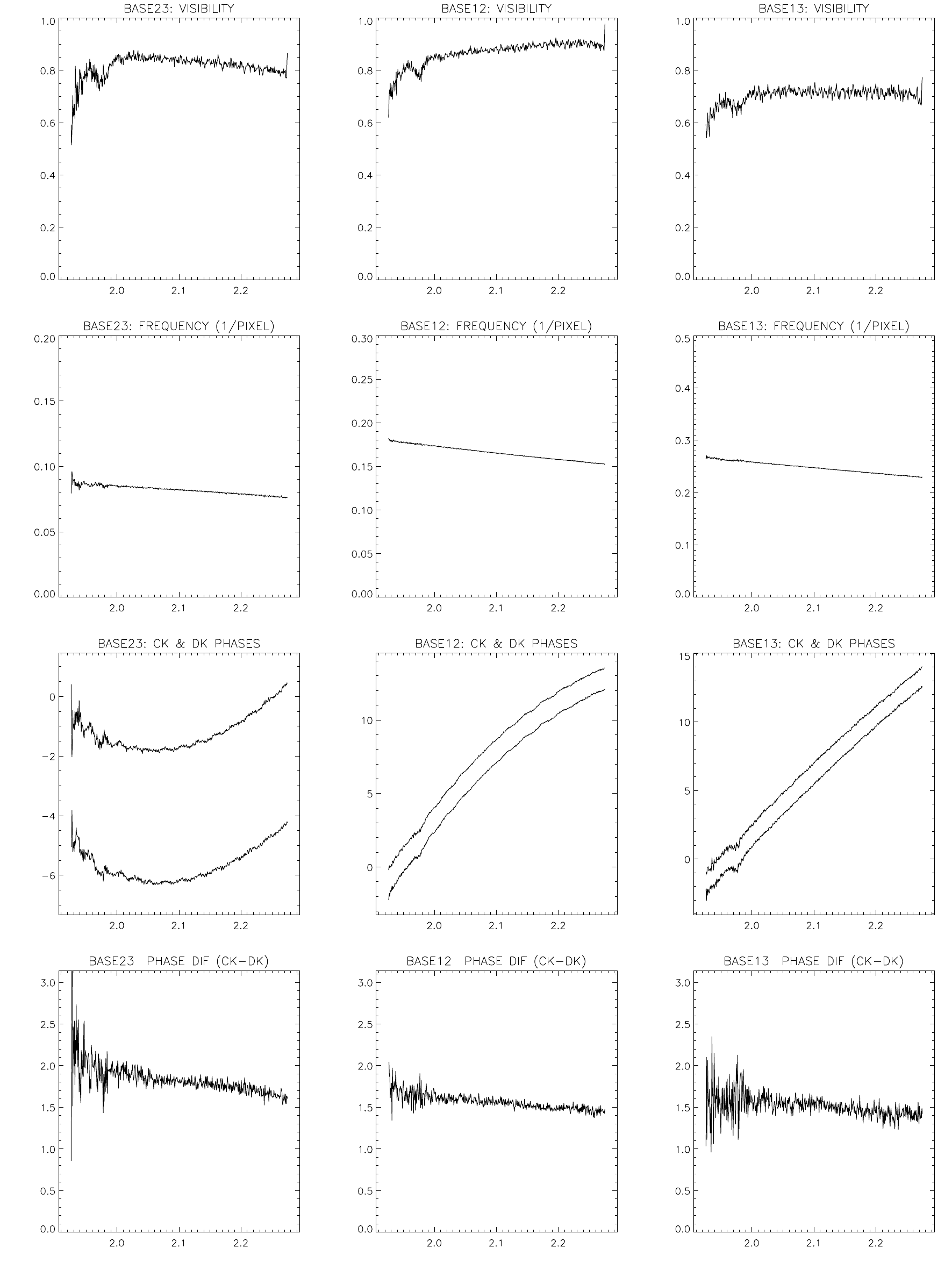}
  \caption{Characterization of the medium resolution P2VM in the K
    band at 3 telescopes. From top to bottom, as a function of the
    wavelength: P2VM visibility, frequency of the carrying waves (in
    pixel$^{-1}$), phases of the carrying waves (in radians), phase
    difference between $d_k$ and $c_k$. P2VM data set acquired for
    observations on $\alpha$ Arae obtained in Feb 2005 during SDT.}
\label{fig:ptvmmr}  
\end{figure}  
\begin{figure}[p]
  \centering
  \includegraphics[width=0.85\hsize]{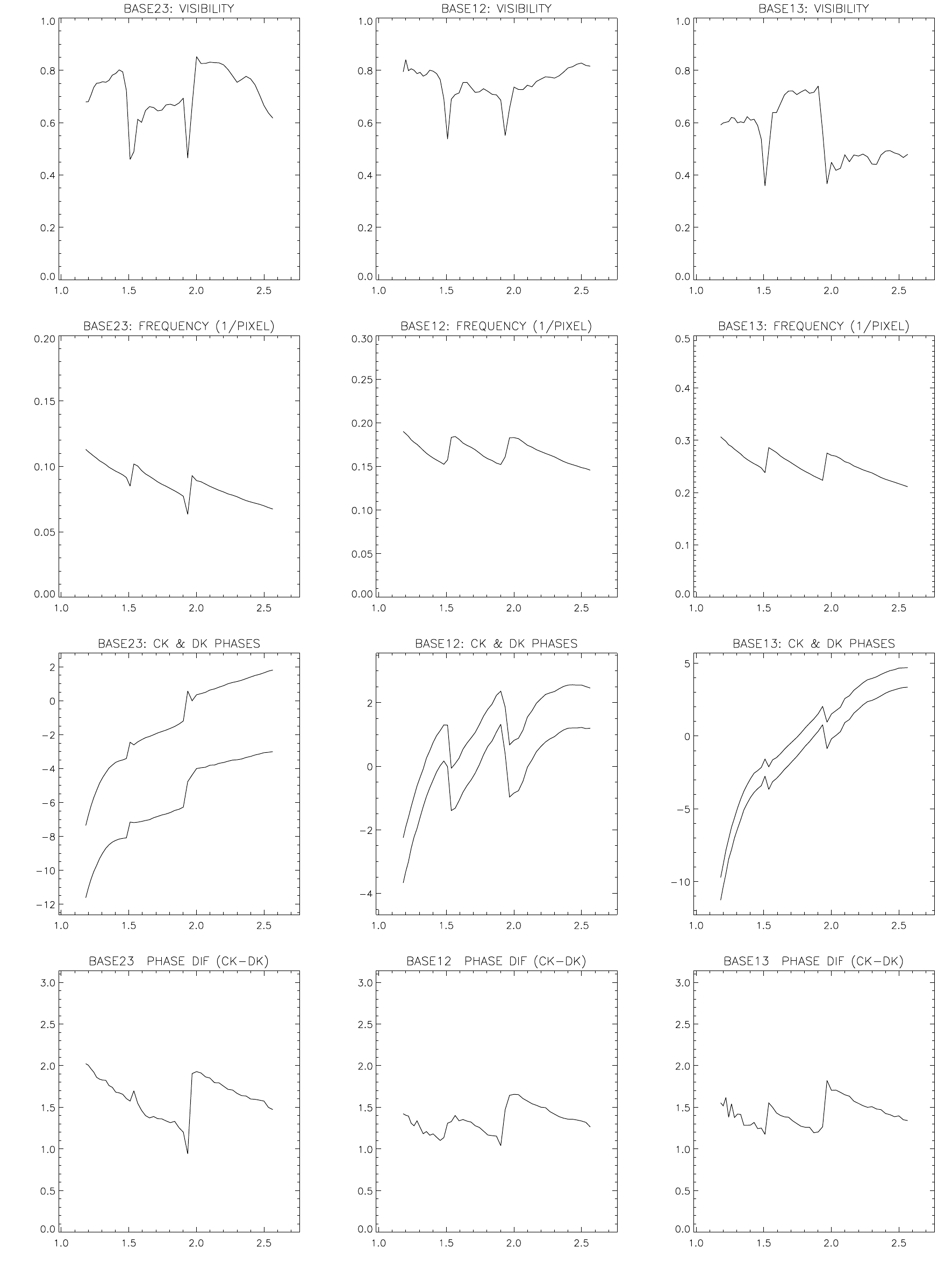}
  \caption{Characterization of the low resolution P2VM at 3 telescopes.
    From top to bottom, as a function of the wavelength: P2VM
    visibility, frequency of the carrying waves (in pixel$^{-1}$),
    phases of the carrying wave (in radians), phase difference between
    $d_k$ and $c_k$. P2VM data set acquired for private
    observations obtained in Dec 2007.}
\label{fig:ptvmlr}
\end{figure}

We present the results of the P2VM modelling for 2 sets of data.  The
first set of data is the P2VM of the Alpha Ara data test set obtained
in Feb 2005 during SDT at medium resolution in the K band. The results
are plotted in Fig.~\ref{fig:ptvmmr}, for each baseline, as a function
of the wavelength. At the top we see the system visibilities, with
a mean value of about 0.8 beyond 2$\mu m$. Below, we find the spatial
frequencies (expressed in pixel$^{-1}$) whose careful examination (at
higher magnification) shows that the spectral dispersion is not
constant, but is a linear function of the wavelength. Then, we see the
phases of the two carrying waves whose curvatures indicate that
interferometric pattern is also curved along the spectral axis. At the
bottom, the phase differences between the $c_k$ and $d_k$ are
reasonably well within the $\pi/2$ specifications.

The second set of data is a P2VM acquired on December 29th, 2007, in
low resolution mode. The results are plotted in Fig.~\ref{fig:ptvmlr}.
The low resolution P2VM visibility is in the range 0.5 to 0.8, with
large variations from one spectral band to the other.  The decrease
around 1.5 and 2 $\mu m$ is simply due to the superposition of two
fringe systems at different spatial frequencies. Also, we may observe
some structures within each band. Below, we find the 3 frequency
systems and the phases of the two carrying waves. The phase
discontinuities between the bands correspond to the observed shifts
between the 3 fringe systems. The phase differences are plotted at the
botom.
\subsection{An improved data reduction software}
The characteristics of the new data reduction software are as follows:
\begin{itemize}
\item the P2VM is computed as described in the previous section
\item a regularization continuum parameter is applied to each
  interferogram. This parameter is estimated fitting the mean interferogram
  with 7 parameters (3 complex coherent fluxes and 1 continnum
  parameter)
\item A special attention has been given to the calculation of the
  covariance matrix of the measurements. The noise resulting from the
  detector readout noise and the background is estimated from the dark
  data. The covariance matrix takes into account the correlation
  between the measurements, and as a consequence is not diagonal. This
  matrix is inverted spectral line by spectral line.
\end{itemize}

This new data reduction software has been used to reduce all the data
of the ATF run. The detail of the algorithm will be published
elsewhere. 

\begin{figure}[hb]
  \hfill
  \includegraphics[angle=270,width=0.3\hsize]{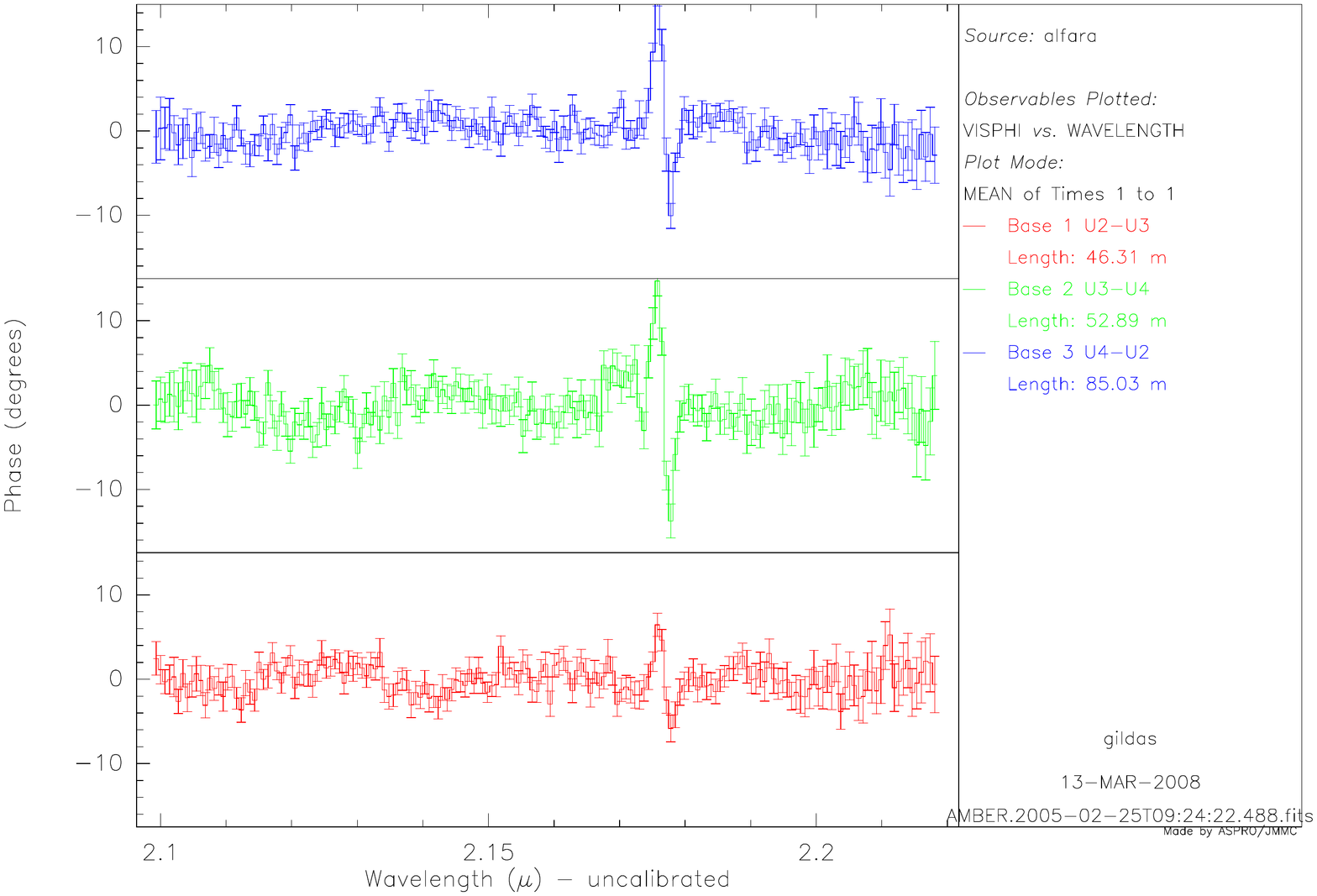}
  \hfill
  \includegraphics[angle=270,width=0.3\hsize]{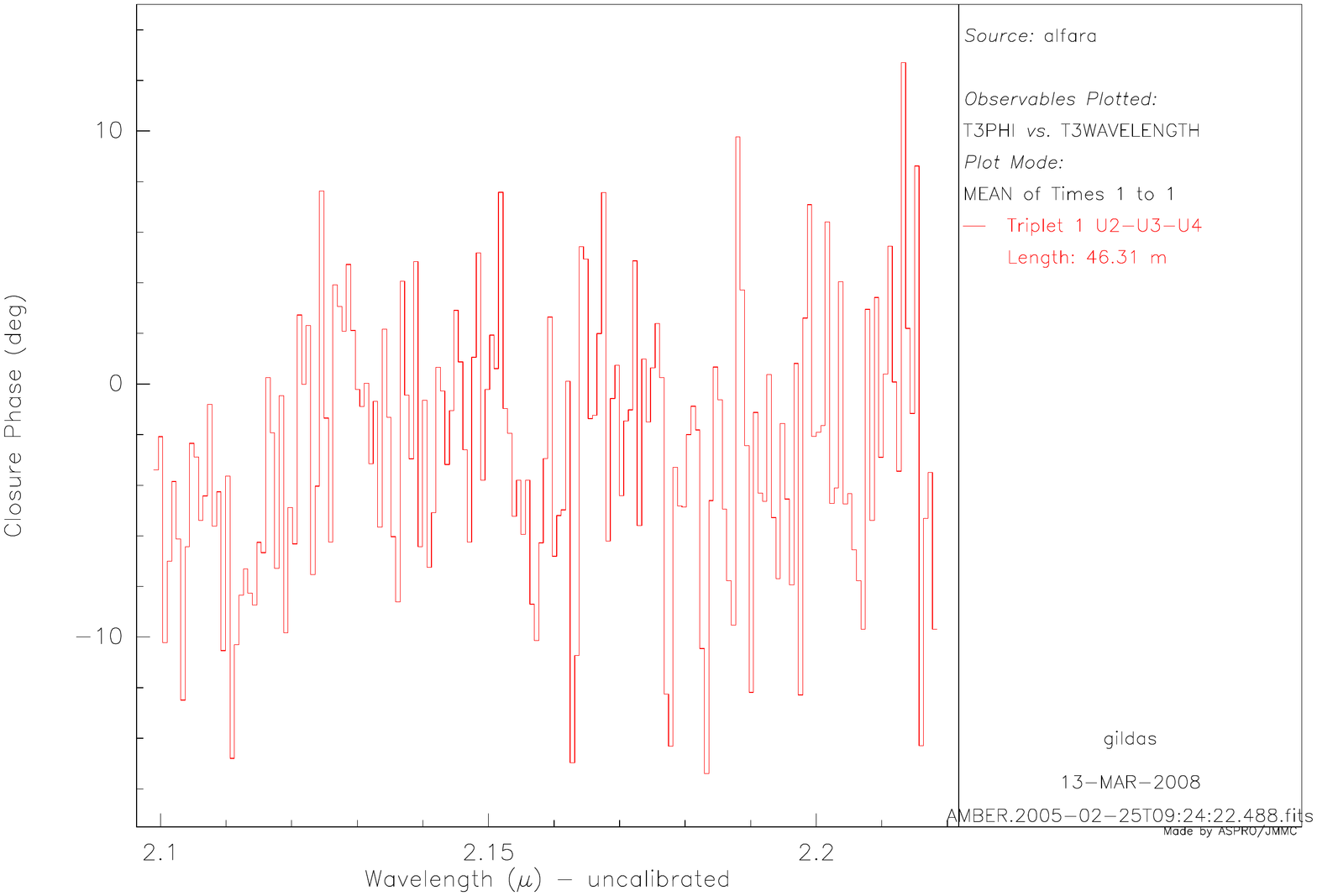}\\
  \includegraphics[angle=270,width=0.3\hsize]{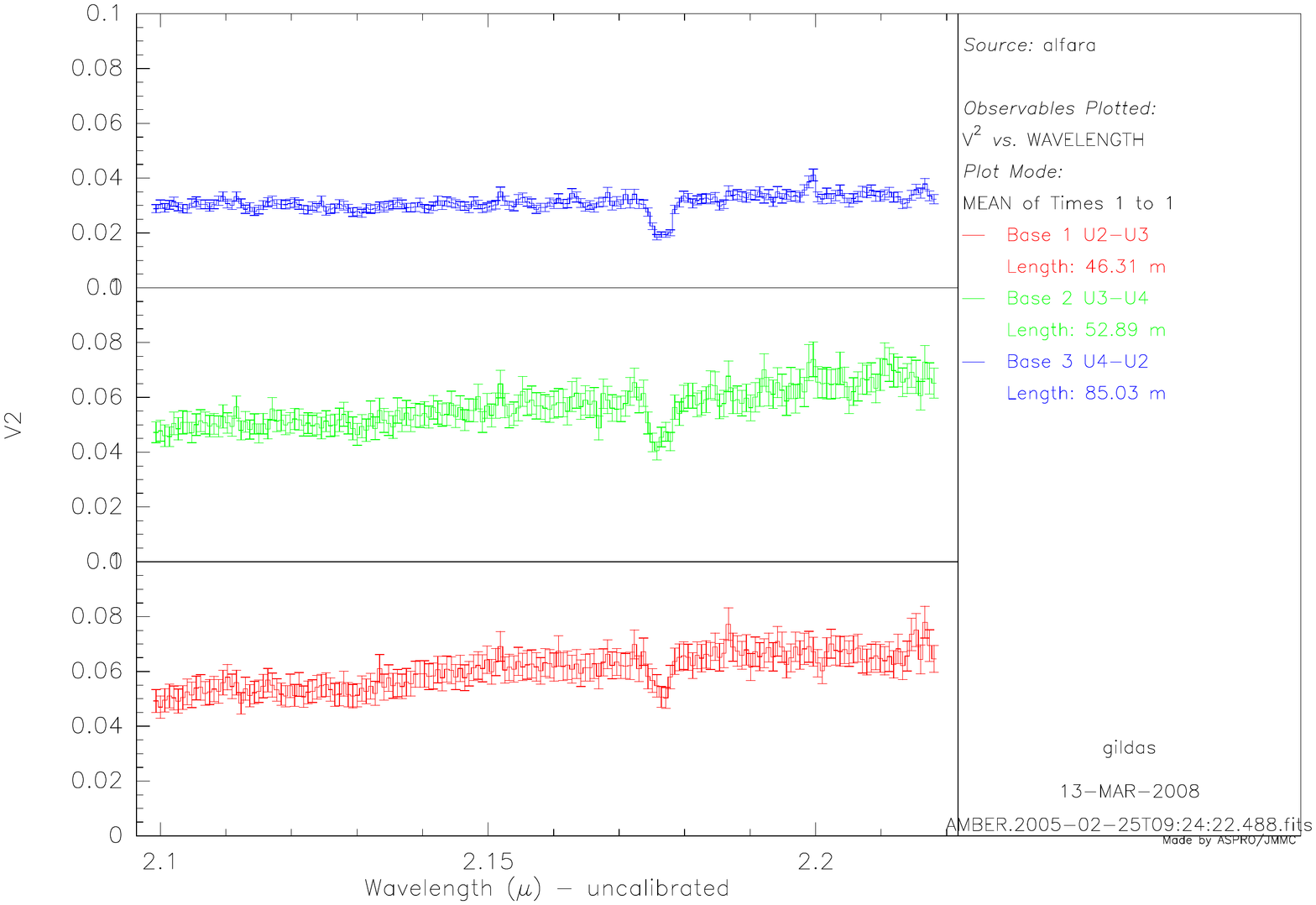}
  \hfill
  \includegraphics[angle=270,width=0.3\hsize]{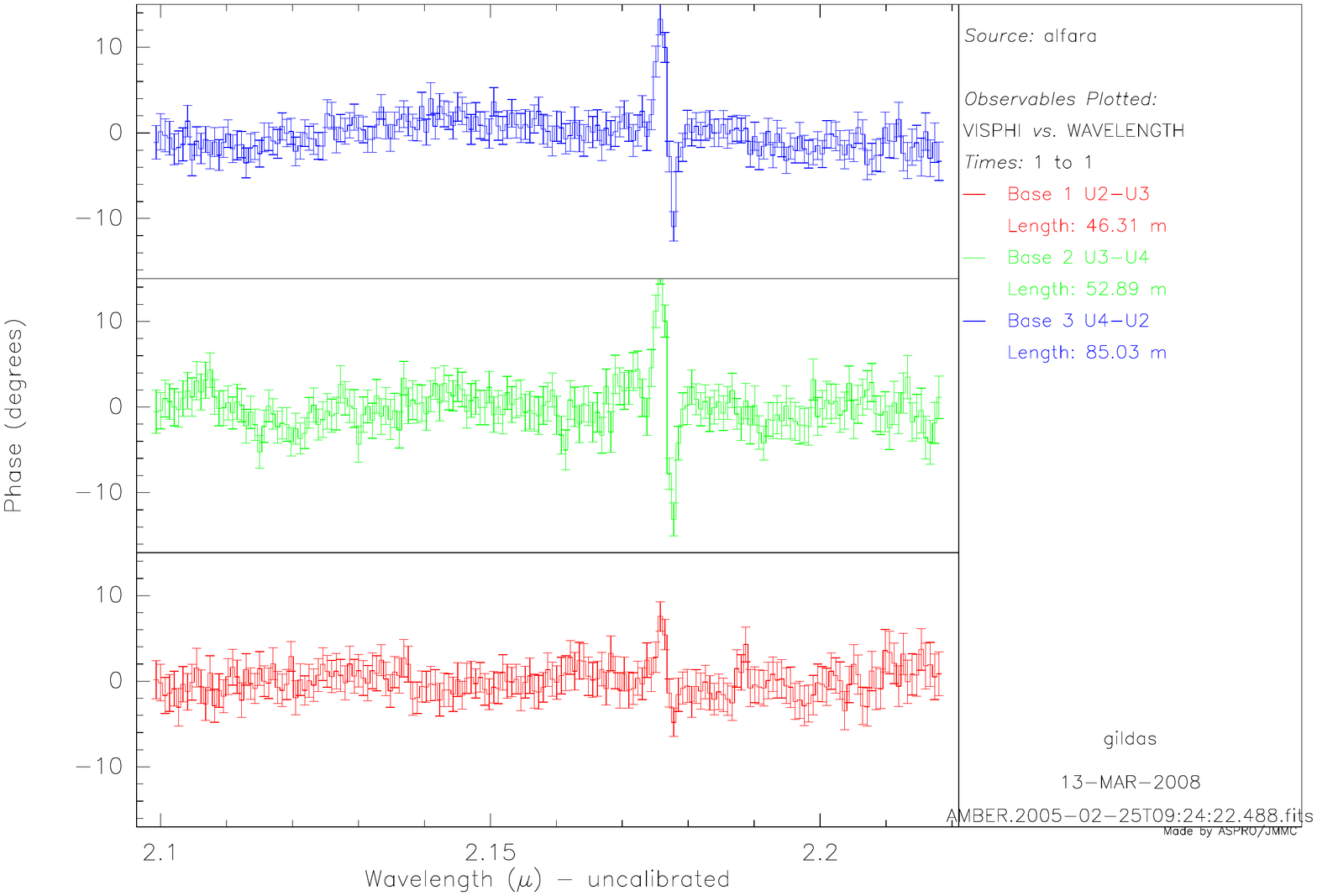}
  \hfill
  \includegraphics[angle=270,width=0.3\hsize]{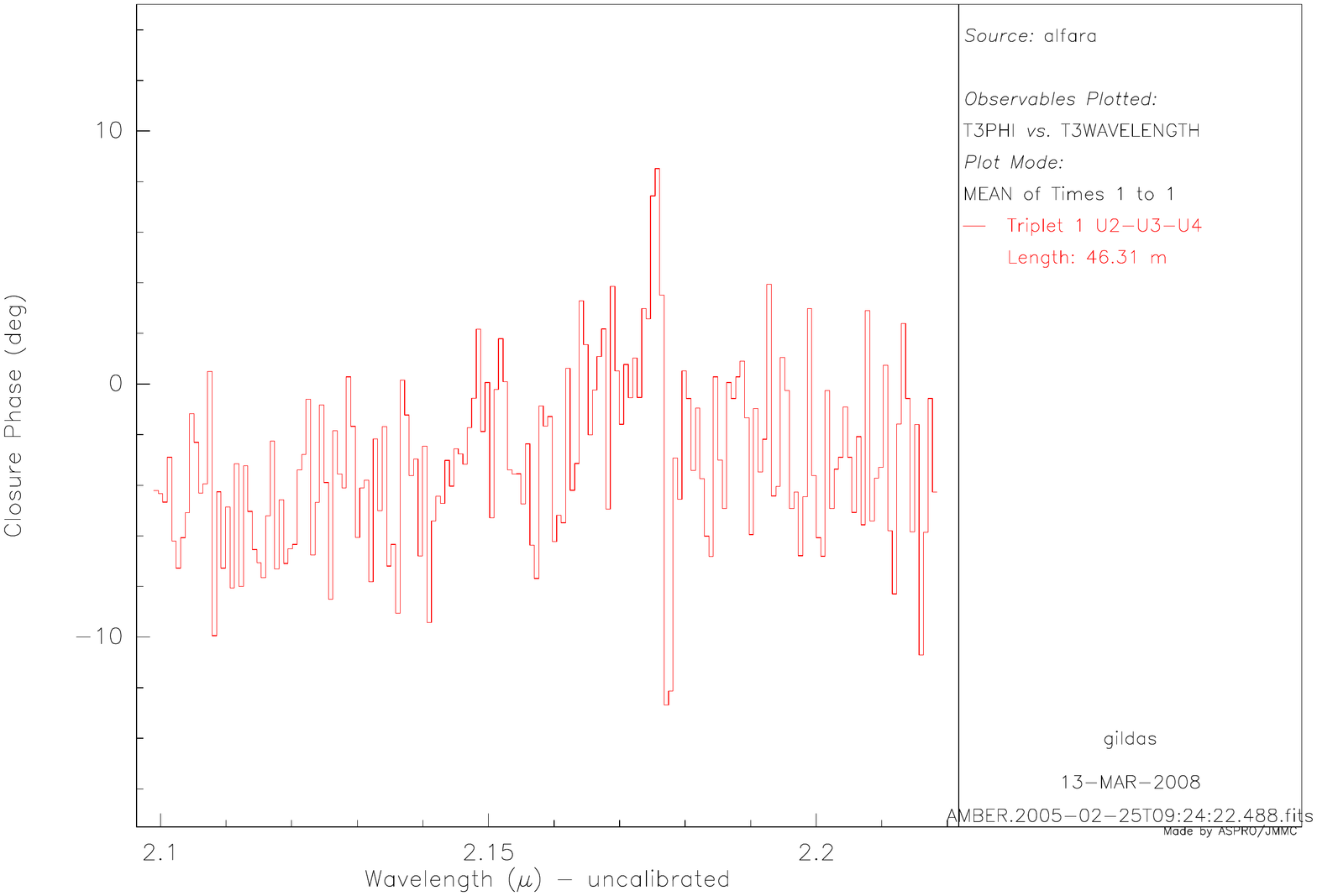}
  \caption{Comparaison of noise levels on $\alpha$\,Arae data, one
    exposure (500 frames of 70 ms). X axis is Wavelength
    (microns). From left to right: Squared Visibilities, Differential
    Phases and Closure Phase. Top panels are amdlib v2.1 results with
    standard frame selection, bottom ones use ATF method, with no
    selection. }
  \label{fig:alphara-noise-comparison}
\end{figure}

Figure \ref{fig:alphara-noise-comparison} shows the relative improvement in S/N
for an observation in the test dataset available with amdlib v2.1.  It consists
of high S/N data taken in good observing conditions in 2005, with the UTs.

The results need a few comments. This is one case where the new algorithm should
not produce markedly different results, since the S/N in each frame is strong
and statistics are dominated by the vibrations of UT+VLT presents at this
date. We forced the spectral displacement values obtained in
sect.~\ref{sec:spectral-shifts} for both data reductions, since this is the good
value to use, whereas the standard (pipeline) reduction would have used quite
different values. The prescribed standard data reduction, in particular in this
UT case where vibrations were present, calls for frame selection, however
results with the new software are obtained without frame selection (hence the
lower visibilities). The gain in S/N is partially due to this, and is more
striking in the phase closure.

\subsection{On the P2VM spatial extension}
\label{app:largep2vm}
The P2VM, as its name suggests, is the mean to convert pixels
intensities to correlated fluxes. Its spatial extension on the camera
needs not, and should not, be restricted to the 32 pixels wide strips
used for the observation.

We have checked that restricting the P2VM extension to the 32 pixels
is not harmful in terms of loss of information, since the 6
independent parameters of the 3 simultaneous complex correlated fluxes
are more than sufficently fitted on 32 independent pixel values (24
would suffice in theory).

However, we found that in the current AMBER setup where the
characterization of $C_k$s and $D_k$s depends on the phase induced by
an air blade of variable properties (temperature, humidity,
repeatability of piezo), one needs the best possible determination of
this phase. Errors on this phase measurement will convert into power
wrongly transferred between the real and imaginary parts of the
instantaneous complex coherent flux proportionally to the piston and
the piston jitter.

To minimize the errors in this phase measurement, one needs to measure
by an integral method the phase between the $C_k$s and $D_k$s with the
maximum possible spatial extension, at least 64 pixels instead of the
current 32 (TBC).

\clearpage

\section{Optical alignment}
\label{app:optical-alignment}

\subsection{Adjustment of the DET/SPG focus}

While the dewar was open for maintenances on DET and SPG, we had to
adjust properly the focus and tilt of the camera with respect to the
spectrograph.  A first alignment done during the cooling of the
spectrograph was not sufficient.

\begin{figure}[hb]
  \centering
  \includegraphics[width=0.7\hsize]{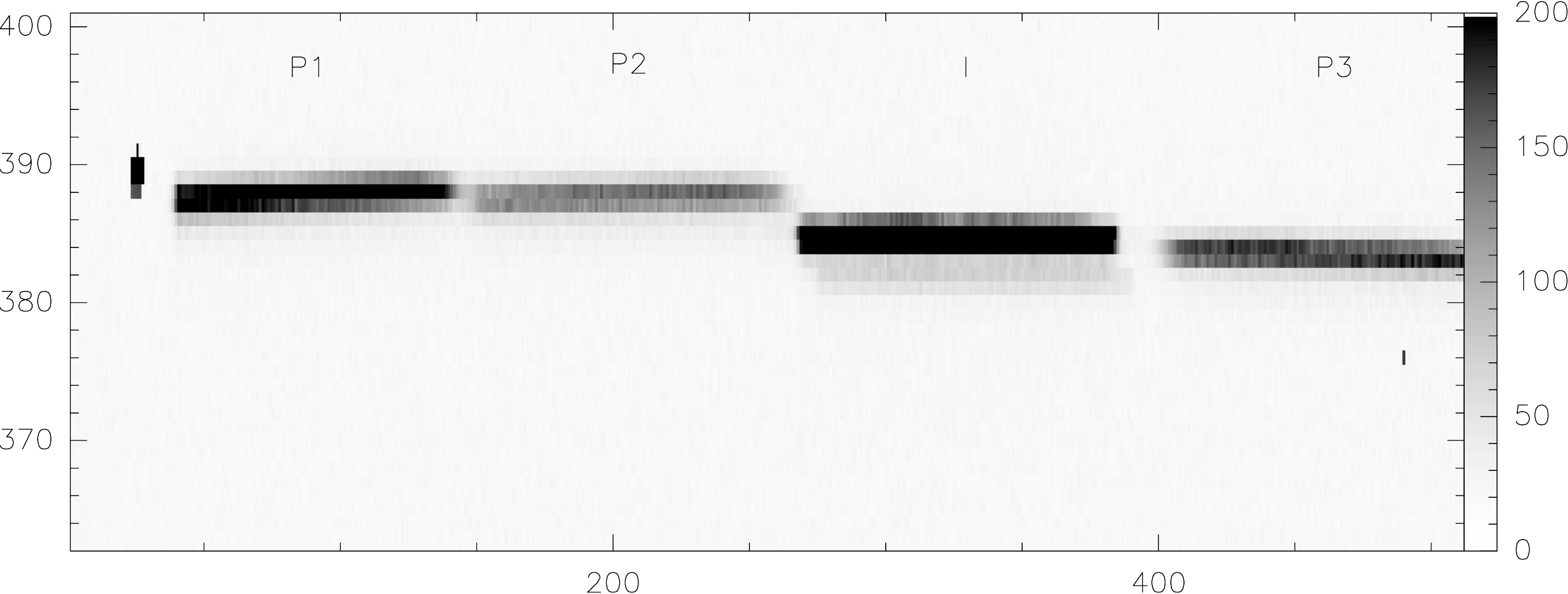}
  \caption{Image of the smaller slit (COLD stop) on the detector. Note
  that interferometric part of the image (third beam) is duplicated by
  2 pixels down and right. One can see the ghost reported in appendix
  \ref{app:char-ghosts} in the interferometriuc channel.}
\label{fig:cold-stop}
\end{figure}

This alignment is done using the Cold Stop slit at the entrance of the
spectrograph, imaged on the camera in normal cold conditions (see
Fig.~\ref{fig:cold-stop}).  Before starting the alignment, the two
brakes of the tuning device must be released (4 screws each beneath
the support). The brakes are located on the bottom of the cryostat
support on each side.  It is better to completely release also the two
guiding spheres (3 set of large guiding screws) that allows the three
rotation around a point close to the detector.  The upper sphere
guides the rotations around the optical axis (horizontal axis driven
by the 2 vertical screws) and the tilt of the detector (around the
vertical axis using the 2 horizontal screws).  The lower sphere allows
the tip around the horizontal axis within the plane of the detector.

The focus is performed using the long screw located beneath the
support (using a key \#7 (tbc)) .  The focusing is monitored while the
slit is imaged in the middle of the detector (0 order / cold stop /
Pupil Technics) with a flat illumination (a flash light).  The
orientation of the slit on the detector is adjusted to be parallel to
the columns of the detector.

The tip-tilt is monitored while moving the slit image from one side of
the focal plane array to the other one using the movement of the
grating.  The slit image must be lower than 2 pixels at the center,
and must be enlarged by the same amount on the two sides.

In the present case, the previous alignment, due to the lack of time
after the maintenance of the grating turret, was probably done before
the whole assembly was properly cooled down.

We adjusted the setting to achieved the required tuning, and achieve a
2 pixels large focusing of the cold stop.

Note: doing this measurement one can observe a duplication of the
interferometric beam slit, corresponding to a parasitic image (see
Fig.~\ref{fig:cold-stop}). Latter measurements showed that this ghost
induce fringes in the beam 2. The translation of the ghost with
respect to the interferometric beam is 2 pixels down and around 2
pixels to the left side.

\subsection{Pupil alignment and focaliser adjustment}

The focaliser mirror was completely re-aligned, considering the wrong
position of all of its tuning knobs and its obvious misalignment. The
alignment start by putting all setting in medium position, which
consist in removing all apparent angle of the parabolic mirror with
respect to the optical beam. The tilt of the beam image on the camera
(spectro set at zero order) can be due to the tilt of this parabolic
mirror. It must be noticed that this mirror must NOT be used to adjust
the position of overall beams on the SPG entrance slit. This
setting must be done using the 2 translation stages of the periscope, one on
each mirror that correspond to the 2 directions. 

During this alignment we corrected pupil vigneting on the previous 45°
mirror where beam 2 and 3 were obscured by its mechanical mount.  To
correct this vigneting it was necessary to adjust the dichroic angles
(output K dichroic in this case).  The correction was performed using
the dichroic tip-tilt of the beam K2, and then all dichroics where
adjusted with respect to the new position of the output K2 dichroic.
It was performed an adjustment of the beam print on the 3TK cold pupil
inside the cryostat in order to reach the higher signal.  In the mean
time we determine the correct position of the 3TK pupil mask (-44804
position).  It was determine by a maximization of the flux in the
image.  At the end of the procedure:
\begin{itemize}
\item Image position is tuned using the angular tuning of the dichroic, and maximizing the image flux
\item Pupil position is tuned using the dichroic translation and maximizing also the image flux
\item Check of the image position (and adjustment with angular adjustment)
\item Check of the pupil position using the Pupil imaging lens of the
  detector (manual knob on the detector). The pupil image through the
  mask is not overlapping the pupil print as seen with a wide field
  illumination of the spectrograph (flash light)
\end{itemize}

Doing the alignment of the focaliser we improved a lot the image
quality of all the beams.  For all J H and K beams the images on the
camera are lower than 25x2 pixel with a mean spot angle of 179°
(between 178 and 180).

All H and K pupils were then adjusted to fit the 3TK and 3TJHK pupil
mask using the adjustments of the H dichroic beam splitters for H
band, and the adjustments of the J mirrors for J band (tilts for image
positions and translations for pupil adjustments while checking the
flux level).  In all cases the flux reduction while introducing the
pupil mask do never exceed 5\% with respect to the images without any
pupil mask (see tabled values).

The optimized value for the two masks are now entered in the ICS.
\begin{itemize}
\item 3TK  mask position: -44804,
\item 3T JHK mask position: -27330
\end{itemize}

Doing the pupil adjustments, most of the translation stages acting on
the OPD were released toward their medium position and no translation
stages are not any more in extremal position.

\begin{figure}[p]
  \centering
  \includegraphics[width=0.95\hsize]{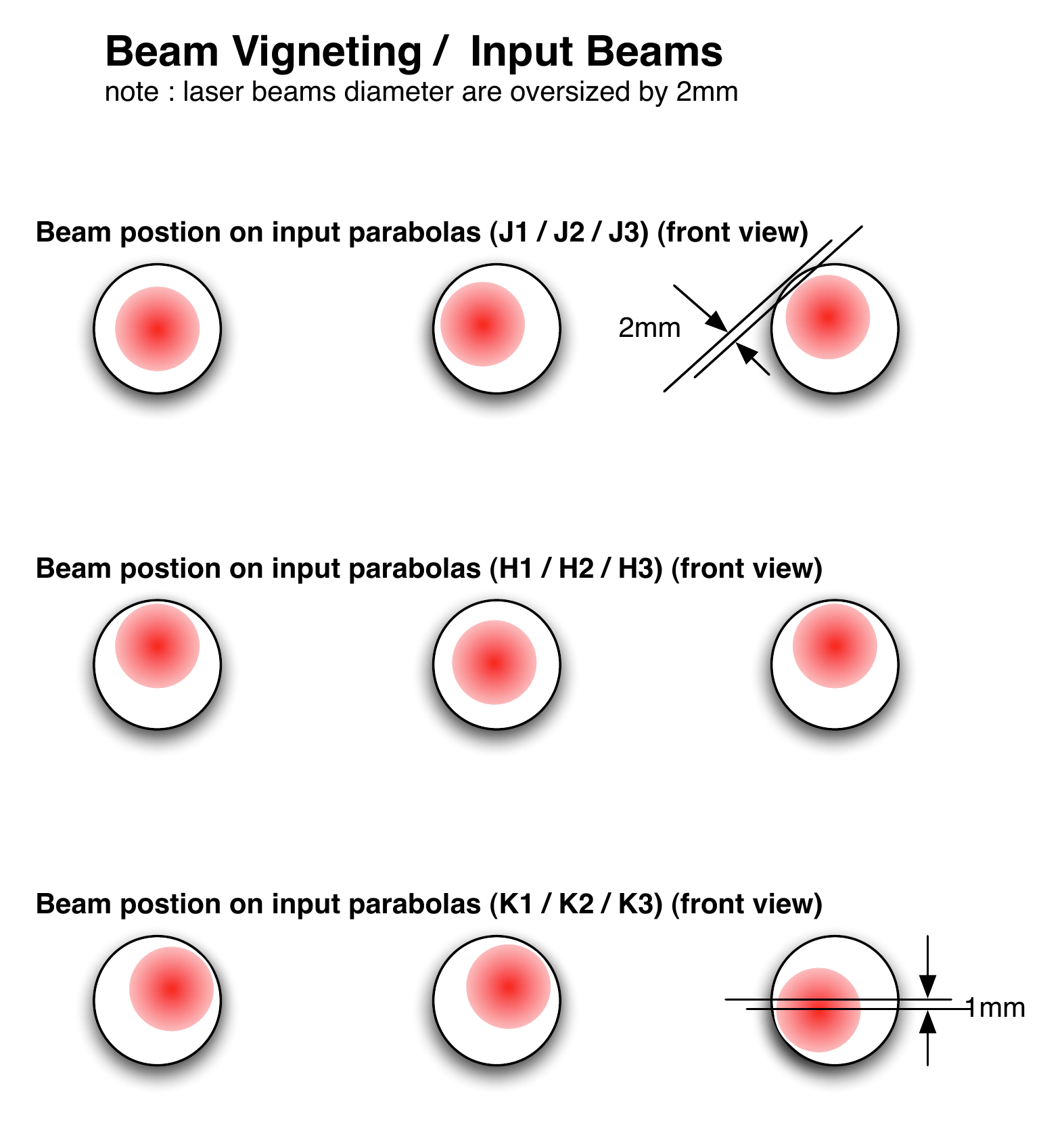}
  \caption{Beam positions at the input parabolae as reported on
    February 6th 2008.} 
  \label{fig:input-beam}
\end{figure}

\begin{figure}[p]
  \centering
  \includegraphics[width=0.8\hsize]{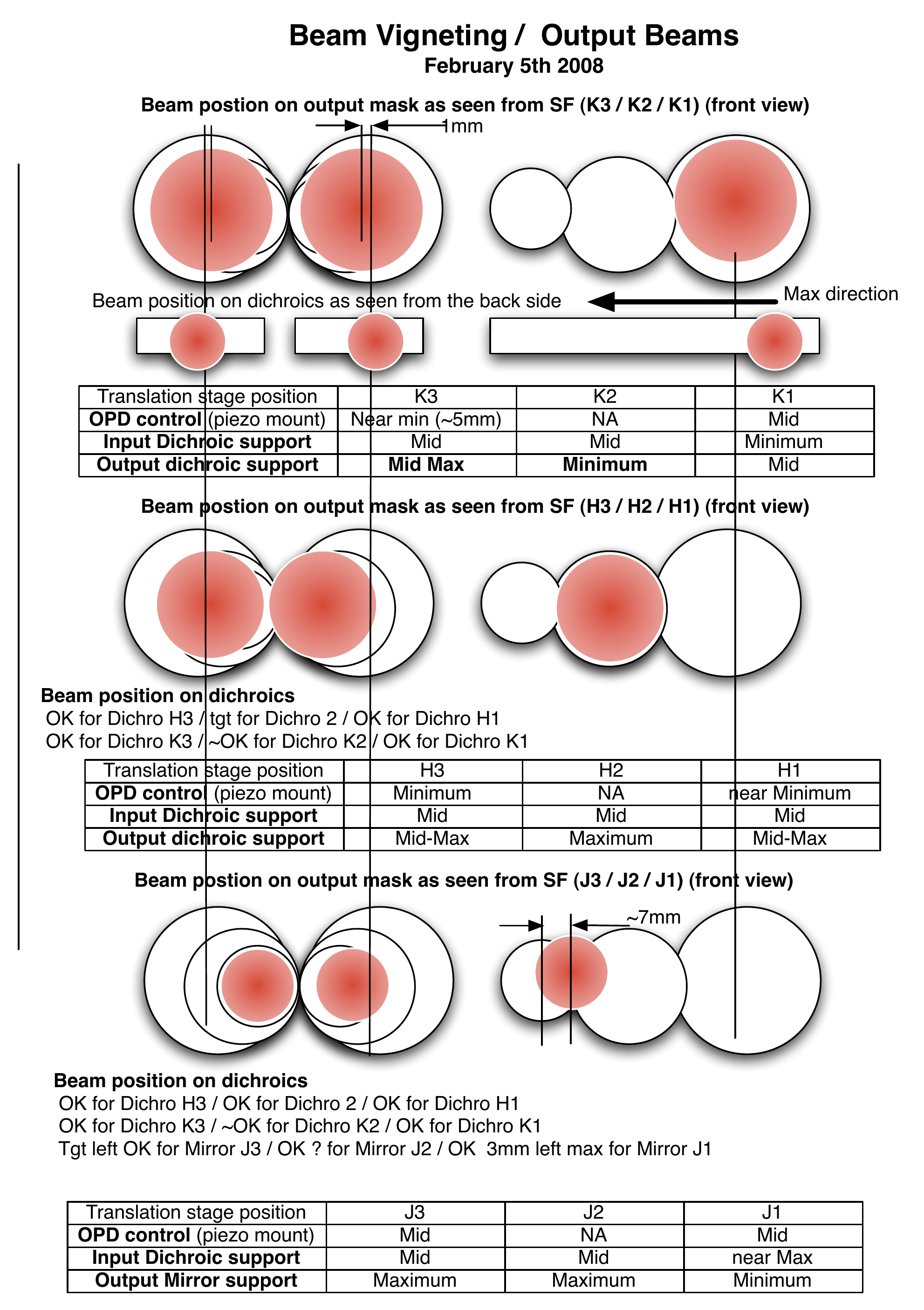}
  \caption{Summary of the beam positions at the output of the spatial
    filters, as reported on February 6th 2008.} 
  \label{fig:beam-vigneting}
\end{figure}

The figures report the positions of the beams on the the input
parabola (see Fig.~\ref{fig:input-beam}) and on the pupil
alignment mask located at the output of the Spatial Filters (see
Fig.~\ref{fig:beam-vigneting}).

It must be mentioned that this alignment requires to proceed accurately
taking into account all related parameters, using the
input and output dichroic translations and the piezo actuator support.
\begin{itemize}
\item  Piezo support is an independent support, but its translation
  may strongly affect the beam injection, due to its poor guiding
  capability. Be aware that if the flux is lost in the fiber, the
  process is heavy to ge it back. 
\item Input dichroic translation affect the beam position on input
  parabolas 
\item output dichroic translation affects the pupil position on pupil
  mask in the cryostat. It may also introduce vigneting for related
  beams (the beam itself, the adjacent ones, and the transmitted beams
  (for instance H and J on the K dichroics). 
\end{itemize}
For SFJ, it was noticed that the J1 mirror does not allow to put the beam in the theoretical position of the design. If one need to put it in a correct position in order to achieve the correct arrangement of the J pupil, the mirror mount must be translated horizontally along the mirror surface plane. This adjustment requires to drill a new hole for the fixation of this mirror.

\clearpage
\section{Characterization of ghosts}
\label{app:char-ghosts}

\subsection{Mismatch of photometric calibration}
\label{sec:mismatch calibration}

The acquisition of the P2VM (pixel-to-visibility matrix) is pivotal in
the observing and data reduction scheme of AMBER. Immediately after
P2VM calibration, every possible calibration parameters have been
obtained. The quality of the raw data observations needed to acquire a
P2VM do not obey the prerequisites implied by the concept of a P2VM,
described by its mathematical formulas.

\begin{figure}[hb]
  \centering
  \includegraphics[width=0.7\hsize]{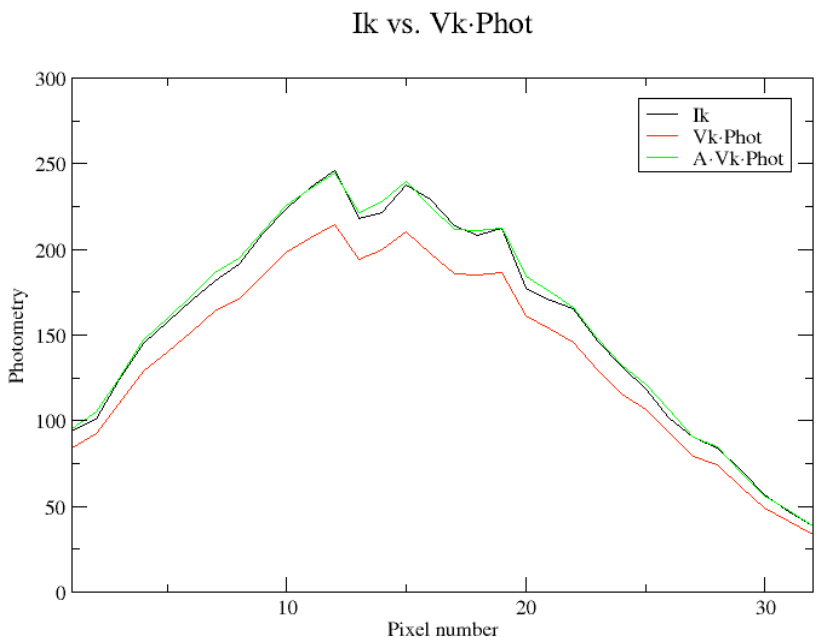}
  \caption{Inteferometric channel: continuum measured by averaging
    1000 interferograms (black line) compared to the continuum computed by the
    software for one frame with the photometric calibration (red
    line). The green line corresponds to the red line multiplied by a
    constant $A$.}
  \label{fig:cont-mismatch}
\end{figure}

In brief, the photometric calibration allows to substract the
continuum from the detected fringes so that the continuum-corrected
interferogram is on average equal to zero.  The first problem was
found when we compared in the continuum obtained actually measured in
the interferometric channel from the one which is being
substracted. Fig.~\ref{fig:cont-mismatch} shows an intereforgram which
has been averaged over 1000 frames. Since the fringe phase
fluctuations are larger than a fringe period, the fringe modulation
disappears and we got a good estimation of the fringe continuum (black
line). On another hand, using the P2VM calibration files which allow the
ratio between each photometric beam and its counterpart in the
interferometric channel to be measured, one see that the computed
continuum (red line) does not correspond to the actual continuum. 

The reason for this mismatch is that the assumption that there is no
leaks or ghost in other channels is no correct. One way to compensate
this is to introduce a multiplicative factor $A$, or even better to
take into account the cross-talk between the channels in the
interferometric/photometric ratios (see Sect.~\ref{sec:dataproc} and
appendix \ref{app:dataproc}).

\subsection{Investigating P2VM calibration files prior to Feb 2008}

\begin{figure}[p] 
  \centering
  \begin{tabular}{cc}
     \includegraphics[width=0.3\hsize]{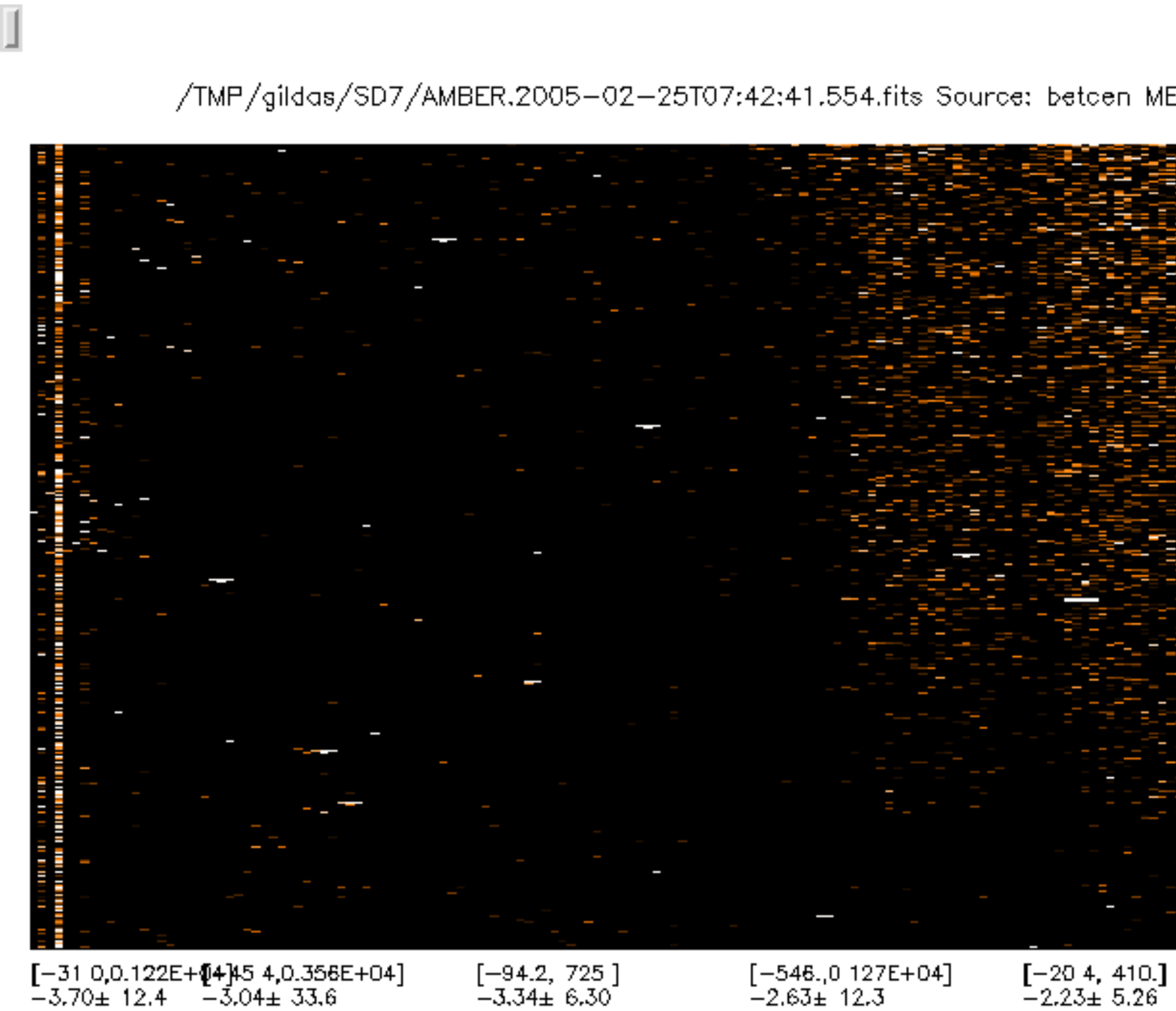}
    &\includegraphics[width=0.3\hsize]{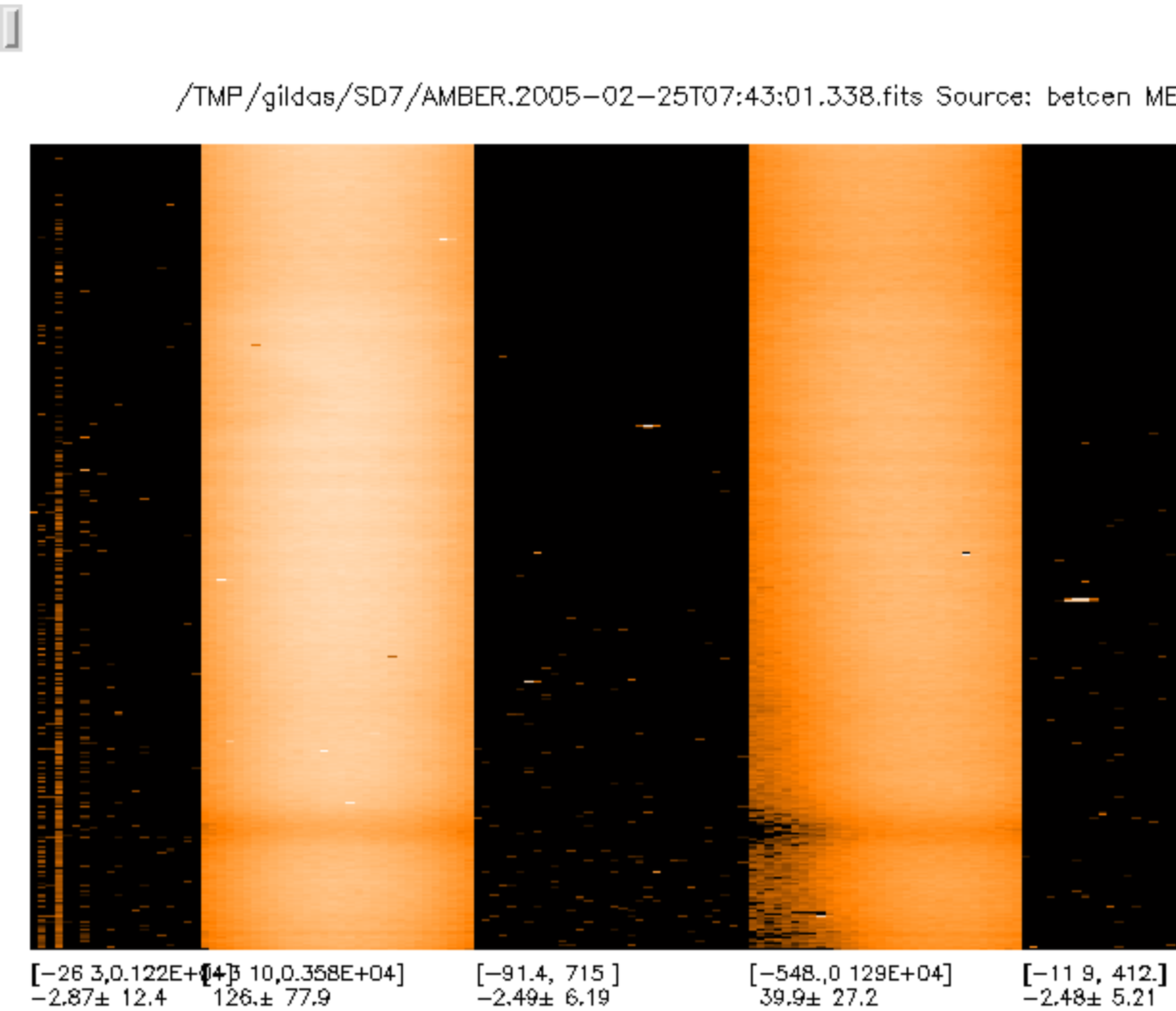}\\
     \includegraphics[width=0.3\hsize]{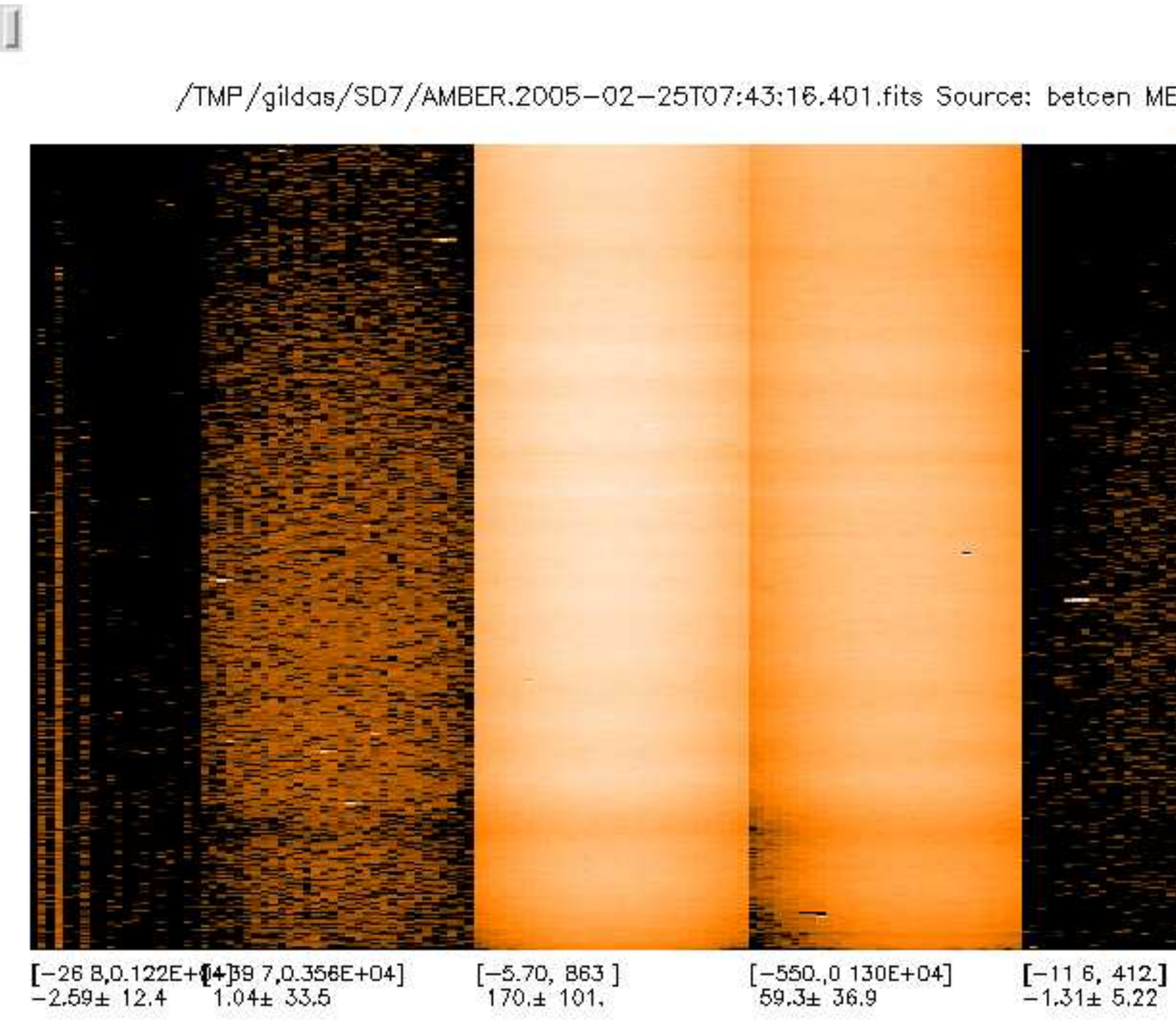}
    &\includegraphics[width=0.3\hsize]{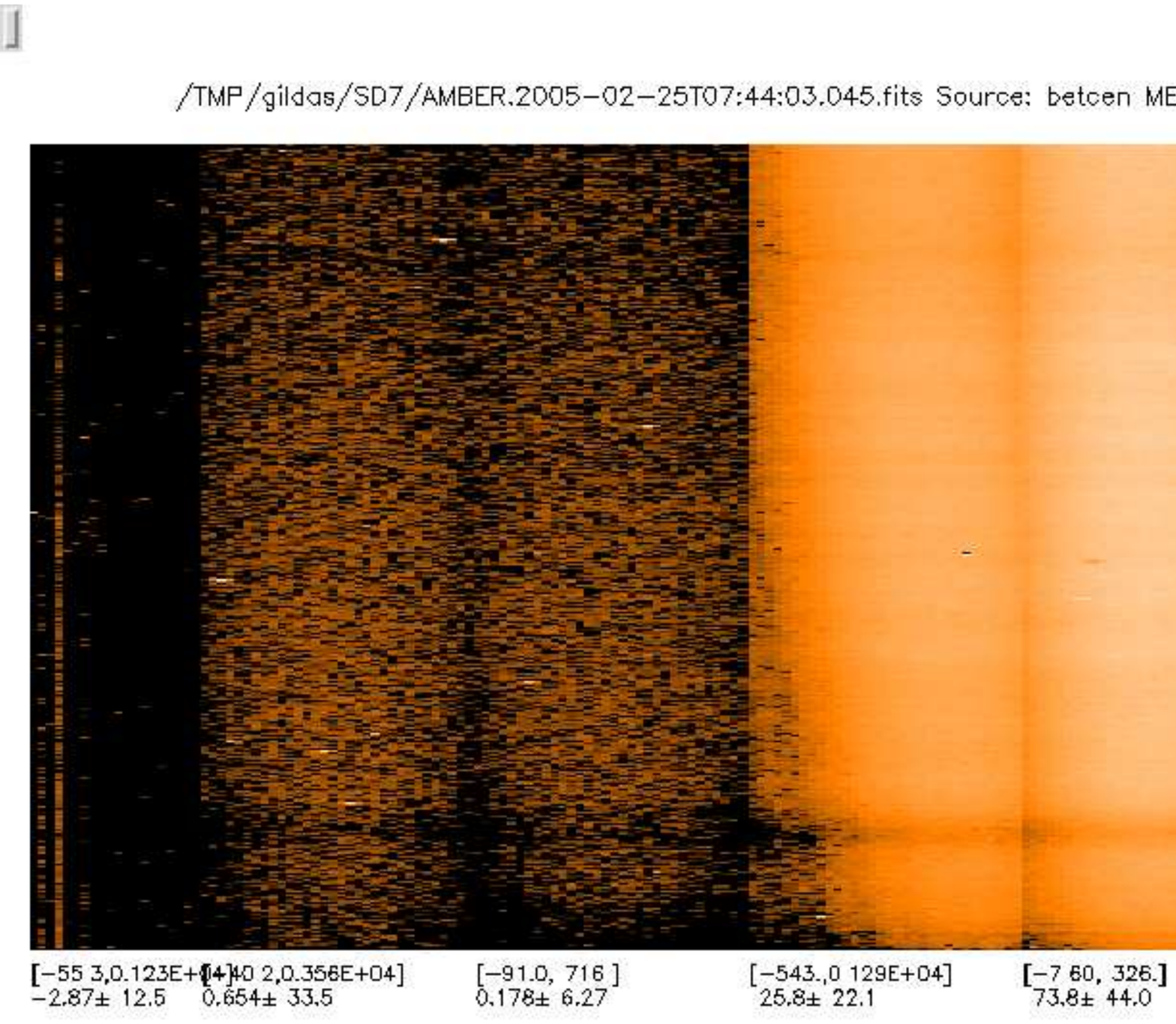}\\
  \end{tabular}
  \caption{\label{fig:1-1234} P2VM calibration files obtained in Feb
    2005. \textbf{Top Left:} DARK. It is not entirely dark on beam 3 :
    (this does not seem to be remanence of the previous
    observation---spectral calibration of beam 3). \textbf{Top Right:}
    BEAM \#1 . No ghost visible (note that beam 3 is now dark).
    \textbf{Bottom Left:} BEAM \#2. Pollution at 0.5\% of beam 1 by
    beam 2 (but this can be remanence) , a hint of something in beam 3
    (so this is not remanence). \textbf{Bottom Right:} BEAM
    \#3. Pollution at 1\% of beams 1 and 2 by beam 3 (if remanence,
    should have disappeared in beam 1, proving that in left image it
    was not remanence either)}
\end{figure}

\begin{figure}[p] 
  \centering
  \begin{tabular}{cc}
     \includegraphics[width=0.3\hsize]{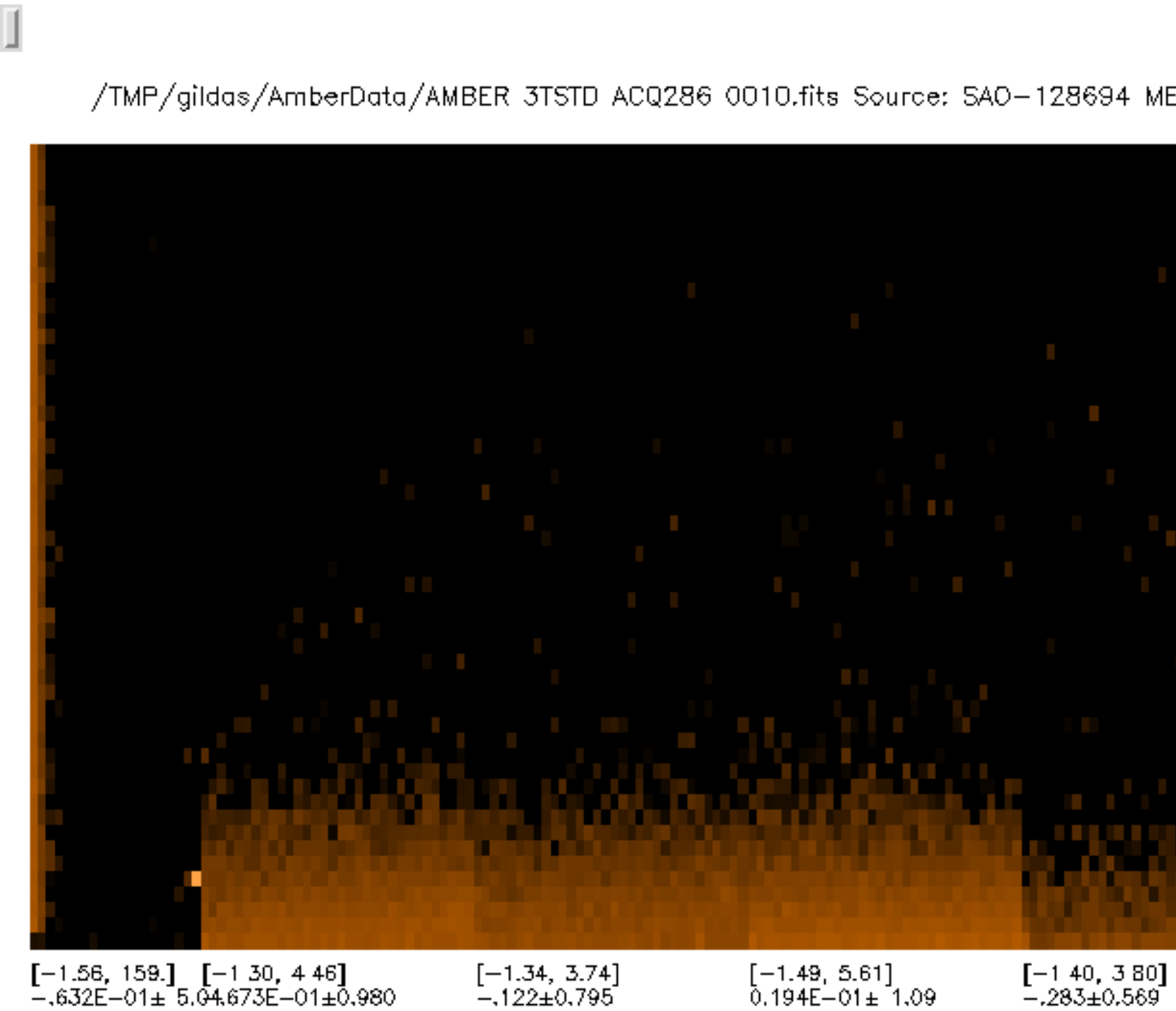}
    &\includegraphics[width=0.3\hsize]{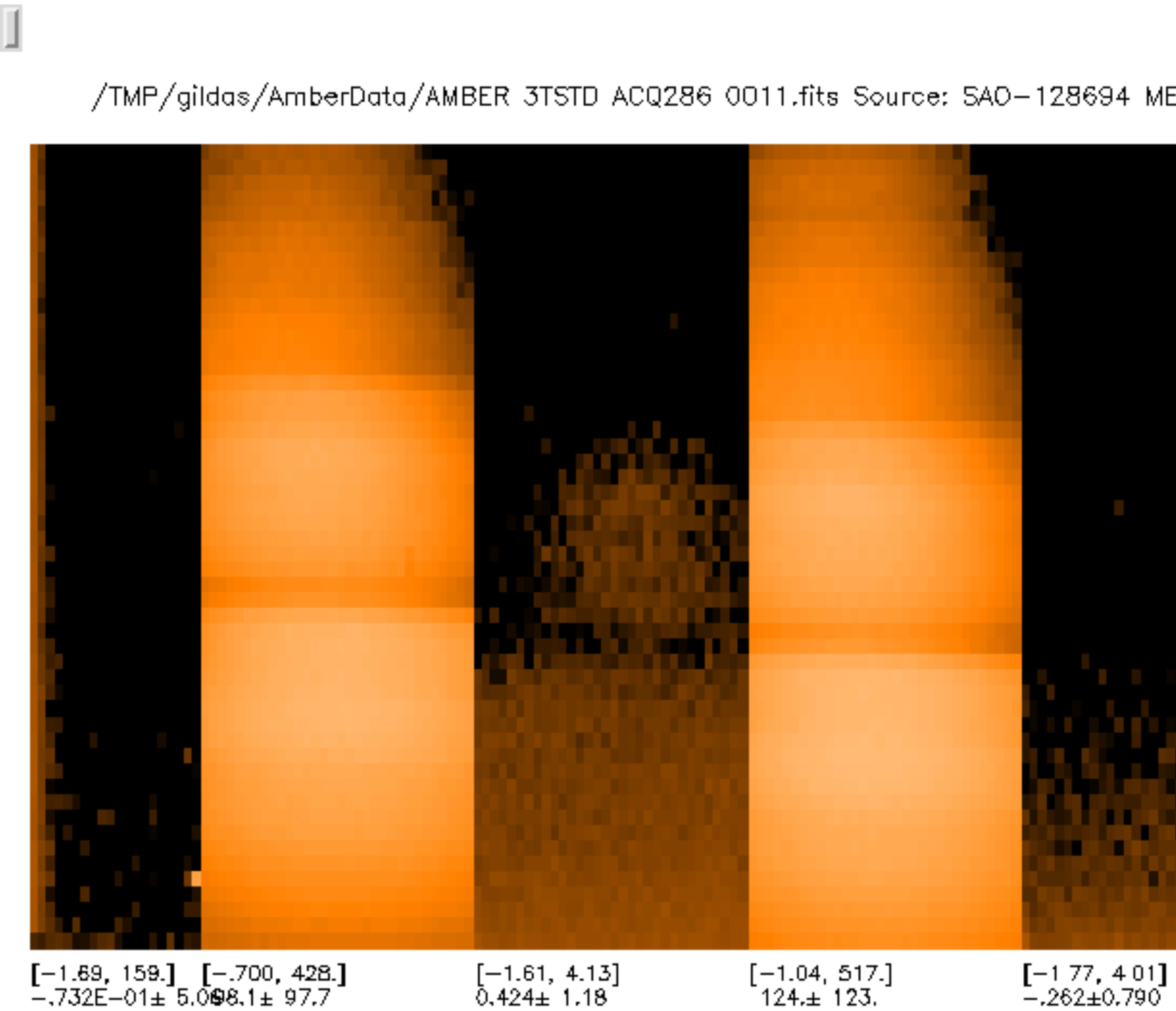}\\
     \includegraphics[width=0.3\hsize]{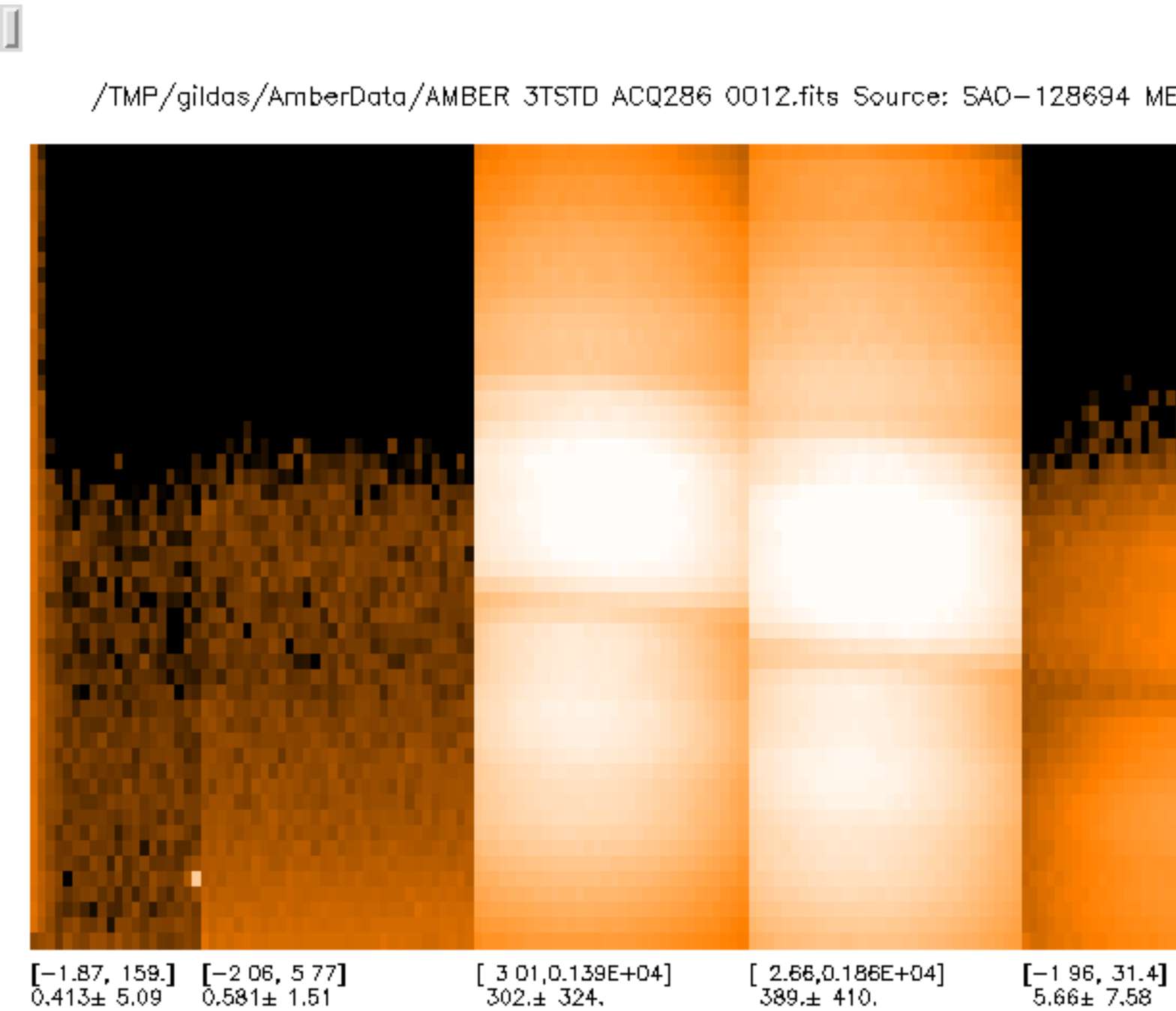}
    &\includegraphics[width=0.3\hsize]{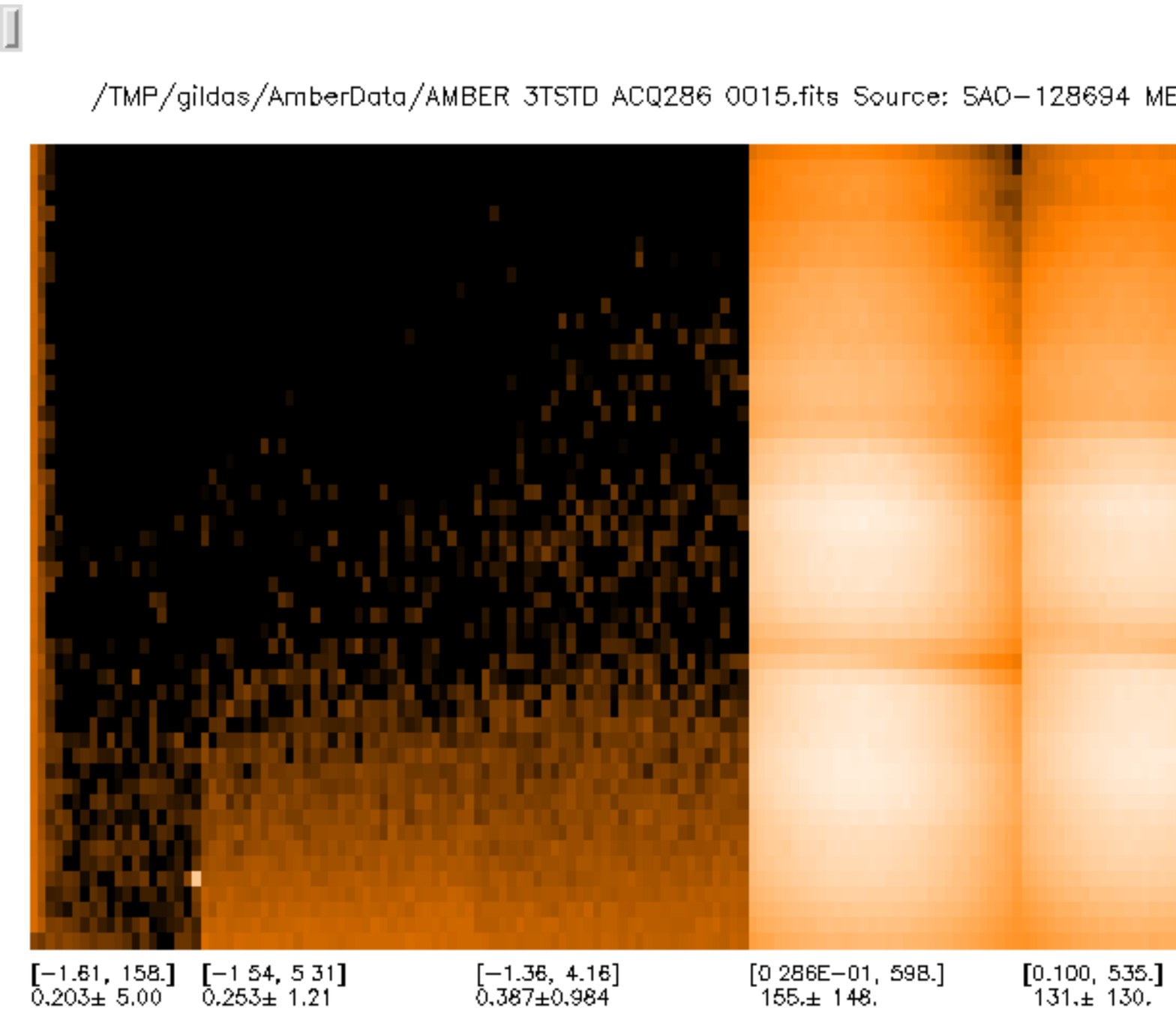}\\
  \end{tabular}
  \caption{ \label{fig:2-1234} P2VM calibration files obtained in Sep
    2007. \textbf{Top Left:} DARK. It \emph{is} dark (except on bottom
    above 2.3 microns where thermal background emission of the fiber
    heads and of the lab is present). \textbf{Top Right:} BEAM
    \#1. Beam 1 pollutes in H and K beam 2 (0.3\%).  \textbf{Bottom
      Left:} BEAM \#2. Beam 2 pollutes beam 3 very strongly (2\%) and,
    in a much smaller measure (0.2\%), beam 1.  \textbf{Bottom Right:}
    BEAM \#3. Beam 3 does not pollute beam 1 and only very marginally
    beam 2. (See additional note in text).}
\end{figure}

We have analyzed historic data set to investigate if it possible to
detect this cross-talk or ghost in the calibration files produced by
the P2VM procedure.  Figures \ref{fig:1-1234} and \ref{fig:2-1234}
shows two data sets taken respectively in Feb 2005 and Sep 2007 just
after the SPG intervention.

In the following, all images were taken on a mean of all frames in the
files. All plots are done with a logarithmic transfer curve to flatten
the dynamics. Black is 0.1 ADU, white is 1000 ADU.
On all presented images, from left to right we have :
\begin{enumerate}
\item masked channel (20 pixels wide, never illuminated)
\item Photometric Channel 1 (32 pixels wide)
\item Photometric Channel 2 (idem)
\item Interferometric Channel (idem)
\item Photometric Channel 3 (idem)
\end{enumerate}
The vertical axis shows the spectral signal from shorter wavelength
(bottom) to longer wavelengths (top) for LR mode, the reverse for MR
mode. Due to a bug in the gildas interface to the \texttt{amdlib}
software used here for displays, the wavelength information on the
right of the boxes is incorrect for LR mode.

In Feb 2005 data (Fig.~\ref{fig:1-1234}), one detect a pollution of
beams 1 and 2 by beam 3 at the level of 1\%. The example given is done
with the ``Old'' Camera, in mode 3T~Medium~K. Relevant comments are
made in the label of each figure. We note that since this dataset was
used for the P2VM of the successfull Alfa Arae data, the effects
depicted here were not seen at the time and thus acceptable.

In Sep 2007 data (Fig.~\ref{fig:1-1234}), beam 2 pollutes beam 3
clearly at the level of 2\%. Theses calibrations were done with the
``new'' Hawaii camera (no remanence) in mode 3T Low\_JHK. Note also
the position of the dark limit between H and K, and its displacement
in beams 1 and 3 wrt the position in beam 2 (enormous with today's
tilted slit). This displacement is not the same in all beams, so the
crosstalk between beams is also a crosstalk between wavelengths. It
will be difficult to get accurate visibilities with such a crosstalk.

\begin{figure}[t]
  \centering
  \includegraphics[width=0.3\hsize]{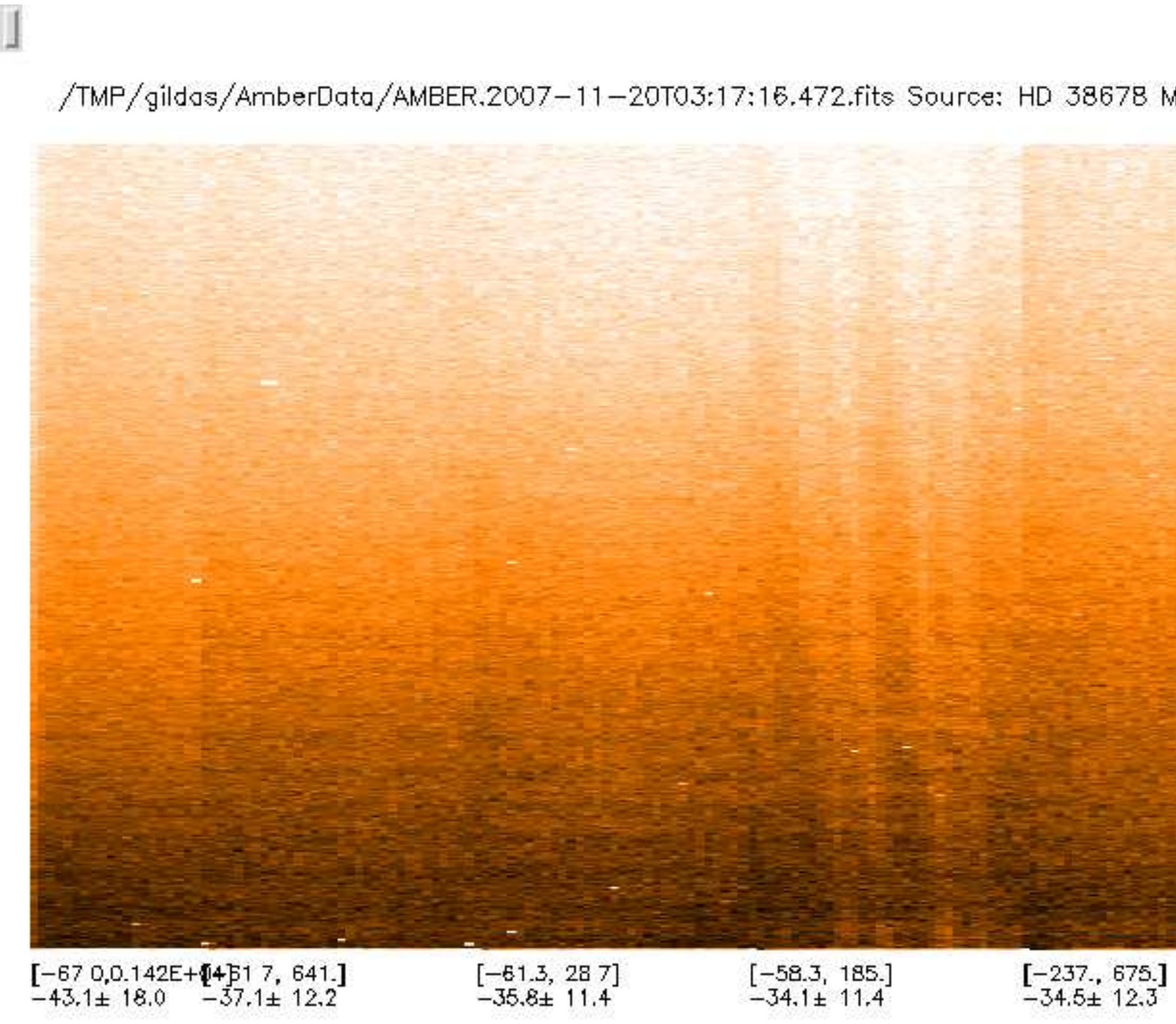}
\hfill
  \includegraphics[width=0.3\hsize]{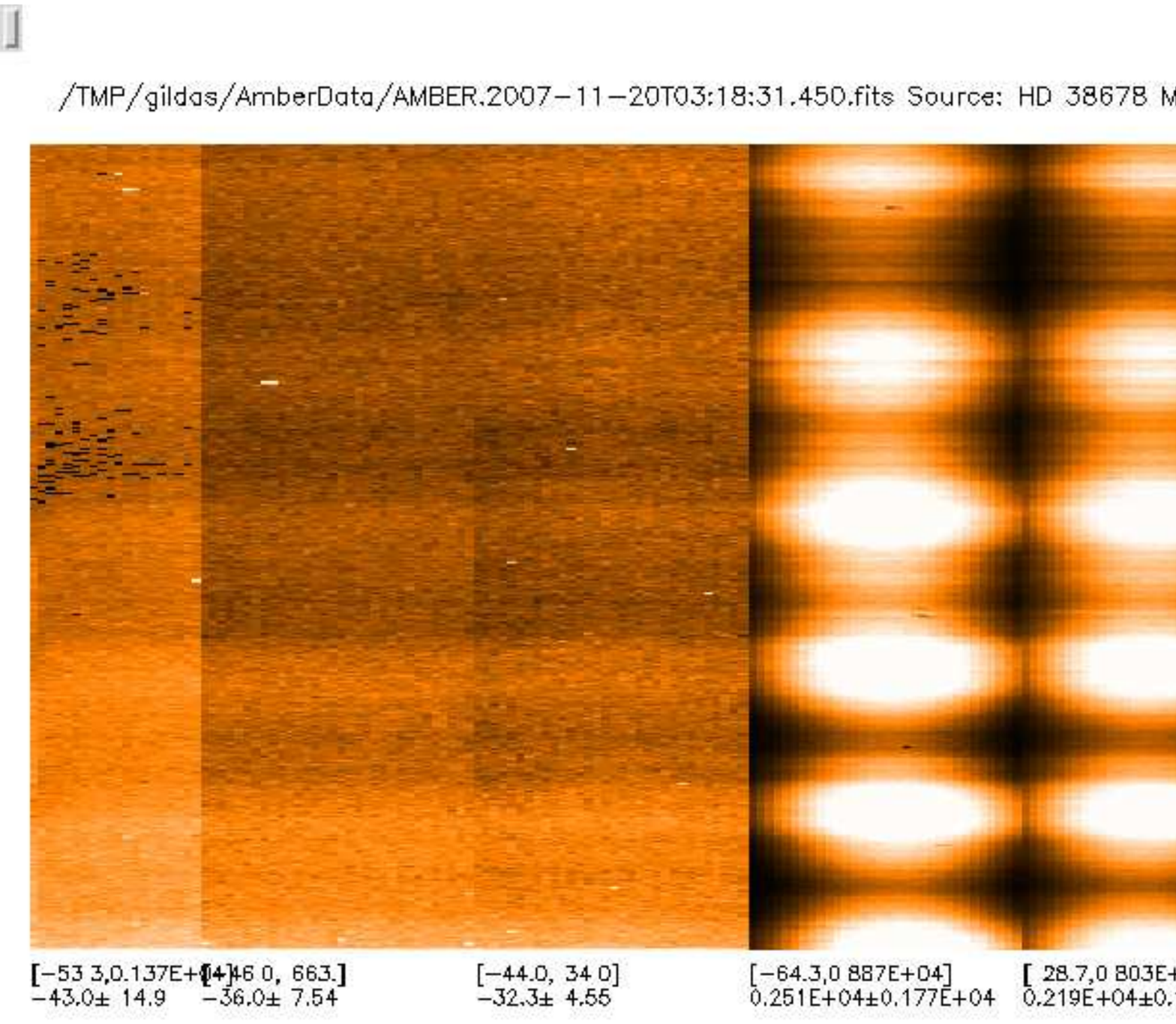}
\hfill
  \includegraphics[width=0.3\hsize]{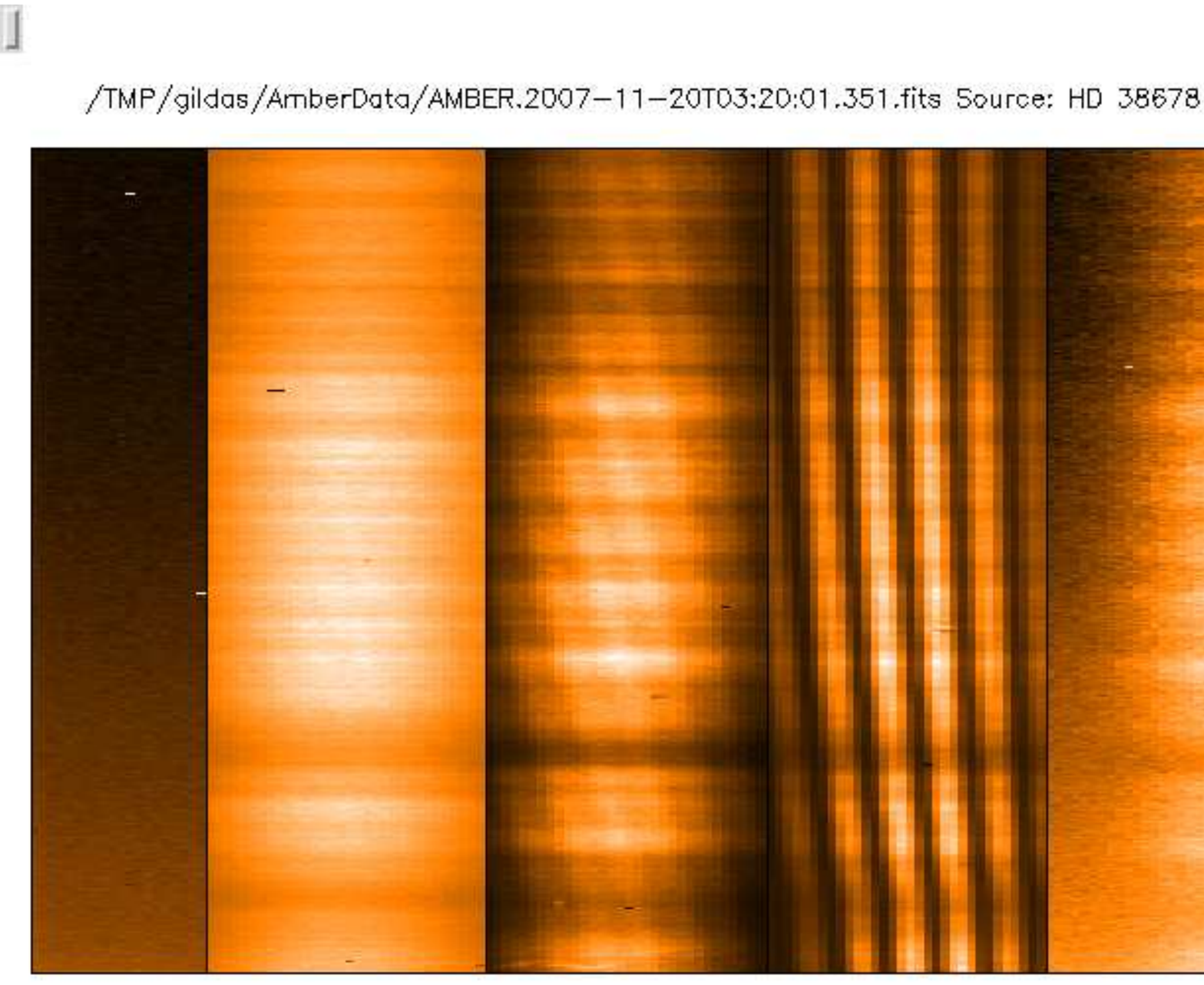}
  \caption{ \label{fig:3-123} P2VM calibration files obtained in Nov
    2007.  \textbf{Left:} DARK. Some fringes are present in the
    interferometric channels probably due to remanence effect.
    \textbf{Middle:} spectral file BEAM \#3. Frame taken for the
    spectral displacement calibration. The scale is histogram
    equalisation to see subtle effects: influence of enlighted beams
    on the background of dark channels.
    \textbf{Right:} Frame taken for the P2VM Ck value for beams 1--2.}
\end{figure}
A different type of problem is seen in a second set of observations
taken in Medium\_K mode (see Fig.~\ref{fig:3-123}). In the dark
exposure taken before the three files used for computing the spectral
displacement (left panel), we note fringes in the interferometric
channel. These are due to a remnant feature of the previous
observation. It would seem that the camera still has some
remanence. Since this kind of frames is used to compensate for pixel
biases, any degree of pollution by artifacts will have a severe effect
on all the observations. It is advisable to introduce in the AMBER OS
a test on each such frames invalidating the data in this case.

In the middle panel of Fig.~\ref{fig:3-123}, which shows the beam \#3
enlighted with a spectral modulation for teh spectral calibration, the
flux variations in the lighted channels (I and P3) change the
background in all the other channels, even the masked channel. This is
not due to beam pollution but is a know effect of the camera readout,
seemingly unavoidable and possibly not entirely taken in account
properly in the cosmetics performed in the first steps of the data
reduction. 

In the right panel of Fig.~\ref{fig:3-123}, fringes are visible in the
photometric beam 2, and possibly beam 1. Light in beam 3 is due to
pollution by beam 2 as in middle panel. The fringe pattern shifts with
the piston, proving that this effect is either a global interference
pattern in front the camera, or a ghost of interferometric beam seen
in beam 2.

\subsection{Ghosts measured during ATF}

\begin{figure}[t]
  \includegraphics[width=0.95\hsize]{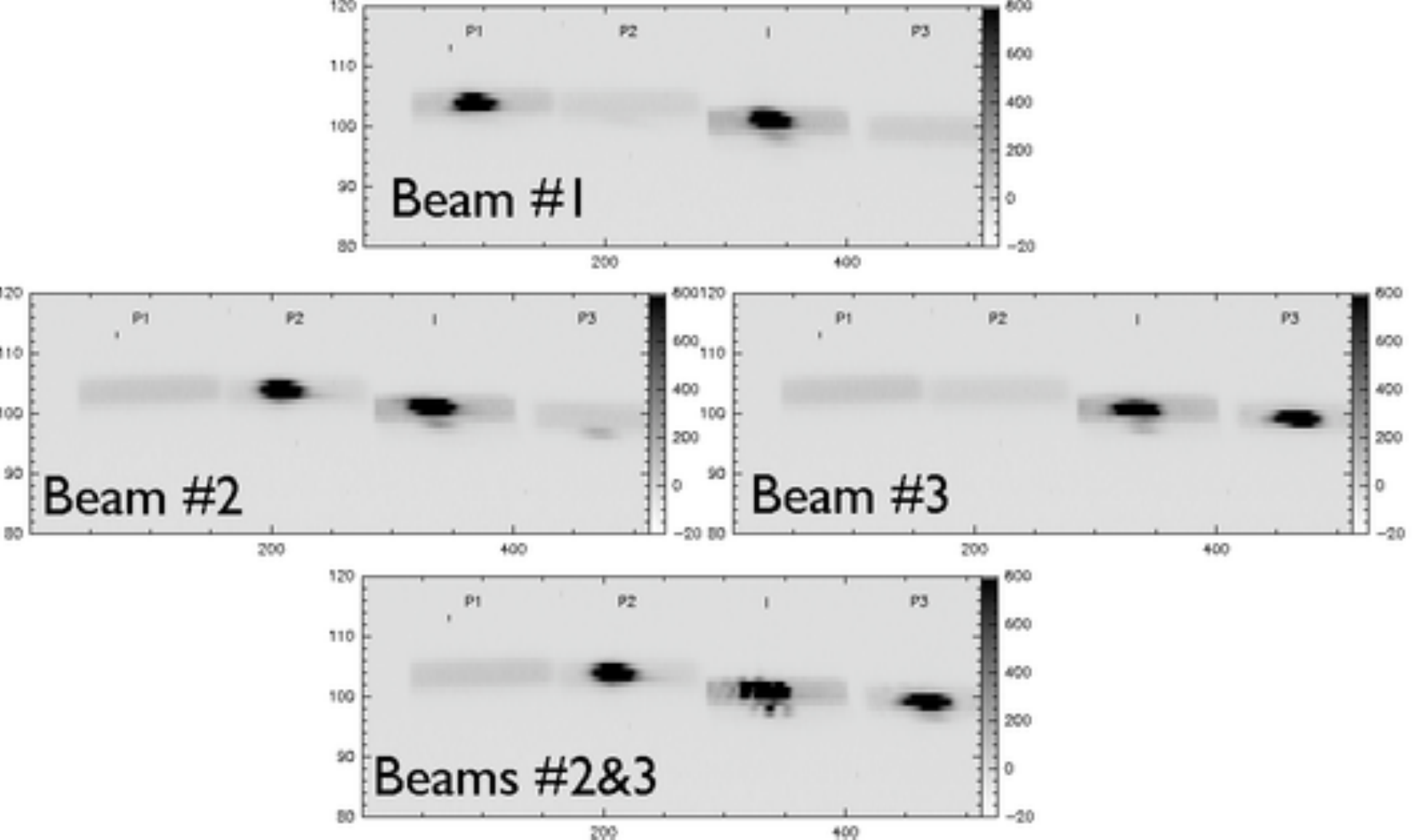}
  \caption{Images of the narrow slit taken at the $0^{\rm th}$ order of
    the grating showing the shape of non-dispersed beams. From top to
    bottom, left to right, we have: the image with only beam \#1 lightened, same
    thing with beam \#2 and beam \#3, then beams \#2 and \#3 lit
    simultaneously. Notice the fringe pattern at zero dispersion in
    the I channel, the fringed ghost two pixels below.}
  \label{fig:ghosts}
\end{figure}

All images have been taken with the narrow slit, in the $0^{\rm th}$
order of the grating and a neutral density 10 on the CAU lamp to
enable a longer integration time. This long integration time enables
the visualisation of the slits itself, even in beams where the shutter
is closed, thanks to the background thermal emission of the lab going
through the slit.

The images of Fig.~\ref{fig:ghosts} show that the slit is very well
imaged on the camera, and that the $0^{\rm th}$ dispersion position
enable to measure very accurately the beam displacement between the
photometric and interferometric beam. 

Beam \#2 has a clear ghost (6\% of flux) in beam \#3 (displacement: 2 pixels below).
Beam \#1 may have a hint of a ghost in \#2.

All beams induce a 'second image' in the interferometric beam (2
pixels below). This double image being common to all beams must arise
in internal reflections in the beamsplitter. When 2 beams are lit, as
in the \#2-\#3 image we have a second set of (less bright) clear
fringes 2 pixels below the first one. This ghost represents 6\% of the
incoming flux (but its relative importance may change with the
wavelength when the flux is dispersed, as it is in normal AMBER
operation). The differential path of these reflections in the
beamsplitter being small, the fringes are just displaced by lambda/4.

When we add dispersion, the interferometric channel will thus see
superimposed 2 sets of fringes, the normal one plus a second, much
fainter, with a different phase, one spectral resoultion element (2
pixels) below.  Normally this displacement should not change with the
spectral dispersion.

We have been able to take into account quite sucessfully the presence of the
ghosts in the data reduction tools we developed specifically for this
ATF run (see Section~\ref{sec:dataproc}), and these algorithms should
be ported to the standard data reduction tools for optimal use of
AMBER.  However this software compensation cannot reach the same level
of precision that would be attained by replacing the optical element
which creates the ghost if the first place\footnote{which remains to
  be identified.}. In particular the ghost fringes seen above cannot
be compensated by software and will induce at minimum a contrast loss.

\subsection{Conclusions and recommendations}

The most probable explanation of the observed ghosts is the effect of
the device inside the spectrograph dewar that arranges the beams on
the camera. A set of three small adjacent mirrors redirects the 3
photometric beams on the detectors. The widths of these mirrors is
very close to the width of the incoming beams and a small misalignment
of one of the pupils on these mirrors lead to a leak to the adjacent
mirror.  According to the pupil alignment, a different amount of leak
may occur, and it seems that the last pupil alignment in October 2006
introduced a larger leak on the beam 3.  In any case it is not sure
that the beam is not a little bit larger that the element-mirrors
leading to a slight ghost on both adjacent mirrors.  

This ghost effect should be measured and compensated by the amdlib
software for the P2VM calculation. However, the large displacements in
data taken between Sep 07 and Jan08 are not retrievable easily, at
least not automatically (perhaps can be entered by hand), and will
reduce the number of useable data channels.  Removal of these ghosts
and removal of the spectral displacement is badly needed. Besides, we
need to have a value returned by the amdlibComputeP2VM routine used in
the amber observation software to tag invalid P2VM observations which
exhibits ghosts effects above a sensible value (0.5\% for example).

The new alignement performed after the Jan 2008 SPG intervention
allowed to decrease the cross-talk between photometric beams, but
a ghost remains in the interferometric channel which comes from the
interferometric channel itself (probably a reflection on the common
beamspliter used to extract photometric beams).

The camera change is believed to have solved the remanence problem,
although we seem to see the contrary in fig.~\ref{fig:3-123} (upper
panel). At the moment, no remanence is taken into account in the data
reduction program, for lack of removal method (before or after which
cosmetics, with what value, etc...).


\clearpage
\section{Optical stability}
\label{app:optical-stability}

\subsection{Output beams}

\subsubsection{Medium term stability survey during the ATF operation}

A survey of the beam positions and of the beam fluxes was performed
during the last week of the ATF operation. No significant drifts were
observed within a time scale of a few days.  In any case during the
second part of the ATF operation, when the main alignments were
completed, we did not need any more intervention in the laboratory to
re-adjust the beams.  Mainly it must be noticed that for the output
beams the requirement used by the Paranal team for the adjustment is
overestimated. Using the related software alignment tool, the green
dot means that the beam is properly aligned, the orange one means
acceptable, a re-alignment can be required latter, and only the red
dot imposes an immediate re-alignment. It is highly preferable to
perform the verification from a control room and to avoid any
disturbances on the bench.

\begin{figure}[hb]
  \centering
  \includegraphics[height=0.4\hsize]{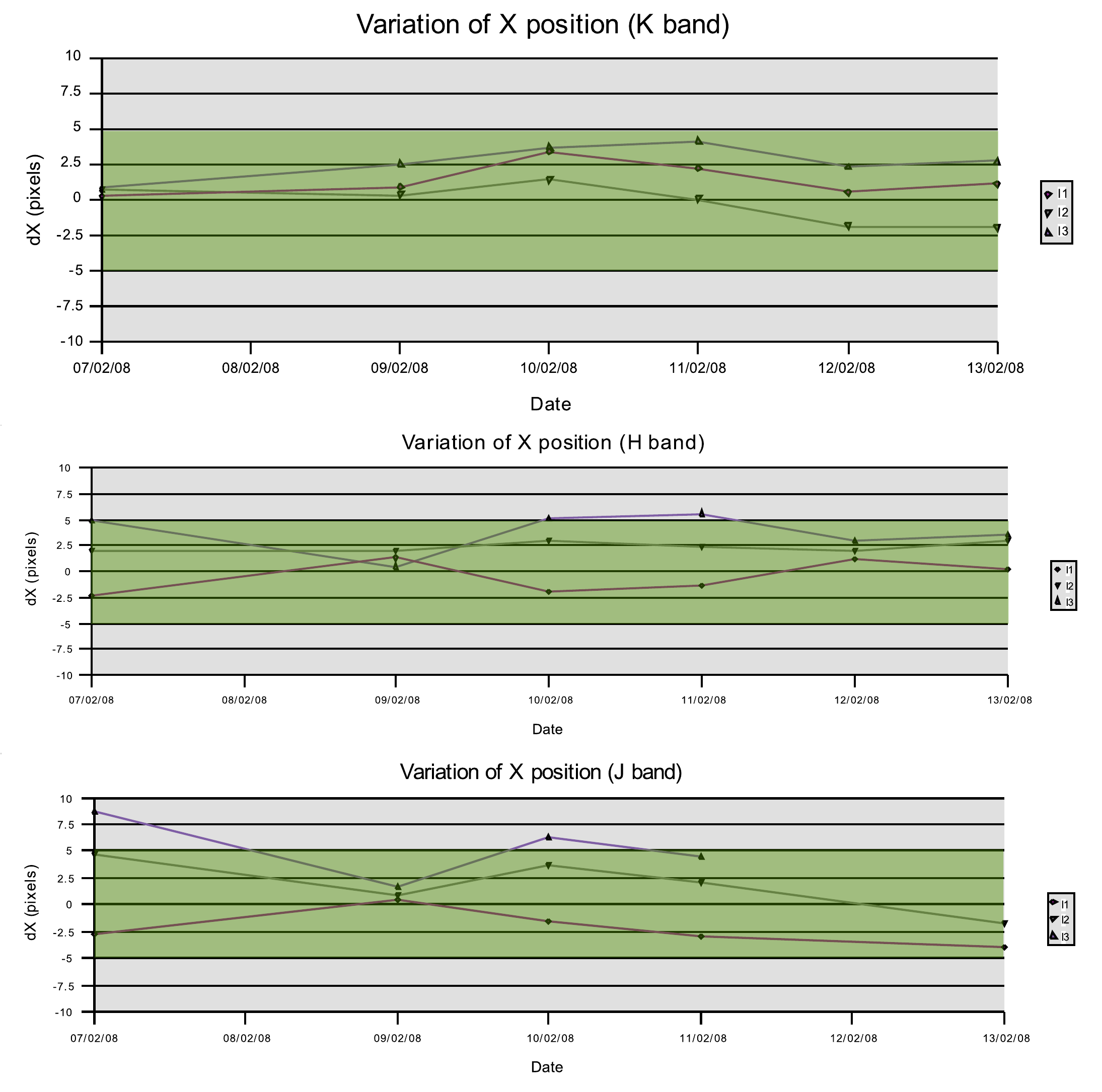}
  \includegraphics[height=0.4\hsize]{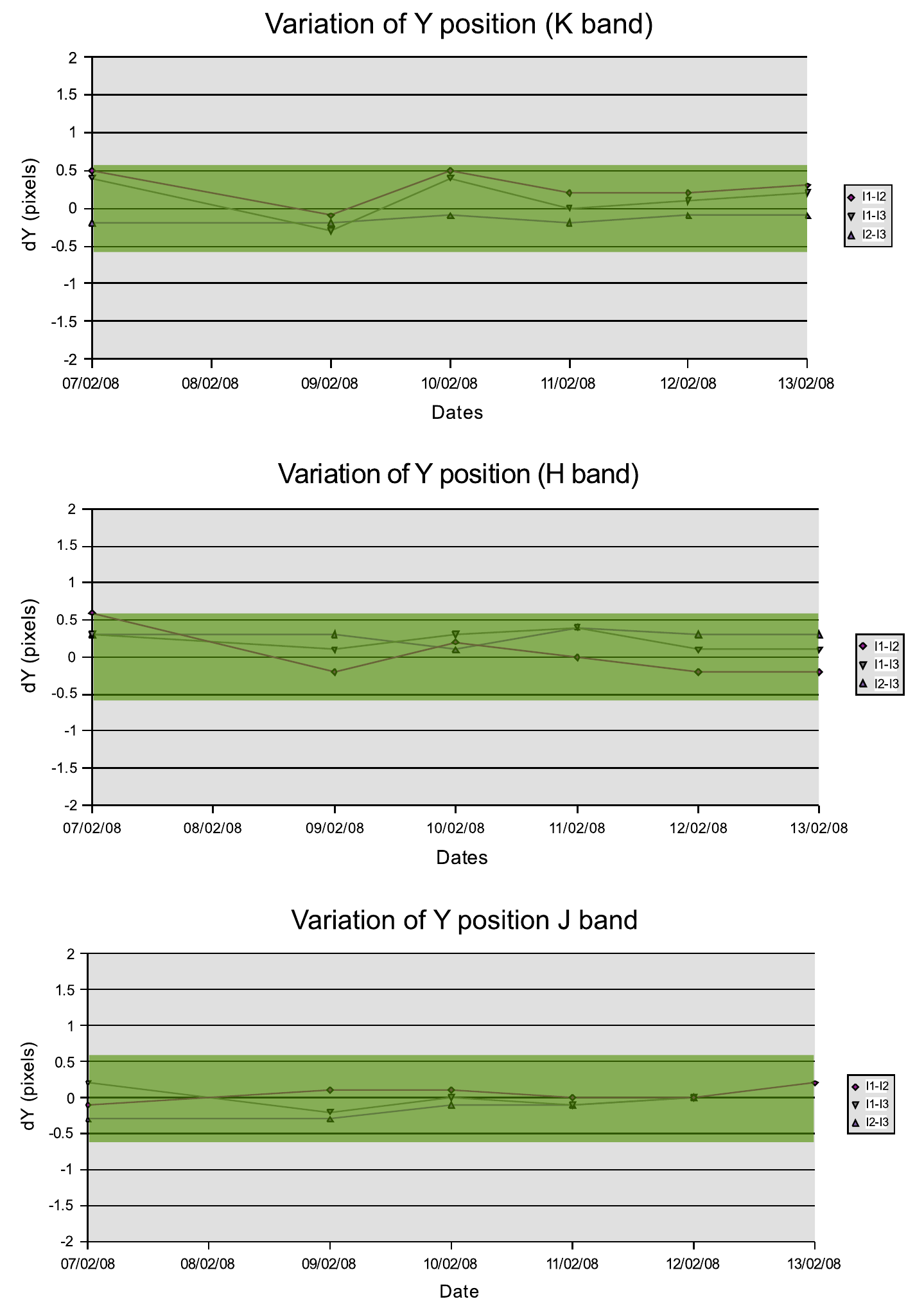}
  \includegraphics[height=0.4\hsize]{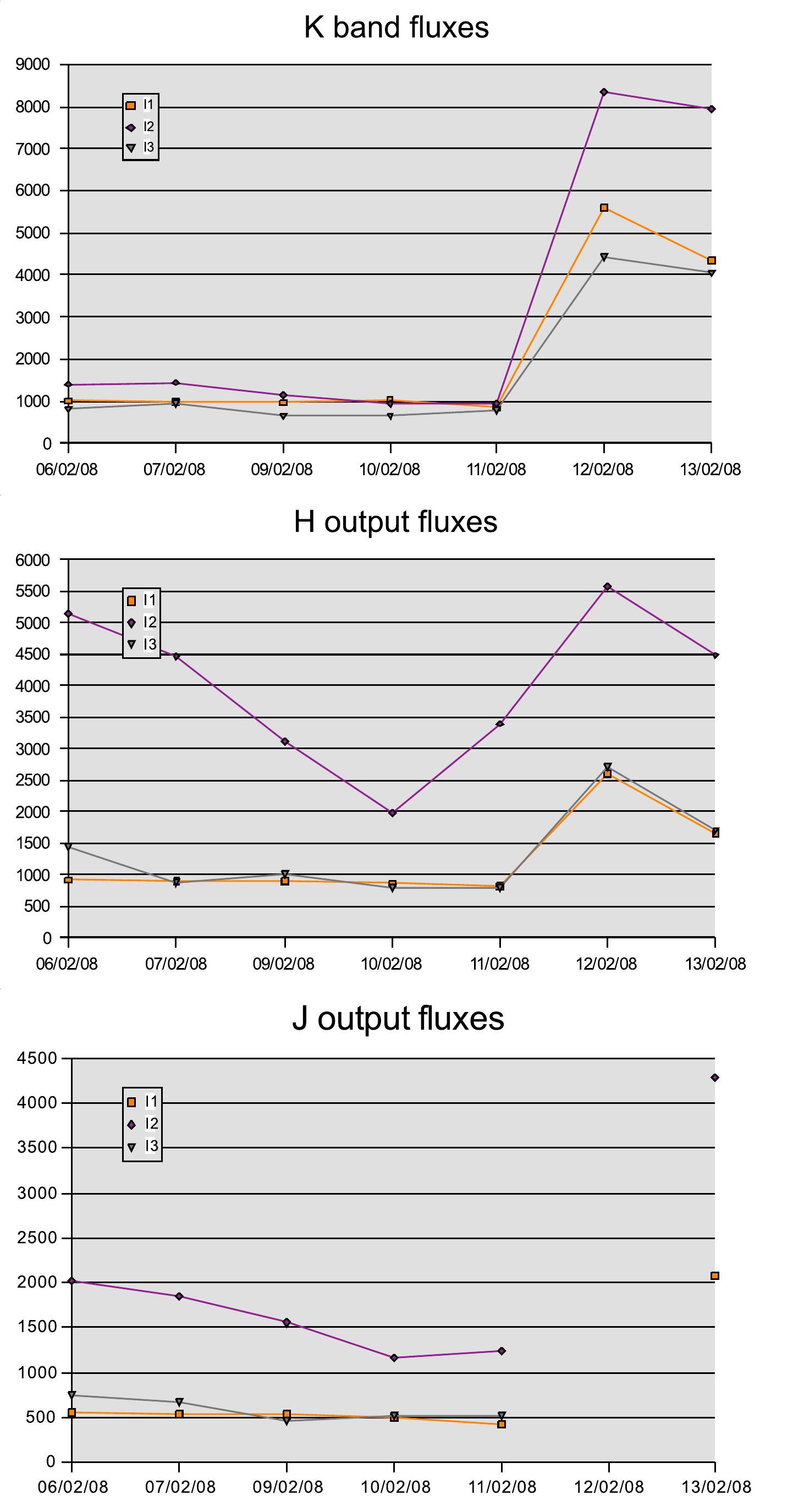}
  \caption{Optical stability of AMBER. Left panels: X-positions of the
    output beams during the ATF week. Middle panels: Y-positions of
    the same beams.  The green zones correspond to a tolerance of 10\%
    contrast loss (see Fig.~\ref{fig:tolerance}).  Right panels:
    fluxes during the ATF week. The flux increase at the end of the
    week corresponds the removal of the polarizers (gain of factor 2
    to 3) and the change of the internal source (gain of about a
    factor 2).}
  \label{fig:optical-stab}
\end{figure}

Left and middle panels of Fig.~\ref{fig:optical-stab} show
respectively the variation of the beam positions in K, H and J in the
X and Y direction, and the right panels display the variation of
output fluxes for the diffrent bands. This survey was performed from
February 6th to February 13th.  On February 12th, the polarizers were
removed explaining the significant improvement of the flux level.

\begin{figure}[p]
  \centering
  \includegraphics[width=0.35\hsize]{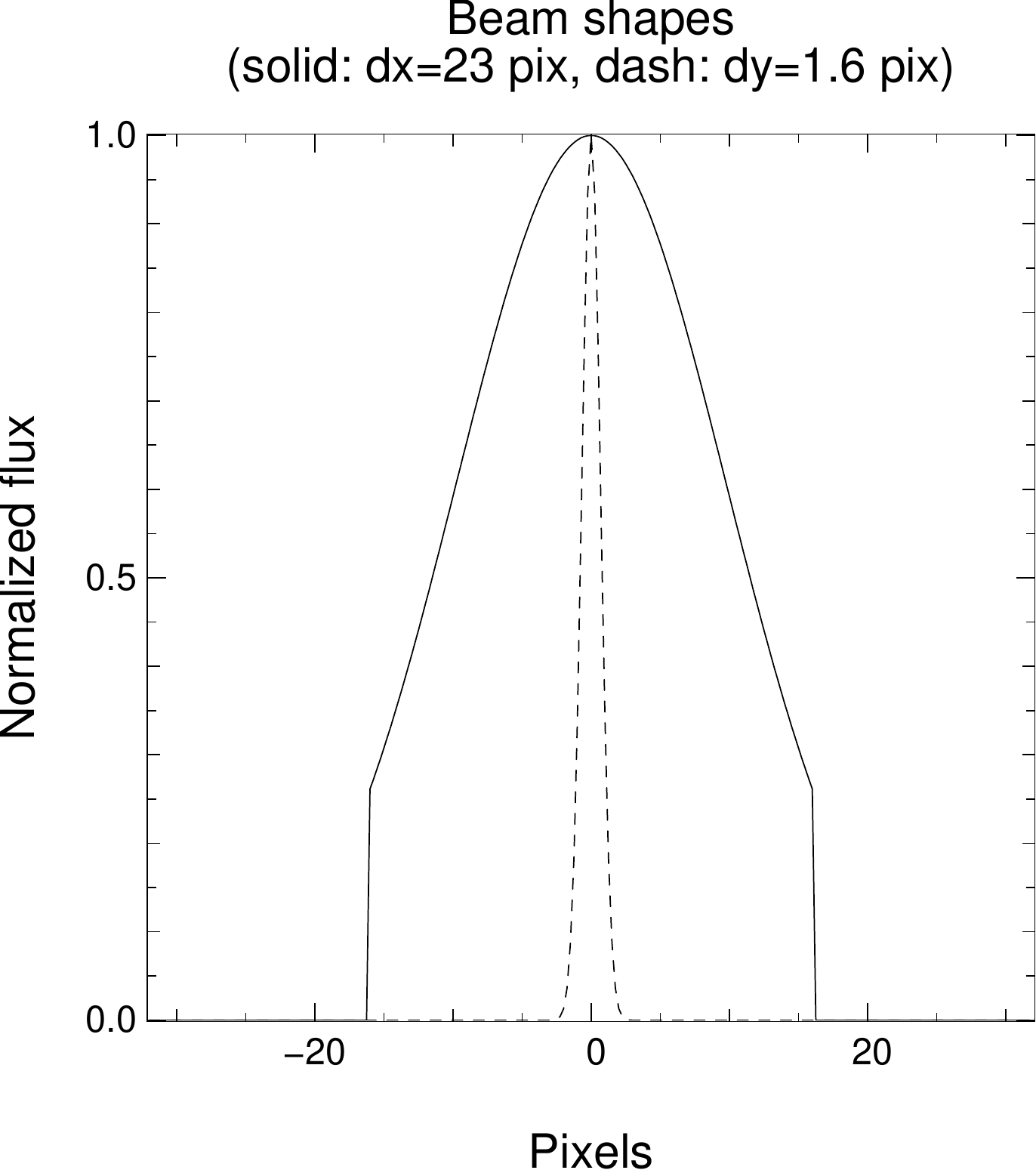}
  \includegraphics[width=0.35\hsize]{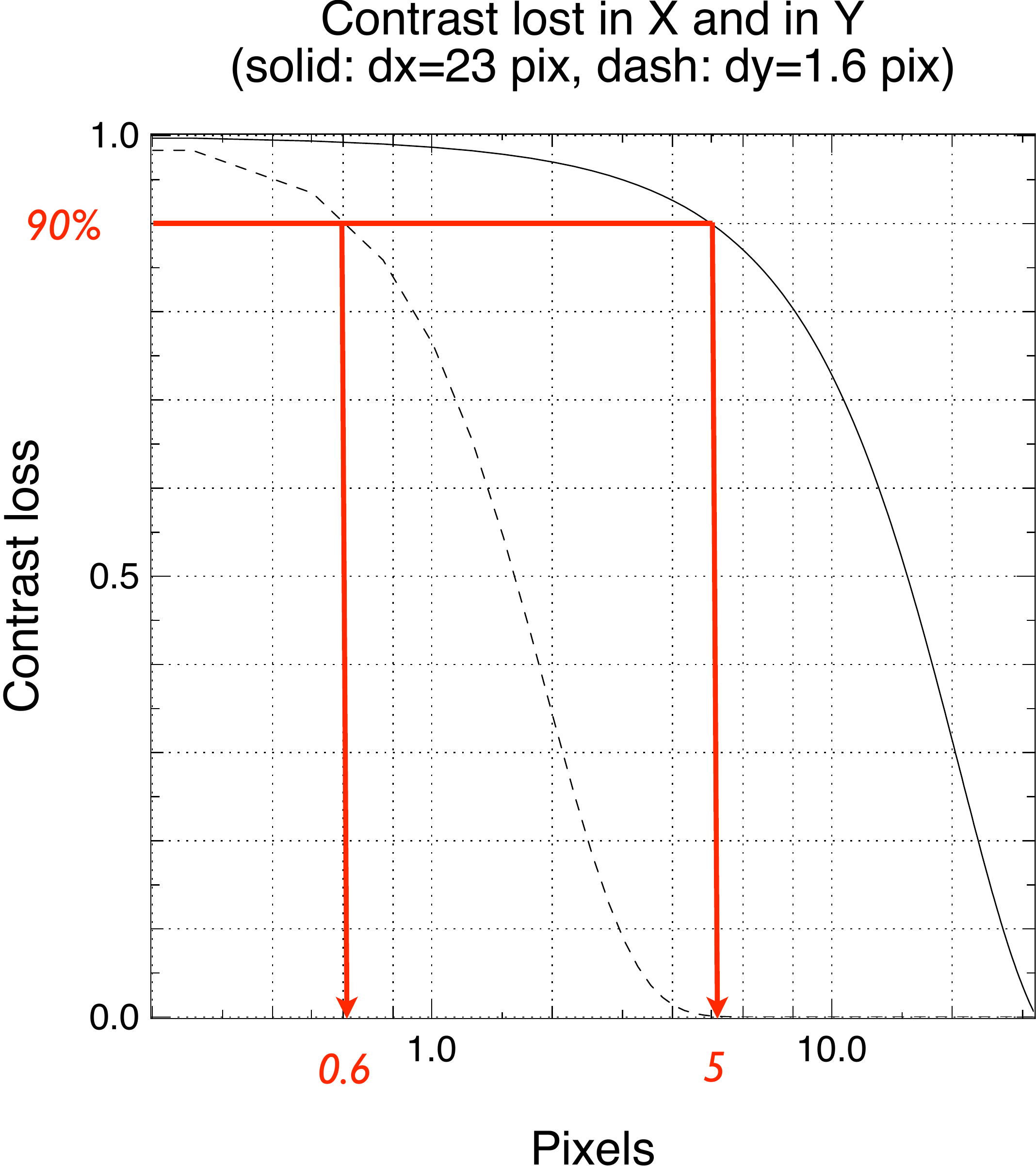}
  \caption{Tolerance of the overlapping output beams. Left: beam
    shapes for the X direction (solid) and Y direction
    (dashed). Right: tolerance for the contrast loss in the X
    direction (solid) and the Y direction (dashed).} 
  \label{fig:tolerance}
\end{figure}

On Fig.~\ref{fig:tolerance}, we plotted on the left the beam shapes
for a typical value of FWHM of 23 pixels in X-direction (solid line)
and of 1.6 pixels in Y-direction (dashed line). On the right part of
the figure is represented the contrast loss due to overlapping
mismatch in pixels. One see that less than 10\% losses requires
overlaping at the level of 5 pixels in X and of 0.6 pixels in Y. The
fact that the positions are within the green zones in
Fig.~\ref{fig:optical-stab} shows that during the ATF run the beams
were always within specifications. 

Our understanding is that the effort done by Paranal to achieve a
better stability of AMBER by cutting and strengthening the
optomechanical elements were fruitful. 

\subsubsection{Sensitivity to vibrations}

\begin{figure}[p]
  \centering
  \includegraphics[width=0.8\hsize]{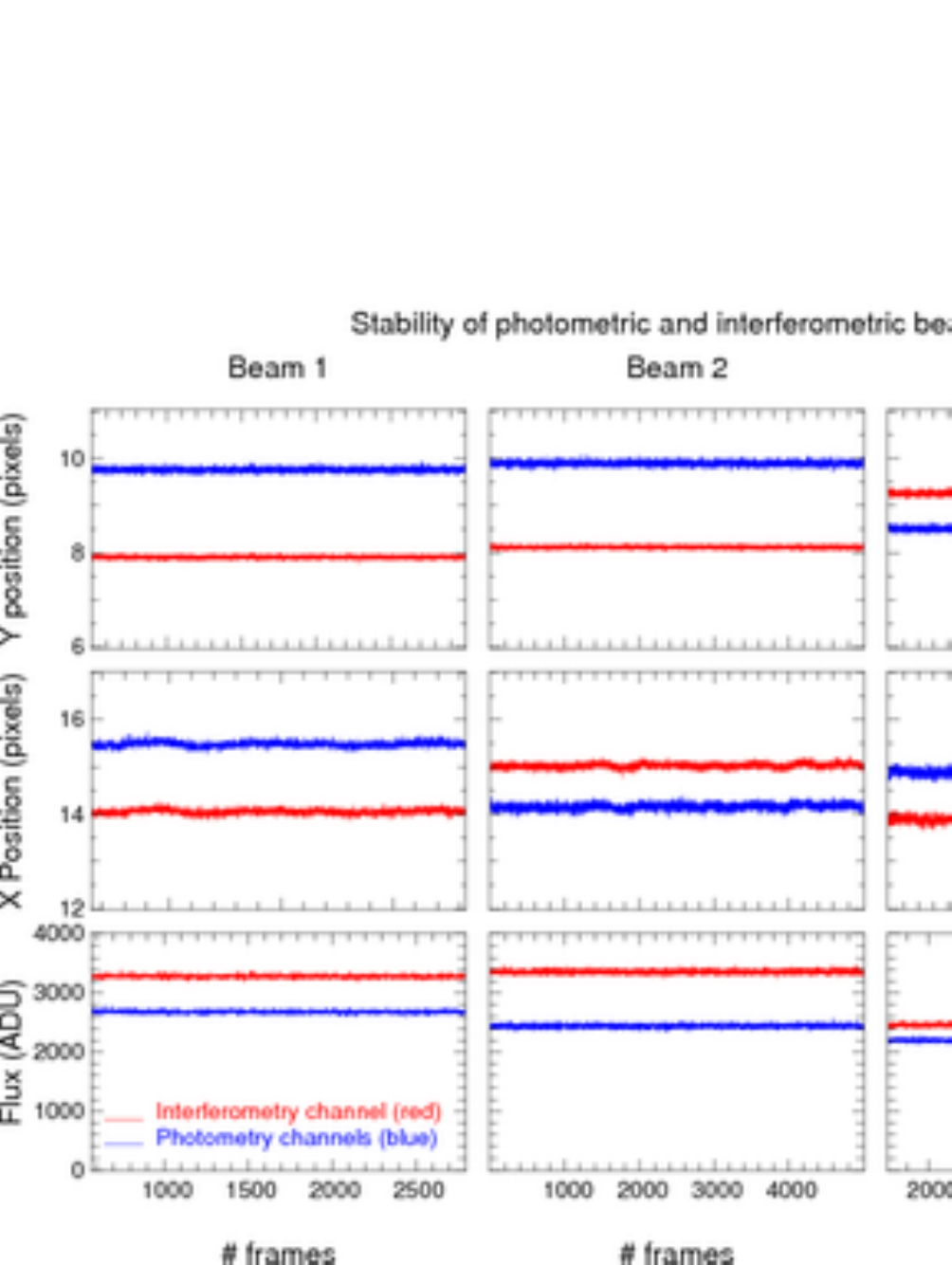}
  \caption{Stability of output beams in X- and Y-position and flux
    with DIT of 20ms. Top and center panels are Y- and X-positions of
    the 3 beams in their photometric channels. Bottom panels are the
    flux levels. The variations are at maximum within a fraction of pixels.}
  \label{fig:optical-fast}
\end{figure}

When we were looking for the origin of the fluctuations of the
visibility (see Sect.~\ref{sec:output-beam-stab}), we have recorded a
set of 1000 frames with the grating in the 0$^{th}$ order in order to
monitor at relatively high frequencies (more than 10Hz) the motion of
the output beams. 

The result of this test is reported in
Fig.~\ref{fig:optical-fast}. The variations are at maximum within a
fraction of pixels, unable to explain visibility variations of a few
percent. We concluded that the output beams are not responsible of the
visibility variations seen.

\subsection{P2VM stability}
\label{sec:p2vm-stability}

\begin{figure}[p]
  \centering
  \includegraphics[width=0.9\hsize]{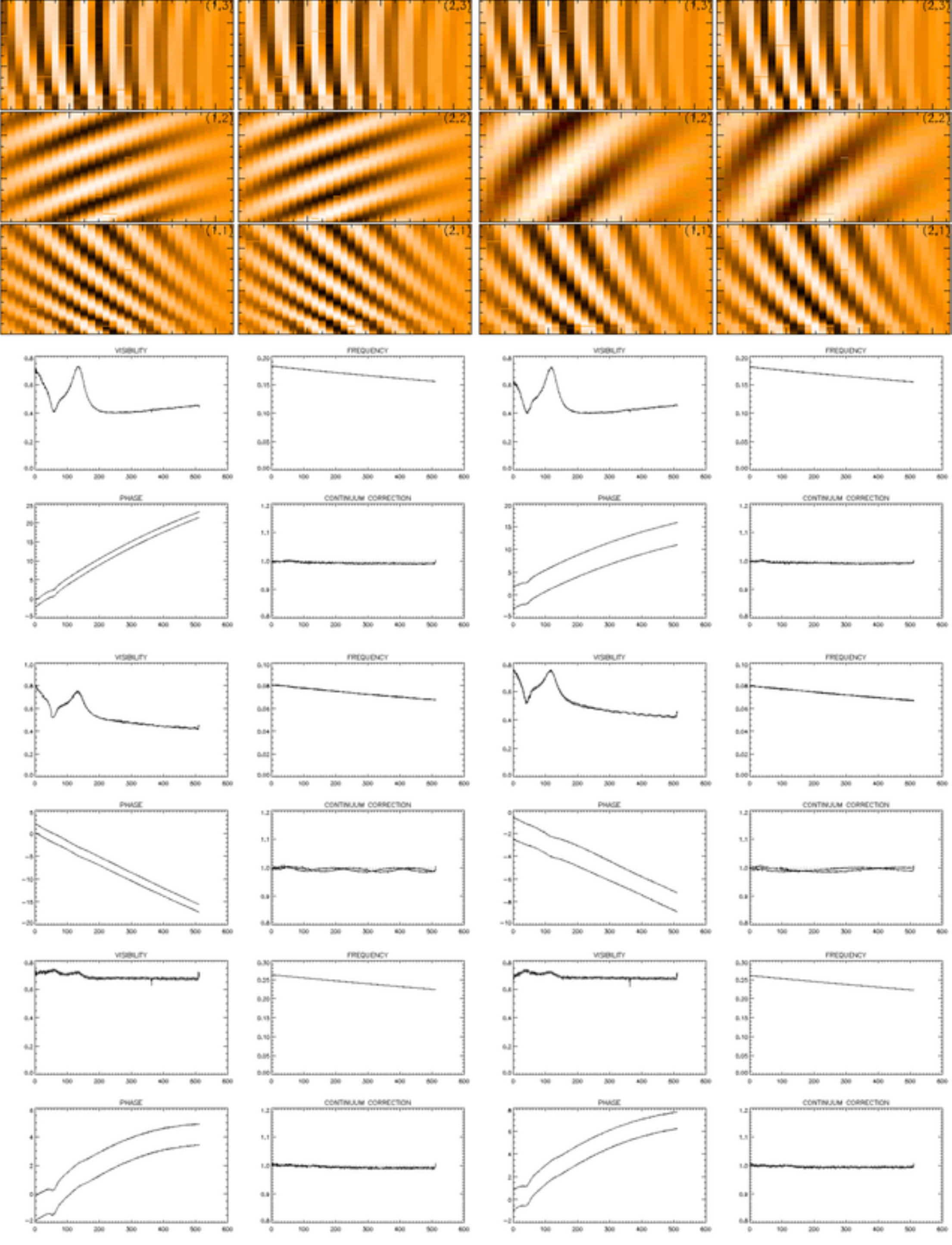}
  \caption{P2VM stability illustrated by the P2VM taken at the
    begining of the sky tests (left) and at the end (right) without
    changing the spectral configuration. See text for details.}
  \label{fig:P2VM-stability}
\end{figure}

The stability of calibrations beams have characterized during the last
ATF night (see Fig.~\ref{fig:P2VM-stability}). Left subfigures
correspond to the P2VM acquired at the begining of the night, whereas
the right ones to the end of the same night. The images at the top of
the figures shows the carrying waves. Fringes of baselines 1-2 and 1-3
have not moved significantly whereas the fringe of baseline 2-3 show
different pistons. However the analysis of the  P2VM using P2VM
modeling reported in graphs below show that the envelope and
frequencies have not changed. Only the phase has changed. 

This is due probably to translation stages of the input dichroics
which have been found very unstable. This is a well known problem
reported in some PPRS. When used too many times, then the OPD is
unstable. However we think it would be possible to make the P2VM
calibration process unsensitive to the piston so that this problem
should not be a problem. This needs further investigation.

\subsection{CAU stability}
\label{sec:bang-test}

We did a series of experiment on the CAU to test its stability. During
the search for the origin of visibility fluctuations, we noticed that
the visibility measurements were very sensitive to some elements of
the CAU, especially the large CAU parabolic mirror. 

\begin{figure}[p]
  \centering  
  \includegraphics[width=0.4\hsize]{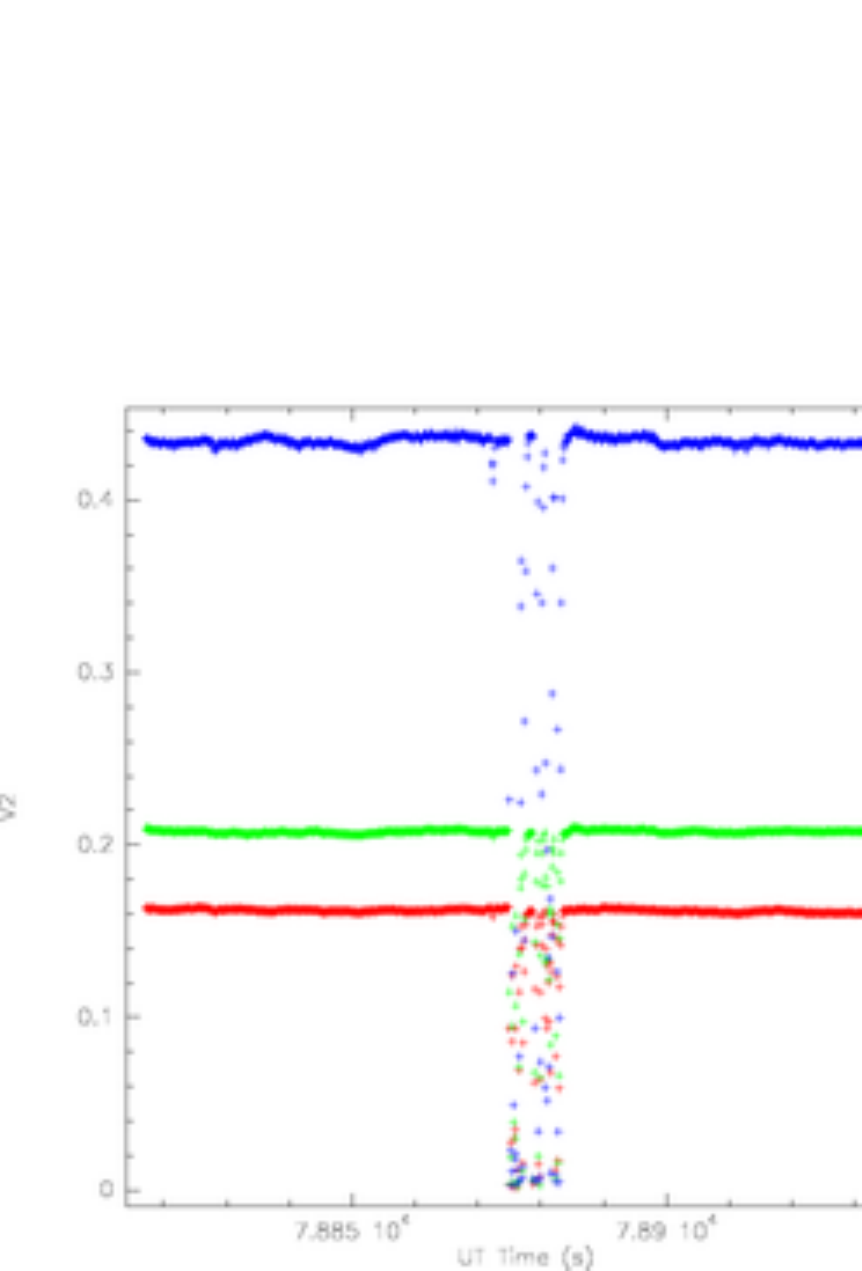}
  \includegraphics[width=0.4\hsize]{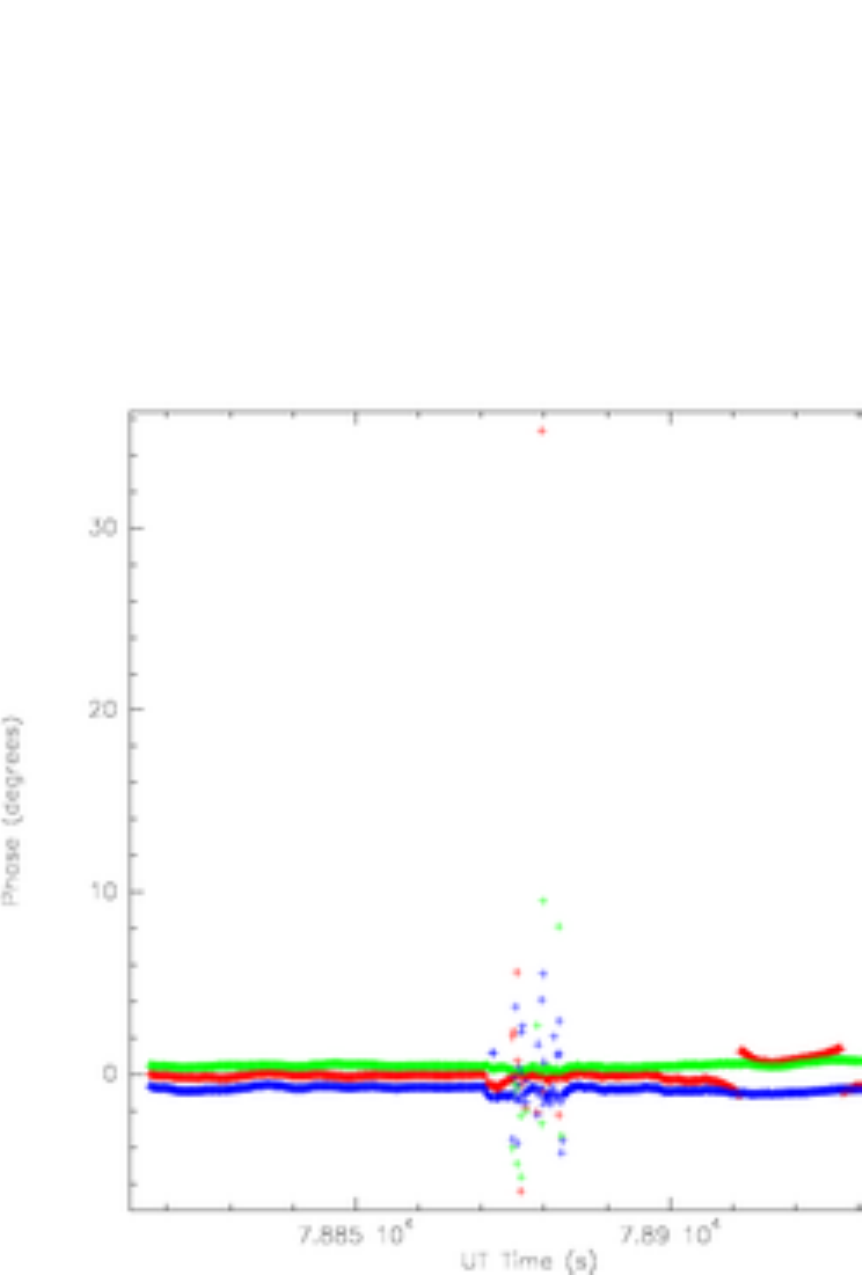}
  \includegraphics[width=0.4\hsize]{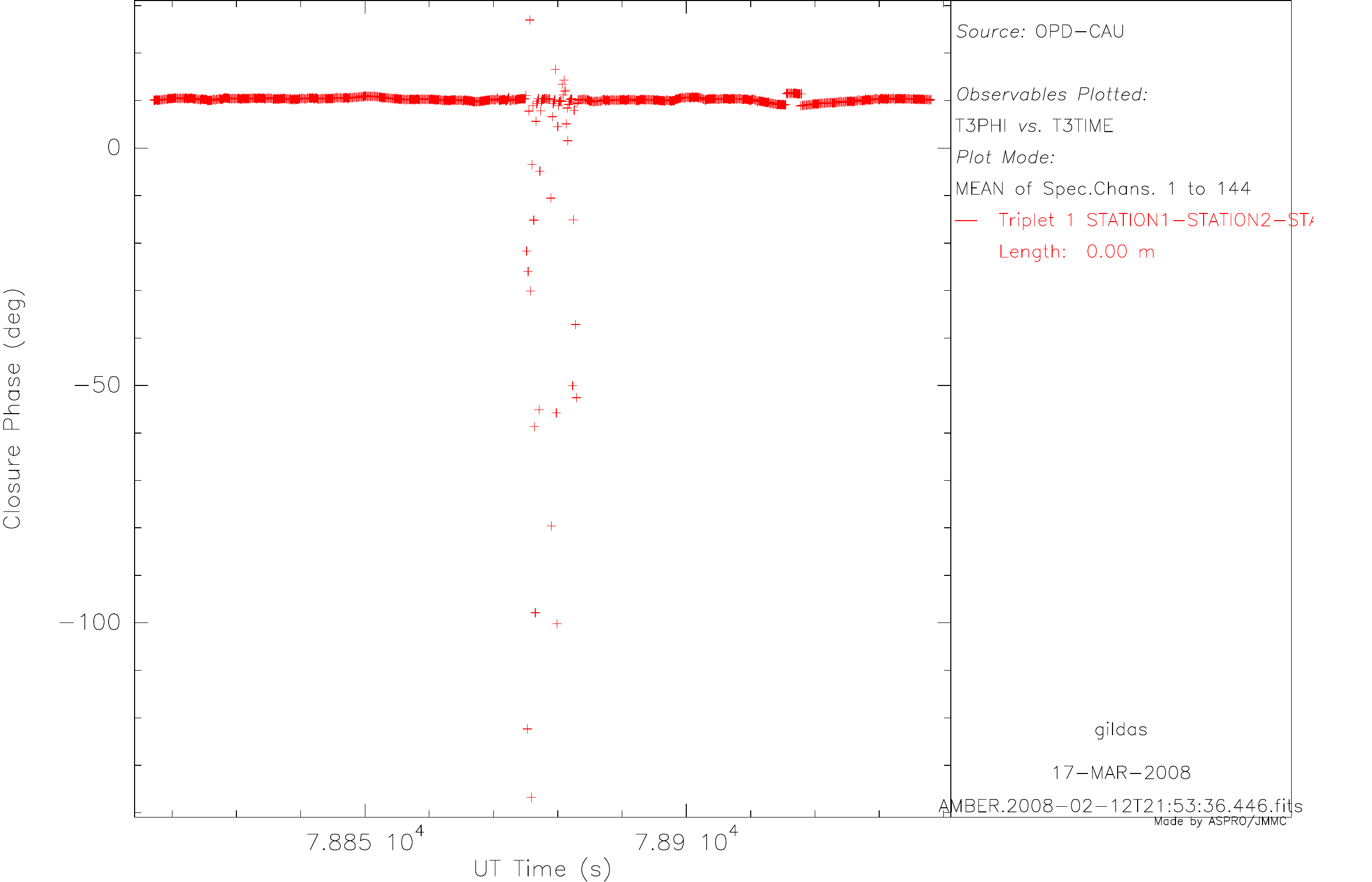}
  \includegraphics[width=0.4\hsize]{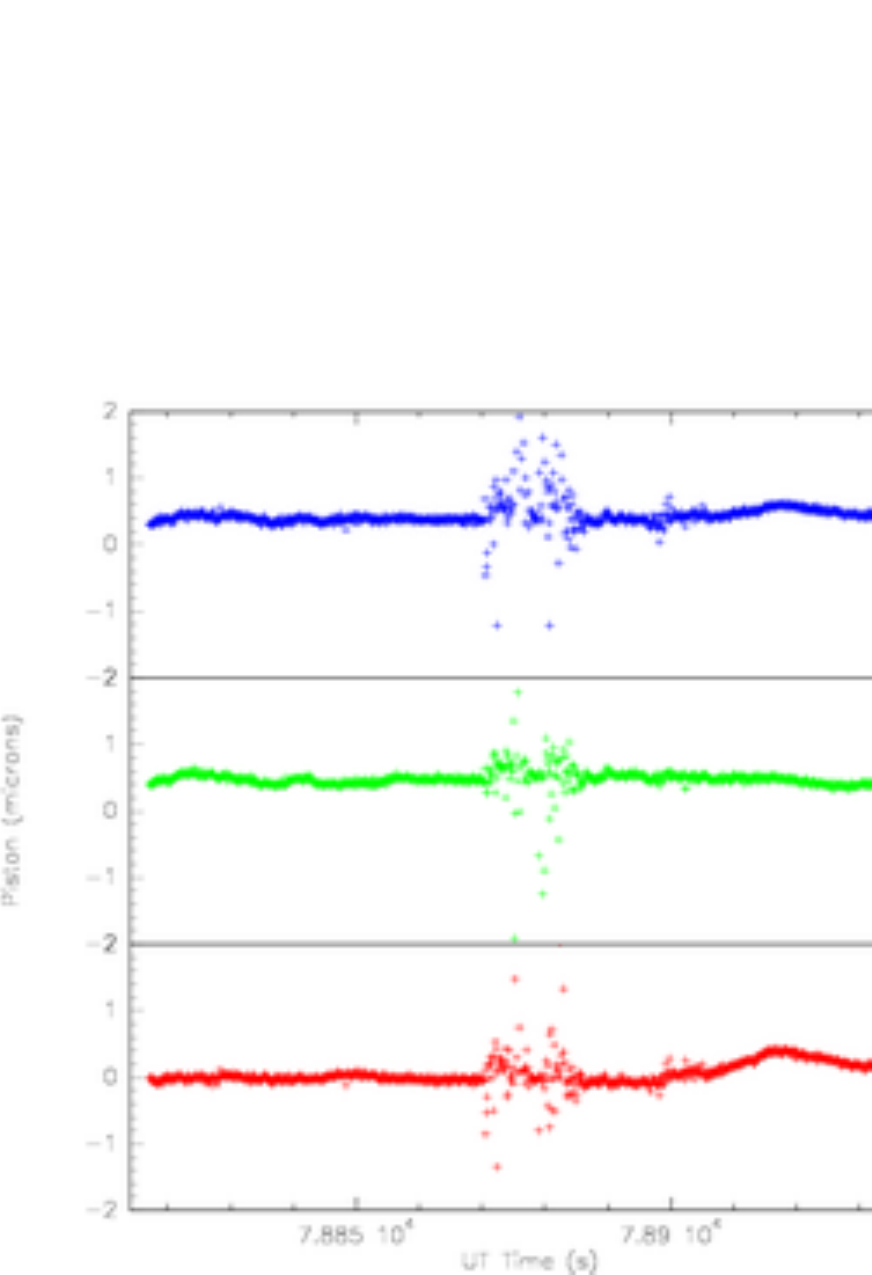}
  \caption{Test of the CAU stability. During observation of the
    internal source, several optical mounts of the
    CAU were touched in order to detect any special sensitivity to
    external stimuli. Upper left: squared visibility; upper right:
    differential phase; bottom left: closure phase; bottom right:
    piston. One sees the impact of touching the CAU large parabolic
    mirror and its relaxation time.}
  \label{fig:cau-bang}
\end{figure}

\begin{figure}[p]
  \centering
  \includegraphics[width=0.8\hsize]{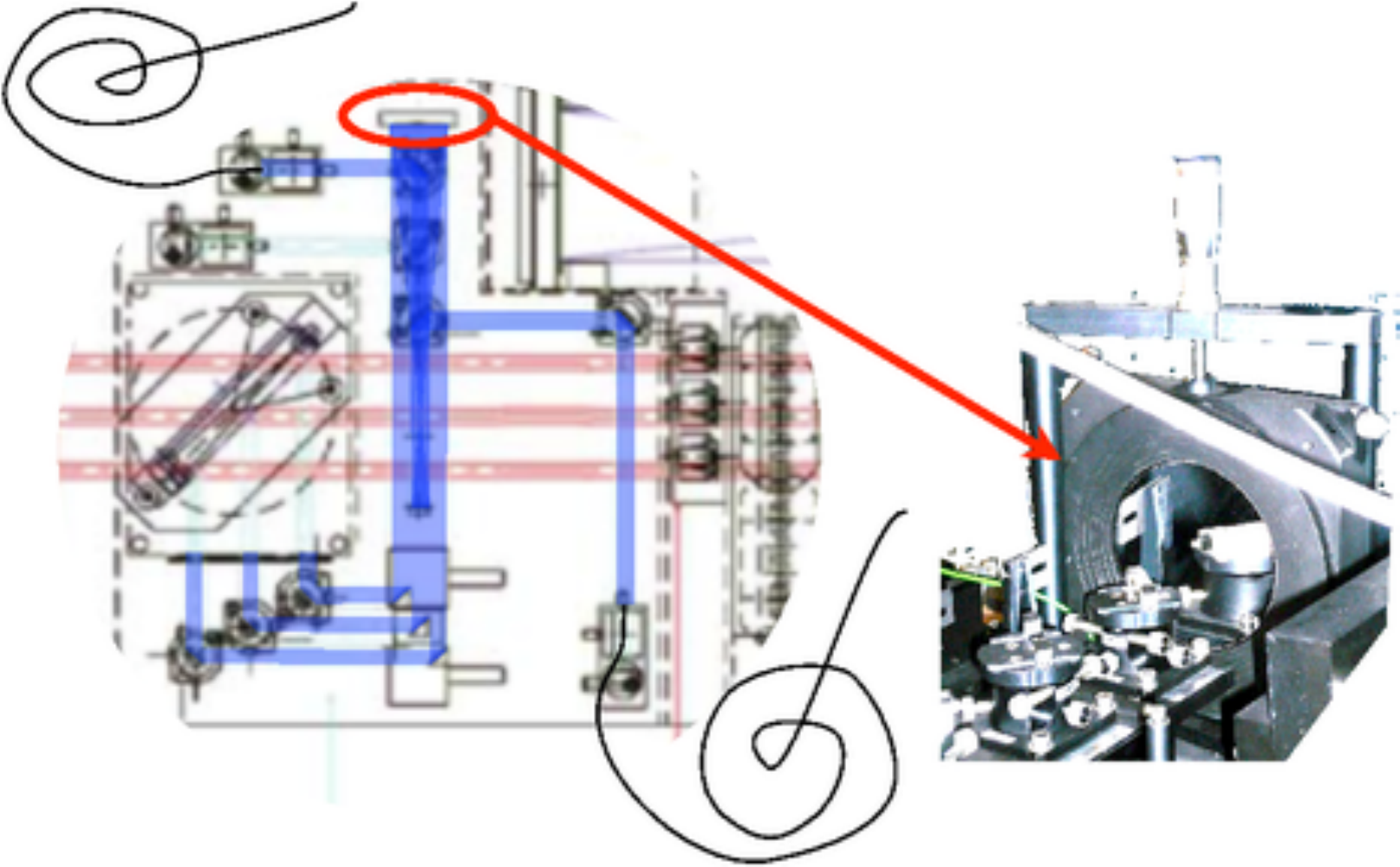}
  \caption{Layout of the CAU (calibration alignement unit) with a
    picture of the CAU large parabolic mirror that we suspect to be
    unstable and to generate OPD instabilities.}
    \label{fig:cau-parabolic}
\end{figure}

In Fig.~\ref{fig:cau-bang}, we have recorded an exposure (1000 frames)
on the CAU source while touching different mechanical supports of the
CAU. We found that the mount of the large CAU parabolic mirror
(identified on Fig.~\ref{fig:cau-parabolic}) was very sensitive to
mechanical impacts. We think that it may be the source of the
instabilities seen when we were looking to the effect of polarizers
(see appendix~\ref{app:POL}). 


\subsection{Conclusion: on alignment requirements and health checking}
\label{sec:alignment-health-check}

We recommend to relax the alignment requirements to 8 pixels (see
fig.~\ref {fig:tolerance}), in order to avoid too frequent
interventions in the laboratory. However, this requirement holds for
each pair of beams, at each wavelength.  We recommend that in the p2vm
computation a fit of the position of each beam, as seen in the
interferometry channel, as a function of lambda, be performed, that
these values be used as health check, and that a warning be issued
each time a significant percentage ($5\%$) of wavelengths are above
these specs.

We also recommend to check the fixation of the large CAU parabolic
mirror.

\clearpage
\section{Analysis of the spectral resolution}
\label{app:spectral-resolution}

\begin{figure}[tbp]
  \centering
  \includegraphics[width=0.6\hsize,angle=270]{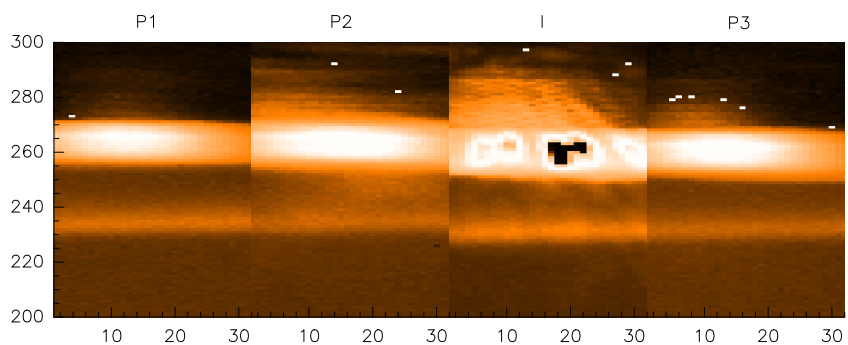}
  \caption{Image of the slit in September 2007. the image is very
    defocussed (4 pixels HPBW instead of less than 2 normally), and a
    ghost images are seen above and below the slit image up to 30
    pixels away.}
  \label{fig:tiltedslit}
\end{figure}

\subsection{Spectral shifts between photometric and interferometric
  channels} 
\label{app:spectral-shifts}

The spectrograph is well-aligned. The spectral shift between
photometric channels and the interferometric ones is characterized
and should not be recomputed each time (gain of observing time).

The spectral shift is primarily due the combination of the position of
the slit (large, narrow, etc) and of a small angle between the optical
components that fold the 3 photometric beams on the camera (3 glued
total-reflexion prisms). The figure \ref{fig:ghosts} show such images
of the slit.  At the worst, this combination changes by a small amount
every time the spectrograph is opened.  To our knowledge, this defines
only 3 periods: 2004--Sep,~2007; Sep, 2007-Jan,~2008 and Feb,
2008--present.

The repositioning of the slit by the slit wheel motor \emph{seems}
(TBC) to be sufficiently accurate to insure that the slit+folding
device setup did not change during these periods.

The displacement of \emph{the image of the slit} on the camera measures
the main displacement.  This can be done only once each time the
spectrograph is opened, and could be done in the following manner:
\begin{enumerate}

\item take 3 exposures full frame images of the narrow slit, one beam
  lit at a time, grating at 0 order. 
\item sum each image on the X axis (taking care of bad pixels. Flat is
  not needed)
\item on these 3 vectors, fit two gaussians of FWHM$\sim 32 pixels$.
  Their position will serve as a marker of the position of each beam;
\item average a few long (1 min) exposures full frame images of the
  narrow slit, \textbf{all shutters closed};
\item extract 4 strips in the Y (lambda) direction, of 64 pixels wide,
  centered on the positions of the beams found before (taking care of
  bad pixels. Flat is not needed);
\item sum each strip on the Y axis, fit a gaussian (of a few pixels
  wide) in each. The displacement of the gaussian positions of strips
  1,2 and 4 wrt. strip 3 is the displacement sought. The width of the
  gaussian is an indication of the quality of imaging of the slit onto
  the camera (hence the spectral resolution). 
\item to evaluate the \emph{quality of the focussing of each beam} the same
  procedure can be reproduced with one beam lit at  a time.

\end{enumerate}

We have done measurements during our ATF stay, and have found old
images of the slit taken in 2004, permitting also this kind of
measurement. Table~\ref{tab:displacement} summarizes these values.

\begin{table}[b]
  \begin{center}
    \caption{Displacement of the photometric channels with respect to the
      Interferometric channel (in pixels).}
    \label{tab:displacement}
    \medskip
    \begin{tabular}{|l|l|l|l|l|}
      \hline
      Period	& P1&	P2&	P3&Notes\\
      \hline
      2004-Aug2007&	+0.873&	+2.064&	+0.472&1\\
      Sep2007-Jan2008&	+2.000&	+2.576&	+0.263&2\\
      Feb2008-Present&	+3.008&	+2.850&	-1.319&3\\
      \hline
      \multicolumn{5}{p{0.5\hsize}}{\footnotesize
        Notes of the table:\hfill~\linebreak 
      1. These 2004 values seem to hold at least until 2005 Feb. It
      is necessary and easy to check that it holds until
      Sep07.\hfill~\linebreak  
      2. Based on Nov07 measurements (SPECPOS frames). 
      Slit was {\bf very} tilted and slit badly focussed
      (See fig~\ref{fig:tiltedslit} and appendix~\ref{app:char-ghosts} comments).\hfill~\linebreak 
      3. Values should be measured again on background-illuminated 
      image of the slit. We believe these are however good values for
      data reduction purposes.}
    \end{tabular}
  \end{center}
\end{table}

\begin{figure}[t]
  \centering
  \includegraphics[width=0.6\hsize]{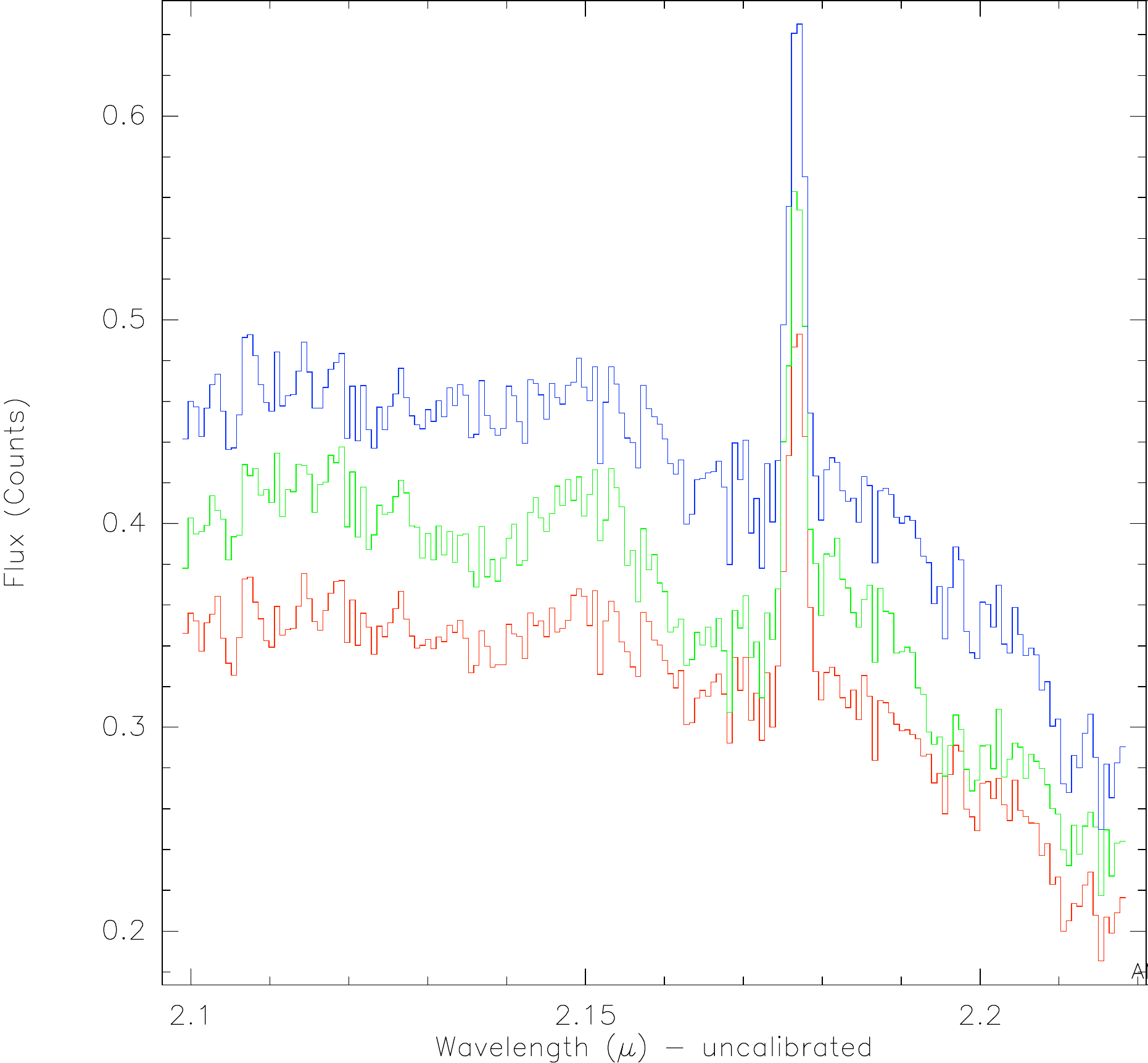}
  \caption{Spectra of $\alpha$ Arae measured in Feb 2005 during Science Demonstration Time. The curves corresponds to beam of each telescopes and have been shifted with a fixed values computed at a different date. }
  \label{fig:shifts}
\end{figure}

The 2004 values seem to hold at least until 2005 Feb, 25 since they
work well with $\alpha$ Arae measurements (see Fig.~\ref{fig:shifts}). It is
necessary and easy to check that is holds until Sep 2007.

During the replacement of the infrared Hawaii detector in Sep, 2007, a
slight defocus and tilt of the slit image was introduced, which was
not corrected until Feb, 2008. During this period, the image of the
slit on the camera worsened severely, as seen in
fig~\ref{fig:tiltedslit}. Such defocussing of the slit induces a loss
of spectral resolution, and the extenuation of the visibility with the piston
(sect.~\ref{app:extenuation} worsens: in LR mode, the maximum tolerable
piston drops to $\pm20$ microns.

Please note, that \textbf{subpixel} shifting, as it is done in amdlib
(and the author does not see how it could be done otherwise)
introduces biases in spectrum and P2VM in presence of \textbf{bad
  pixels}. We recommend to use integer pixel shift for the moment.
($+1$, $+2$, $+0$ for the old data,$+2$, $+2$, $+0$ for the
sep07-Jan08 data ).

In operation, three anamorphed beams pass through the slit. These
beams are, if amber is correctly aligned, highly elongated ellipsoidal
flux distributions tilted with a very small angle ($\leq0.5^\circ$) wrt the
slit, whose shape can be measured at 0th order with a good precision.
Thus, one could in principle apply psf deconvolution on the P1, P2 and
P3 spectra, possibly to gain in spectral resolution in the
\emph{spectrum}. It remains to be proved that the same could be done in the
I channel, taken into account in the P2VM fitting process, and provide
a sensible improvement in the results.

\subsection{Spectral calibration in low resolution JHK mode}

We performed a Fourier Transform Spectrograph analysis of AMBER's Low
JHK spectra by moving incrementally by steps of 0.01 microns the
piezos on beams 1 and 3 for each band. Each pixel of the
Interferometric region on the camera of
amber (32 by 60 pixels in this LR mode) is thus modulated with a
period equal to the wavelength it sees. Furthermore the visibility of the 1-2 and
2-3 fringes will vary as a function of piston and will change as the
fourier transform of each spectral channel shape.

\begin{figure}[t]
  \centering
  \includegraphics[angle=270,width=0.55\hsize]{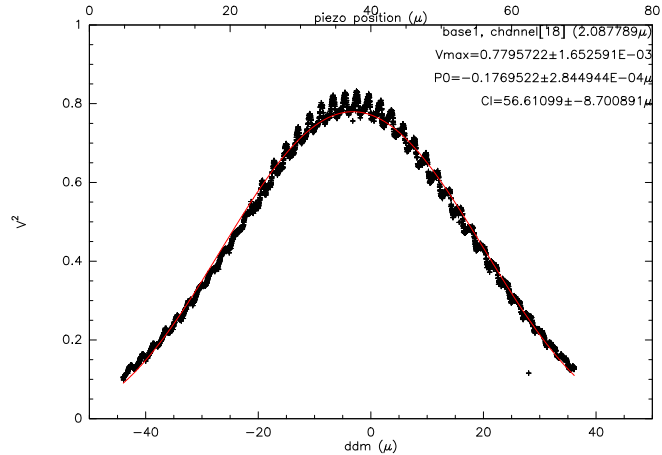}
  \caption{Typical extenuation figure obtained in the K band region
    during our piezo scan. The red line is the fit of
    the extenuation function whose parameters are available in the
    upper right corner of the box. }
  \label{fig:extenuation-scan}
\end{figure}

We acquired four observations of $10^4$ images, covering piezo
displacements of $80\,\mbox{$\mu$m}$ with $0.01\,\mbox{$\mu$m}$
resolution, with the polarizer present, so the data is perturbated by
the phase beating effect depicted in Sect~\ref{sec:phase-beating}, and
the dispersion law was thus obtained for the H and K regions of the
spectrum only. A typical result is depicted in
fig.~\ref{fig:extenuation-scan}, where the phase beating perturbation is
clearly visible.
  
The Dispersion law is found compatible with a linear dispersion of
coefficients:
$$\lambda=2.6917-3.2082\cdot10^{-2}\times\mathrm{i}$$
with i as the pixel number.

\subsection{Calibration of visibility losses due to temporal coherence}
\label{app:extenuation}
Table~\ref{tab:coherence-length} presents the 'equivalent coherence
length' $\Delta\lambda$ of the extenuation of $V^2$ as a function of
piston 'opd'. It has been computed through the K band by fitting the
extenuation law in sinc expected from a square spectral channel of
bandwith $\Delta\lambda$ and
$$V^2=V^2_\mathrm{max}\times\{\frac{\sin(\pi\frac{opd}{\Delta\lambda}+P_0)}{\pi\frac{opd}{\Delta\lambda}+P_0}\}^2 $$
Table~\ref{tab:max-piston} is a subproduct of the data reduction
used in the measurement. It shows the maximum piston
measurable with the Low\_JHK spectral resolution in case of good
S/N.  
 
  \begin{table}[p]
    \centering
    \caption{``Coherence Length'' and central wavelength of each
      spectral channel in the K band}
    \label{tab:coherence-length}
    \medskip
    \begin{tabular}{|c|c|c|}
   \hline
LR Spectral&$\lambda$ & $\Delta\lambda$\\
Channel number&(microns) & (microns)\\
   \hline
1 & 2.6596&46\\
2 & 2.6275&51\\
3 & 2.5954&53\\
4 & 2.5633&52\\
5 & 2.5312&51\\
6 & 2.4992&52\\
7 & 2.4671&55\\
8 & 2.4350&58\\
9 & 2.4029&59\\
10& 2.3708&58\\
11& 2.3387&57\\
12& 2.3066&55\\
13& 2.2745&57\\
14& 2.2425&59\\
15& 2.2104&59\\
16& 2.1783&56\\
17& 2.1462&58\\
18& 2.1141&57\\
19& 2.0820&59\\
20& 2.0499&50\\
21& 2.0179&54\\
22& 1.9858&58\\
23& 1.9537&33\\
    \hline
    \end{tabular}
  \end{table}

  \begin{table}[p]
    \centering
    \begin{tabular}{|c|c|}
   \hline
Band&Maximum Measurable Piston \\
Channel number&(microns)\\
   \hline
J&$\pm20$\\
H&$\pm27$\\
K&$\pm43$\\
   \hline
    \end{tabular}
    \caption{Maximum measurable OPD between beams in Low resolution mode.}
    \label{tab:max-piston}
  \end{table}


\clearpage
\section{Internal light sources and effect on closure phases}
\label{app:internal-sources}

\subsection{Analysis of internal sources}

We investigated how the CAU sources, also called RAS, could be
simplified and understood. We explain below some tests which were
performed.

\begin{figure}[hbp]
  \centering
  \includegraphics[width=0.5\hsize]{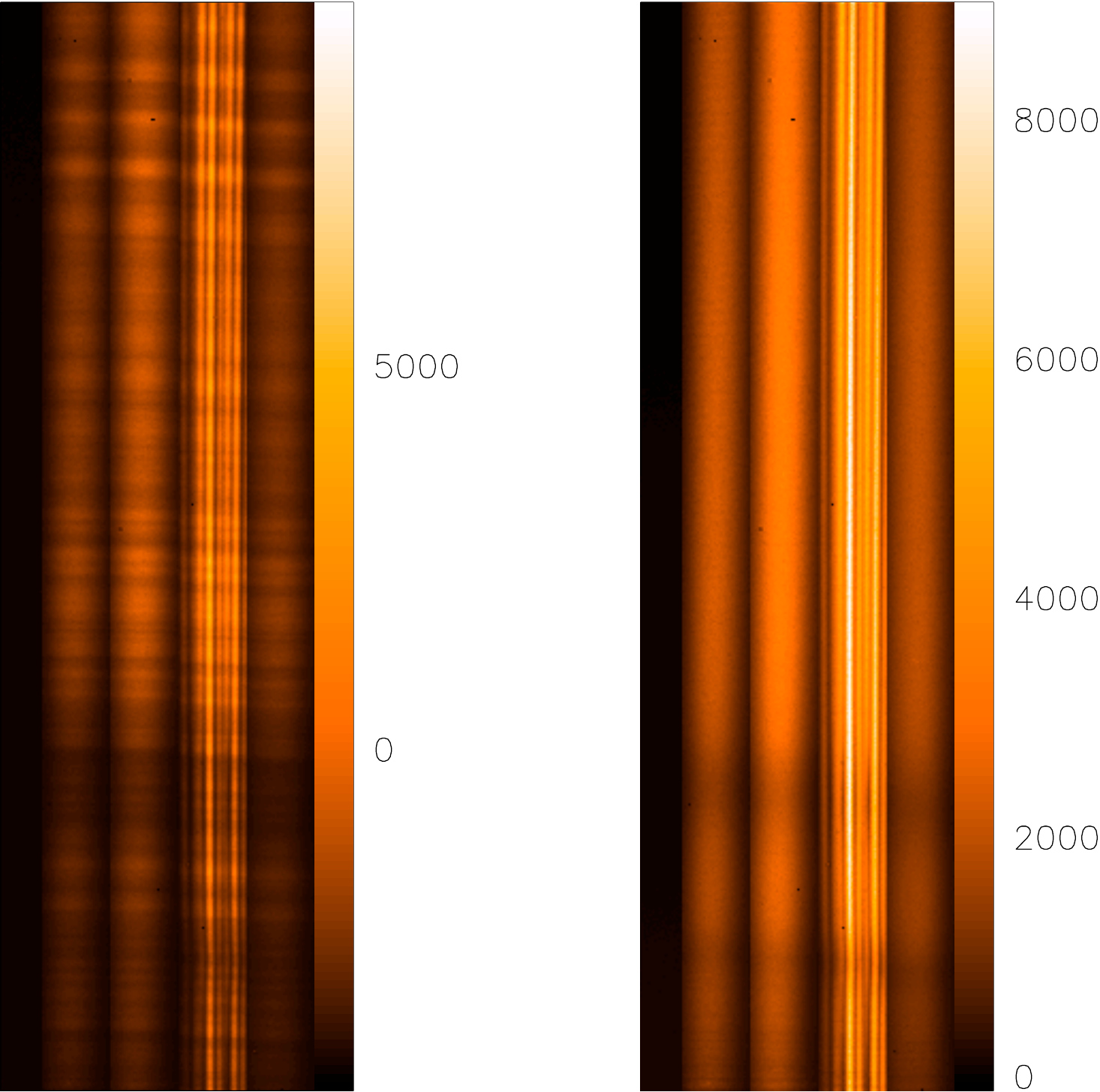}
  \caption{AMBER MR data obtained with the current source (left) and with the new proposed setup (right). See text for details.}
  \label{fig:ras-sources-MR}
\end{figure}

\begin{itemize}
\item \textbf{RAS dichroics effects:} For tests the RAS has been replaced by a
  laboratory source directly connected to the K band fiber of the CAU. In this
  condition, while the K band dichroic of the RAS is shunted, the modulation of
  the flux in all photometric channels is significantly reduced and is not
  detectable any more in the raw data in MR (see
  Fig.~\ref{fig:ras-sources-MR}). LR image in JHK in this condition allows a
  detection for all bands. The signal in J and H bands is then highly affected
  by the fact that the CAU dichroic acts in this test setup, in reflection while
  it is supposed to be operated in transmission for this spectral range. One can
  retrieve the reflection curve of the dichroic presented in
  VLT-TRE-AMB-15830-1010 section 9 (AMBER OPM test report).

\item \textbf{CAU fibers:} It has been observed for a long time by AMBER
  operators than the 2 CAU fibers suffer from incorrect superposition,
  as the H fiber seems to be moving and requires some
  readjustments. This misalignment lead to non-zero phase closure
  measurement on the CAU, with an unstable value. The measurements
  done with a laboratory source, shows that a single K band fiber is
  able to provide the illumination for all wavelengths. Even if the
  core of the K band fiber can be resolved by the AMBER setup, this
  effect is properly taken into account by the calibration procedure
  of the P2VM. If the CAU remaining dichroic is replaced by a mirror,
  then the flux in all bands are sufficiently balanced with a suitable
  transmission. We recommend to replace the CAU dichroic and its
  support by the equivalent mirror and support which deflects the
  light coming from the H band fiber. Then we keep only one fiber (K
  band Verre Fluore fiber) for the CAU and to bypass the 2 dichroics
  of the calibrations sources (RAS and CAU).   
\end{itemize}

\subsection{Effect on closure phase calibration}

\begin{figure}[t]
  \includegraphics[width=0.45\hsize]{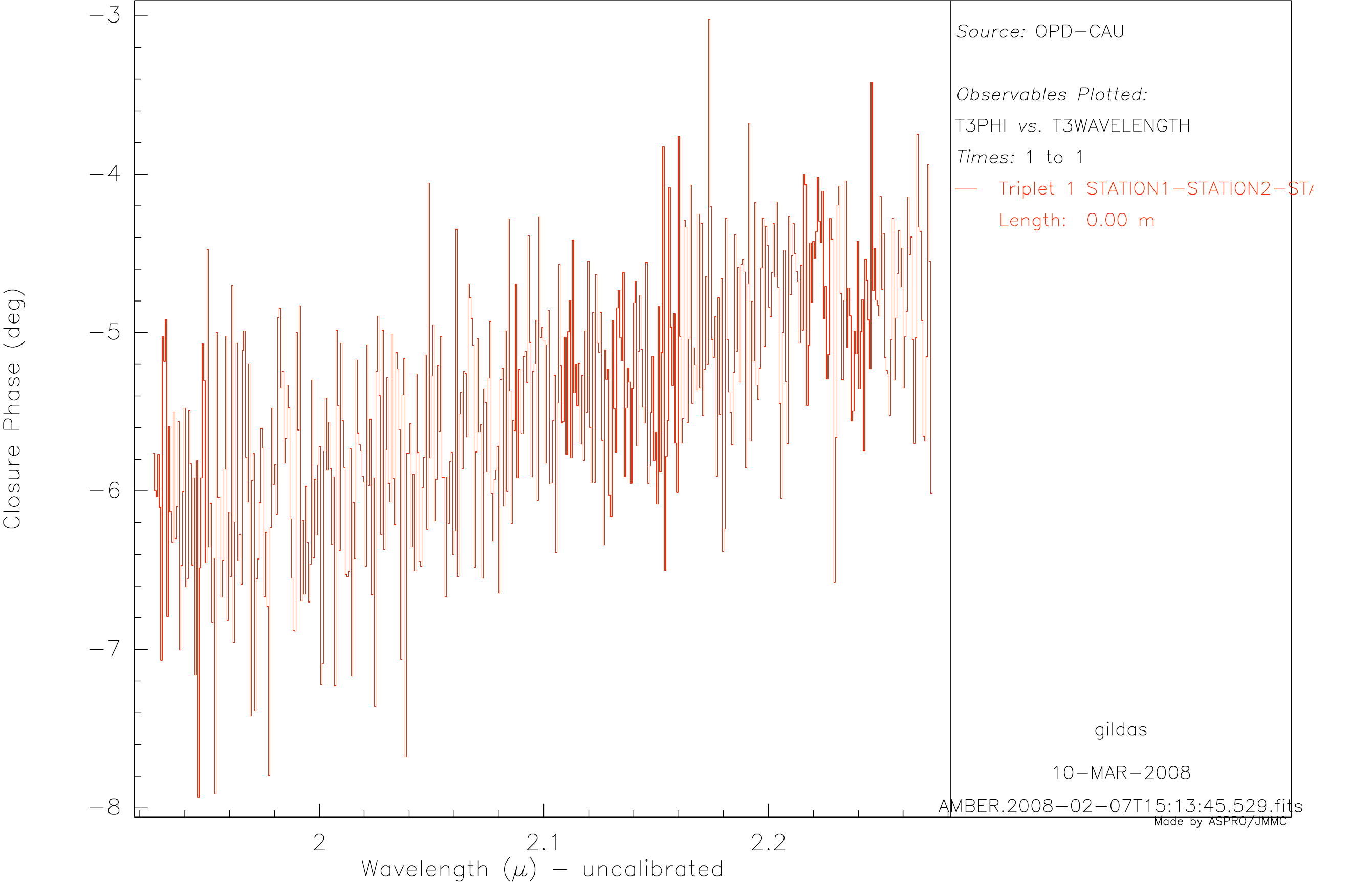}
  \hfill
  \includegraphics[width=0.45\hsize]{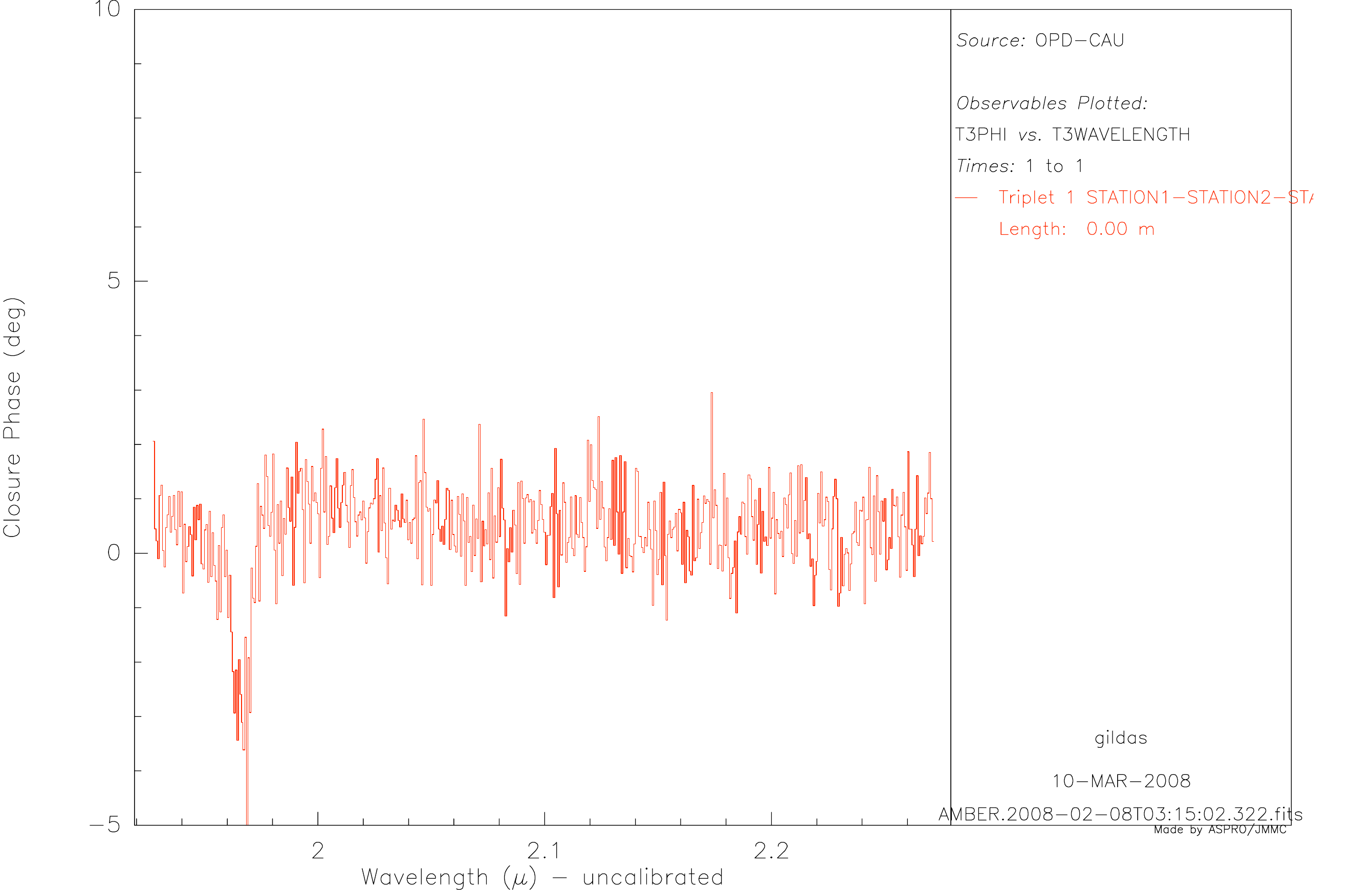}

  \caption{Closure phases measured on the internal source in the K band,
    left leaving the RAS open with residual K light, right with the
    RAS closed.}
  \label{fig:RAS-CP}
\end{figure}

Measurements showed that the closure phase
signal was corrupted. We realize that by shutting down the remote
alignement source (RAS) JH fiber, the K spectrum becomes very
different. Residual K band light the RAS JH fiber polluted the K
source. Therefore the instrument was seeing a binary made by the
strong K fiber and the residual K light going through the JH fiber. By
shutting down the RAS light leads to a closure of 0° all over the K
band (see Fig.~\ref{fig:RAS-CP}).

\clearpage
\section{AMBER polarizers}
\label{app:POL}

The polarizers were found to be at the origin of the phase beating and
partly at the origin of visibility fluctuations of  the AMBER
instrument. In this appendix, we describe the polarizers and present
some measurements made when looking for visibility fluctuations.

\begin{figure}[h]
  \centering
  \parbox{0.6\hsize}{\includegraphics[width=\hsize]{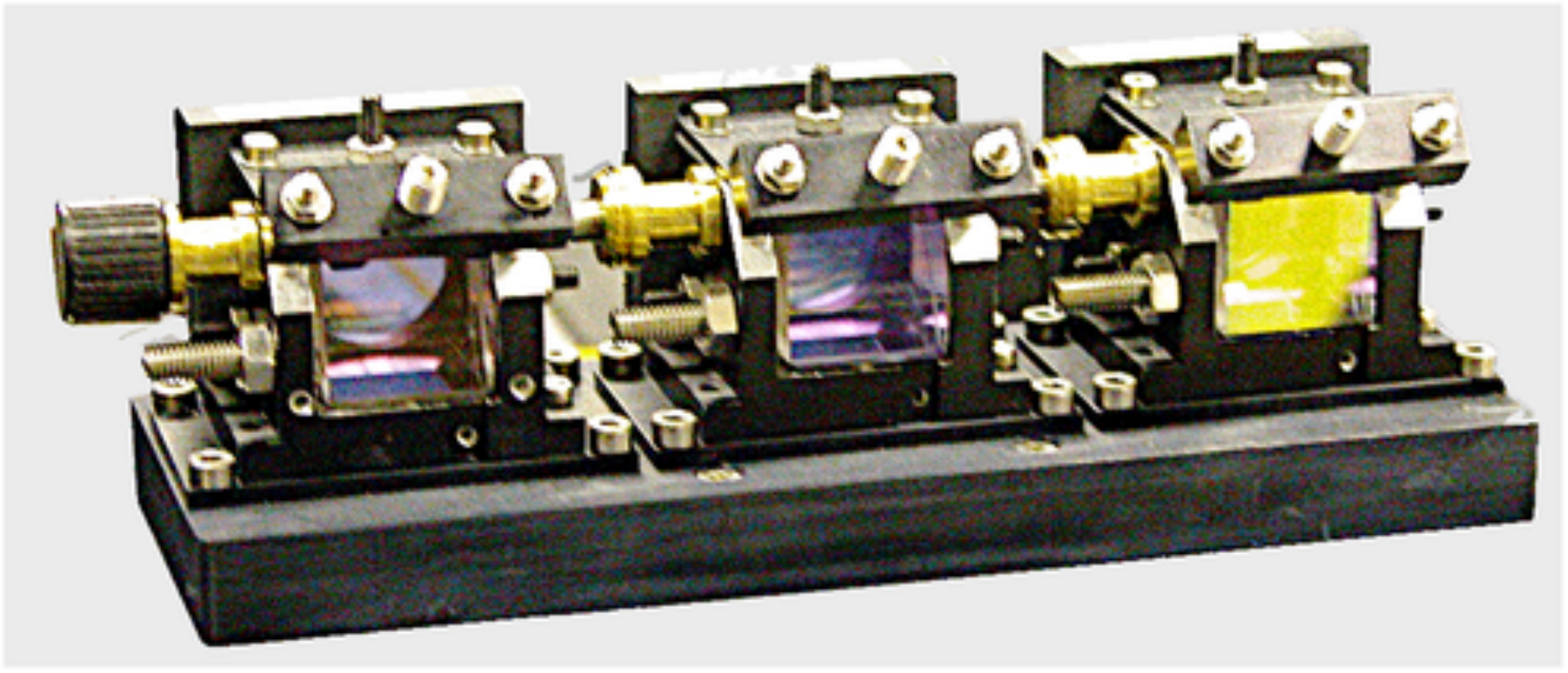}}
  \hfill
  \parbox{0.3\hsize}{\includegraphics[width=\hsize]{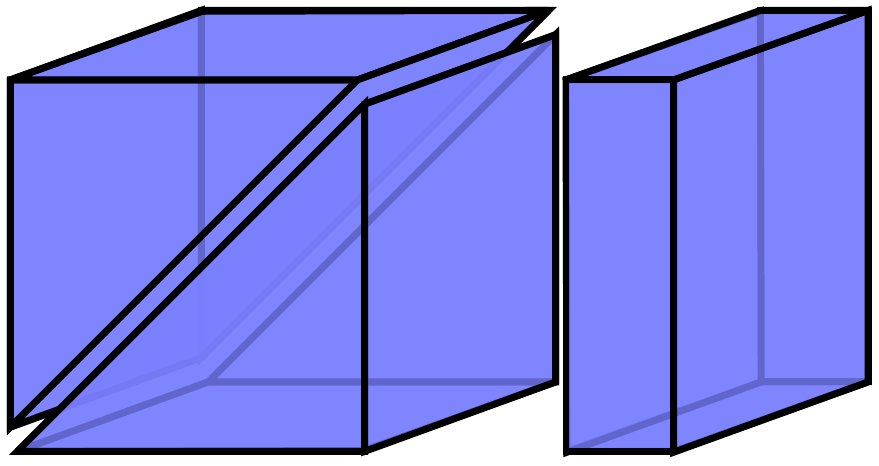}}
  \caption{AMBER polarizers (one for each beam) in their mechanical
    support (left). Each polarizer is made of 2 prisms and a glass
    blade. The air blade between the two prisms is kept parallel by
    the mechanical support. }
  \label{fig:POL}
\end{figure}

\subsection{Polarizers current design}

In the \emph{OPM Warm Optics Design Report} (doc.\
VLT-TRE-AMB-15830-1001, see
[RD~\ref{rd:OPM-LLS}]\footnote{Available at
  \url{http://amber.obs.ujf-grenoble.fr/PLAIN/pae/Documents_PAE/Documents/7_TRE_OPMODR.pdf}})

\begin{quote}
\textbf{3.6 POLarization module}

\begin{verbatim}
Product tree list : AMB-OPM-POL-(PL1,PL2,PL3)-(POL,CBL)

.../...

The maximum defect with respect to the mean thickness of the 3 blades POL-(PL1,PL2,PL3)- CBL
must be lower than +/-100 microns.

The beam deviation of the delivered polarizer (> 1') is not compatible with the +/-5"
specification imposed by the differential observations (see OPM LLS [3]). This deviation is
then minimized by the adjustment of the two prisms constituting the polarizer, the two
latter being dissociated. This allows to control the prismatic air layer at the
interface. In addition, to be compatible with the differential chromatic OPD specifications,
a prismatic blade needs to be added for each polarizer.  The complete definition of the
blades is given in the OPM Test Report.

In order to ensure a good parallelism between the two opposite internal faces after the
dissociation of the prisms which constitute each polarizer, a plastic wedge of 0.12 mm is
introduced between these two prisms. The angle formed by this separation needs to be stable
within 3.5" during the observations.
\end{verbatim}
\end{quote}

In the \emph{OPM Warm Optics Test Report} (doc.\
VLT-TRE-AMB-15830-1010, see [RD~\ref{rd:OPM-TR}]\footnote{\url{http://amber.obs.ujf-grenoble.fr/PLAIN/pae/Documents_PAE/Documents/9_TRE_OPMTR.pdf}}).

\begin{quote}
\textbf{13 Manufacturer delivery of the polarizers POL}

\begin{verbatim}
.../...

13.2.2 After reception of the polarizers

Measurements and simulations were performed by Yves Bresson, OCA. They show that the air
blade located between the two prisms constituents of each polarizer must be parallel in
order for the assembly polarizer and associate compensating blade to respect the chromatic
dynamic OPD specifications. A peeling off of the two prisms constituents of the polarizer
from the initial mounting delivered by the manufacturer was required in order to insert a
plastic shim (with an opening for the optical beam) between them.

.../...
\end{verbatim}

\textbf{14.3 AMB-OPM-POL optical pre-adjustment}

\begin{verbatim}
.../...
14.3.2 Principle of the assembly and of the compensation

The polarizer is composed of two prisms in calcite. In order to ensure a good parallelism
between the two opposite internal faces after the dissociation of the prisms which
constitute each polarizer, a plastic shim of 0.12 mm is introduced between these two
prisms. A first prism leans inside the mount on six iso-static points constituted by 3 tips
on the lower plate, 2 others for the orientation and a reference position [4].

The internal face of the second prism leans on the plastic shim via 3 points constituted by
a tip on the lower plate and 2 lateral tips whose one allows the rotation of the prisms
along the optical axis (range : +/- 0.5 mm). This prism is clamped by a corner plate guided
between two sticks with a spring toe.

Considering the characteristics of the prisms in terms of parallelism and thickness
differences, we have inserted compensating blades in order to correct these defaults and to
be compatible with the OPD specifications. These compensating blades are mounted in
cylinders which can be rotated along the optical axis.

Zemax simulations give the associated chromatic dynamic OPD as a function of the optics
characteristics and positions and of the measured beam deviations.
\end{verbatim}
\end{quote}

\subsection{Influence of polarizers on the AMBER observables}
\label{sec:with-without-pol}

\begin{figure}[p]
  \centering
  \includegraphics[width=0.9\hsize]{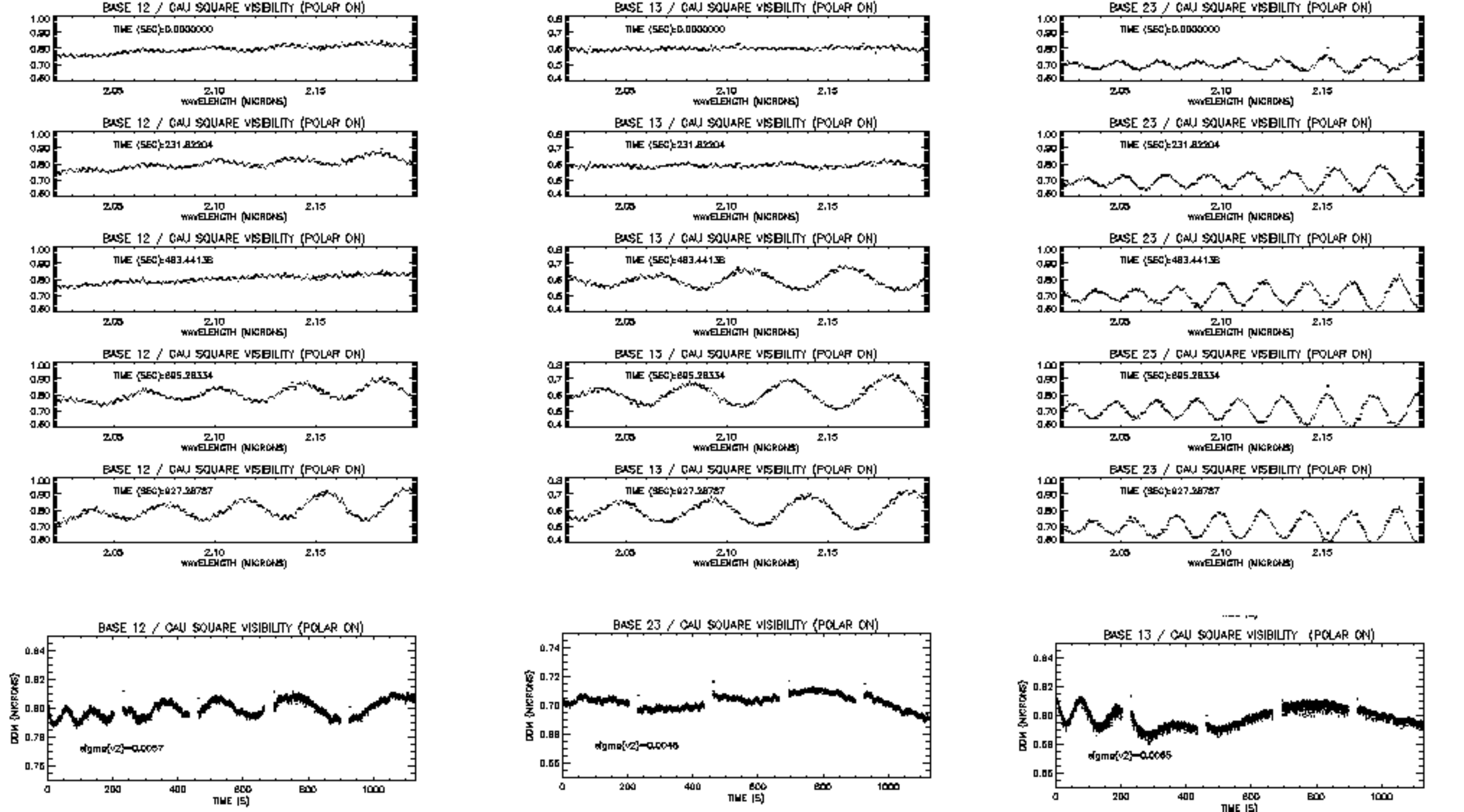}
  \caption{AMBER square visibilities recorded \textbf{with} the polarizers on
    the internal source. From left to right, the different baselines
    and from top to bottom a sequence of 5 consecutive exposures of
    1000 frames lasting each about 50 seconds and separated by 20
    seconds. These exposures have been averaged over the length of the
    exposure. Bottom panels represent the same time series with
    exposures averaged in wavelength but not in time.}
  \label{fig:amber-with-pol}
\end{figure}
\begin{figure}[p]
  \centering
  \includegraphics[width=0.9\hsize]{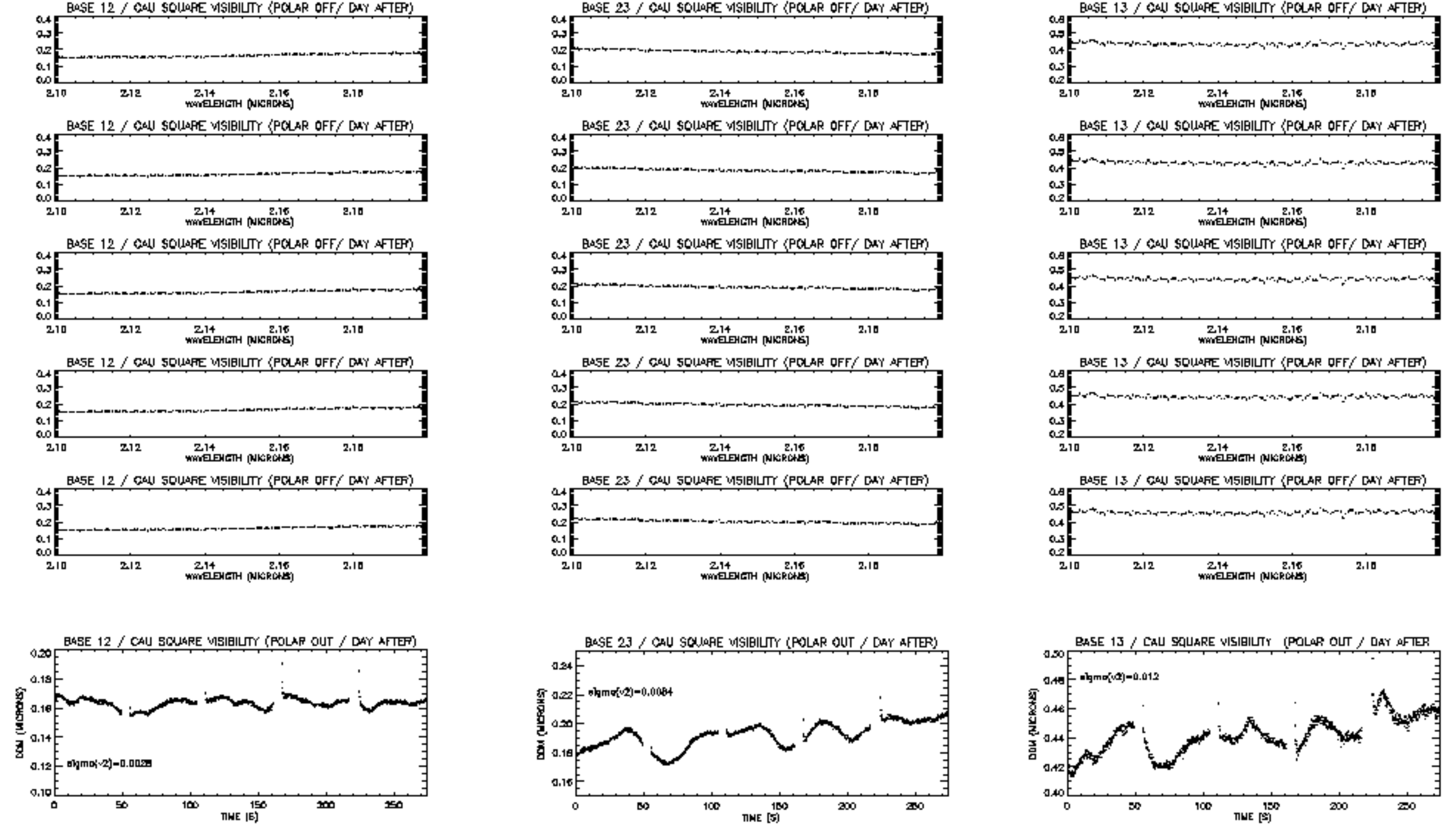}
  \caption{AMBER square visibilities recorded \textbf{without} the polarizers
    on the internal source. Same description as in
    Fig.~\ref{fig:amber-with-pol}. Note that the wavelength modulation
    has disappeared. The visibility fluctuation with time remains
    large, but may comes also partly from a mechanical instability of
    an element of the internal source.}
  \label{fig:amber-without-pol}
\end{figure}

Figures \ref{fig:amber-with-pol} and \ref{fig:amber-without-pol} shows
the square visibilities measured respectively with and without the AMBER
polarizers. It is interestin to note that the fluctuations of the
visibilities seem to come from varying periodic oscillations with
wavelength appearing and disappearing in the visibilities. These
oscillations disappeared totally when the polarizers are taken out of
the instrument. However the averaged visibility decreases to a level
of 0.2 or two baselines and 0.4 for the third one.

Our understanding of this phenomenon is that the air blade between the two
prisms of each polarizer is acting as a Fabry-Perot interferometer. Applying
Eq.~(\ref{eq:Fabry-Perot}) to the number of oscillations seen in a 
Fig.~\ref{fig:amber-with-pol} gives the following Fabry-Pérot thickness:
\begin{itemize}
\item Baseline 1-2: $\approx2.8\,\mbox{periods}$ in
  $\Delta\lambda=0.1\,\mbox{$\mu$m}$ gives a thickness of $58\,\mbox{$\mu$m}$.
\item Baseline 1-3: $\approx2.2\,\mbox{periods}$ in
  $\Delta\lambda=0.1\,\mbox{$\mu$m}$ gives a thickness of $45\,\mbox{$\mu$m}$.
\item Baseline 1-2: $\approx4.9\,\mbox{periods}$ in
  $\Delta\lambda=0.1\,\mbox{$\mu$m}$ gives a thickness of $102\,\mbox{$\mu$m}$.
\end{itemize}
We have performed several measurements and the number of oscillations per
spectral bandwidth is changing with time and veries between $50\,\mbox{$\mu$m}$
and $200\,\mbox{$\mu$m}$, which the order of size of the shim introduced between
the two prisms. As a matter of fact, each baseline sees two polarizers and the
interferences can create beating effects with a large range of frequency and
amplitude. 

The fact that the oscillations disappear when the polarizers are taken out is
the final proof that these feects comes from the air blade present in the
polarizers. 

\subsection{Where to put the new polarizers?}
\label{sec:where-pol}

Polarizers are however necessary to control the polarization. We have dismount
POL3 to try to place it in other part of the instrument. When putting it near
the periscope in a place where the beam propagation was collimated, we obtained
contrast of 0.9, 0.9 and 0.85 instead of 0.4, 0.6, 0.4. Therefore one
should modify the instrument to introduce polarizers without air blades. Unless
we succeed in understanding why the visibility drops to such low value without
polarizers.  

\begin{figure}[hb]
  \centering
  \includegraphics[width=0.65\hsize]{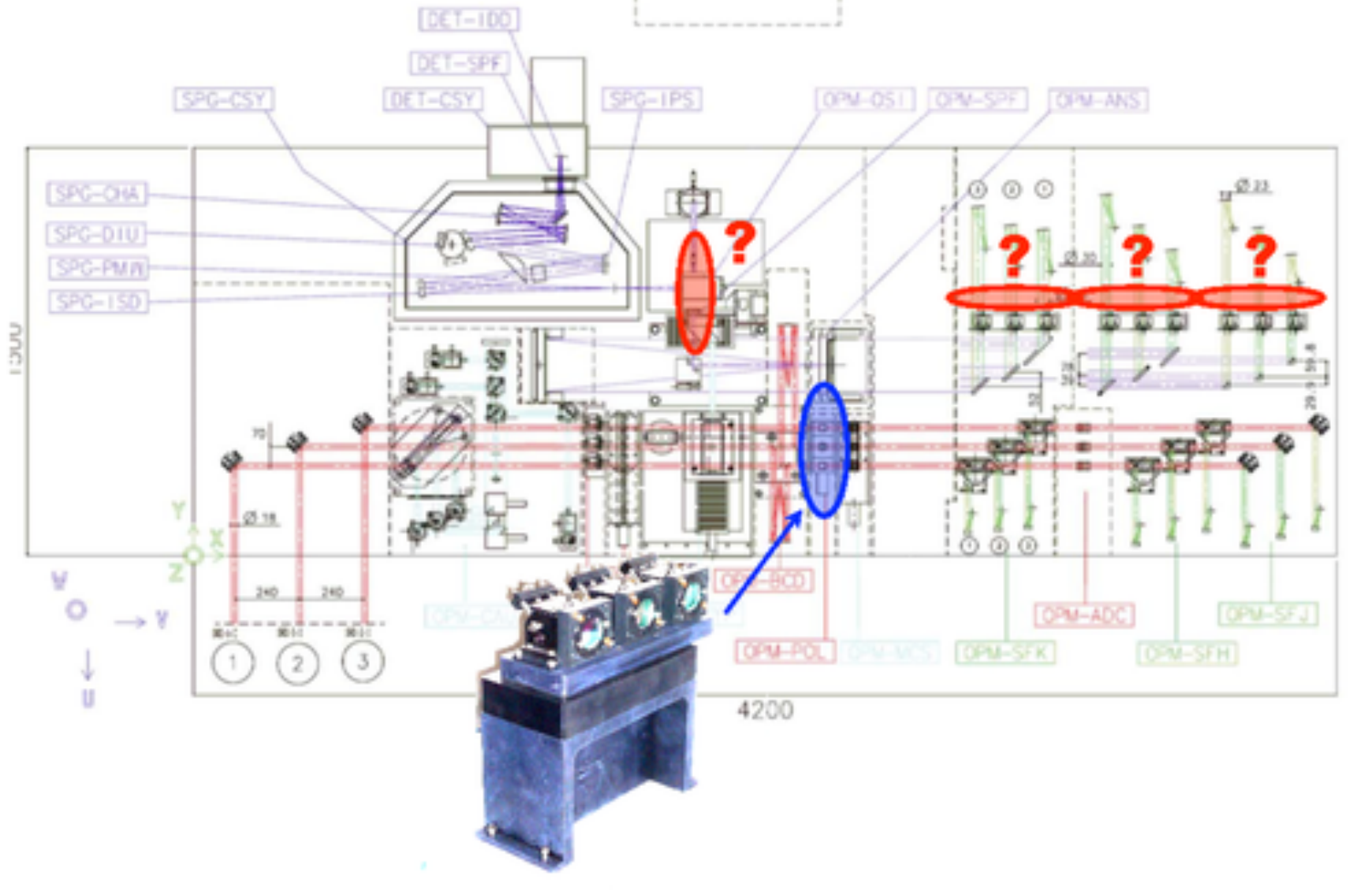}
  \caption{Location of the polarizers on the AMBER table (blue).  Red
    ellipses show the possible locations for new polarizers. The
    position of one unique and common polarizer in the periscope tower
    (upper left) should be privileged.}
  \label{fig:polarizer-table}
\end{figure}

The present location of the polarizers is represented on
Fig.~\ref{fig:polarizer-table}. The location of the new polarizer
should be chosen with great care. On the same figure, we propose two
different locations. However, the position of one unique and common
polarizer in the periscope tower should be privileged rather than
9 independant polarizers located at the output of the fibers. 

\clearpage
\section{Fabry-Pérot interferometer principle}
\label{app:FP}

[From Wikipedia, the free encyclopedia (\url{http://en.wikipedia.org/wiki/Fabry-Perot})]

In optics, a Fabry-Pérot interferometer is typically made of
a transparent plate with two reflecting surfaces, or two parallel
highly reflecting mirrors. Its transmission spectrum as a function of wavelength
exhibits peaks of large transmission corresponding to resonances of
the etalon. 

The resonance effect of the Fabry-Pérot interferometer is identical to
that used in a dichroic filter. That is, dichroic filters are very
thin sequential arrays of Fabry-Pérot interferometers, and are
therefore characterised and designed using the same mathematics.

The varying transmission function of the FP is caused by
interference between the multiple reflections of light between the two
reflecting surfaces. Constructive interference occurs if the
transmitted beams are in phase, and this corresponds to a
high-transmission peak of the etalon. If the transmitted beams are
out-of-phase, destructive interference occurs and this corresponds to
a transmission minimum. Whether the multiply-reflected beams are
in-phase or not depends on the wavelength ($\lambda$) of the light (in
vacuum), the angle the light travels through the FP ($\theta$), the
thickness of the etalon ($l$) and the refractive index of the material
between the reflecting surfaces ($n$).

\begin{figure}[hb]
  \centering
  \includegraphics[height=0.35\hsize]{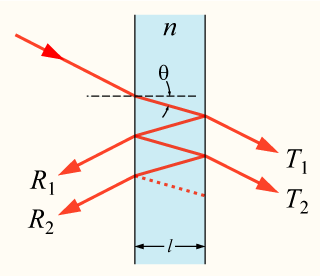}\hfill
  \includegraphics[height=0.35\hsize]{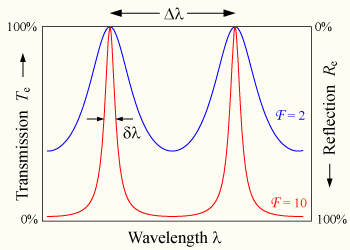}
  \caption{Left: A Fabry-Pérot interferometer. Light enters the etalon and undergoes multiple internal reflections. Right: The transmission of an etalon as a function of wavelength. A high-finesse etalon (red line) shows sharper peaks and lower transmission minima than a low-finesse etalon (blue).}
  \label{fig:FP}
\end{figure}

The wavelength separation between adjacent transmission peaks is called the free spectral range of the FP, $\Delta\lambda$, and is given by:
\begin{equation}
  \label{eq:Fabry-Perot}
      \Delta\lambda = \frac{ \lambda_0^2}{2nl \cos\theta + \lambda_0 } 
\approx \frac{ \lambda_0^2}{2nl \cos\theta } 
\end{equation}
where $\lambda_0$ is the central wavelength of the nearest transmission peak. 

Therefore the thickness of the étalon is given by:
\begin{equation}
  \label{eq:Fabry-Perot2}
       l \approx \frac{ \lambda_0^2}{2n \, \Delta\lambda \, \cos\theta }   
\end{equation}

\clearpage
\section{Piezos controling differential OPDs}
\label{app:piezos}

\begin{figure}[hbp]
  \centering
  \includegraphics[width=0.8\hsize]{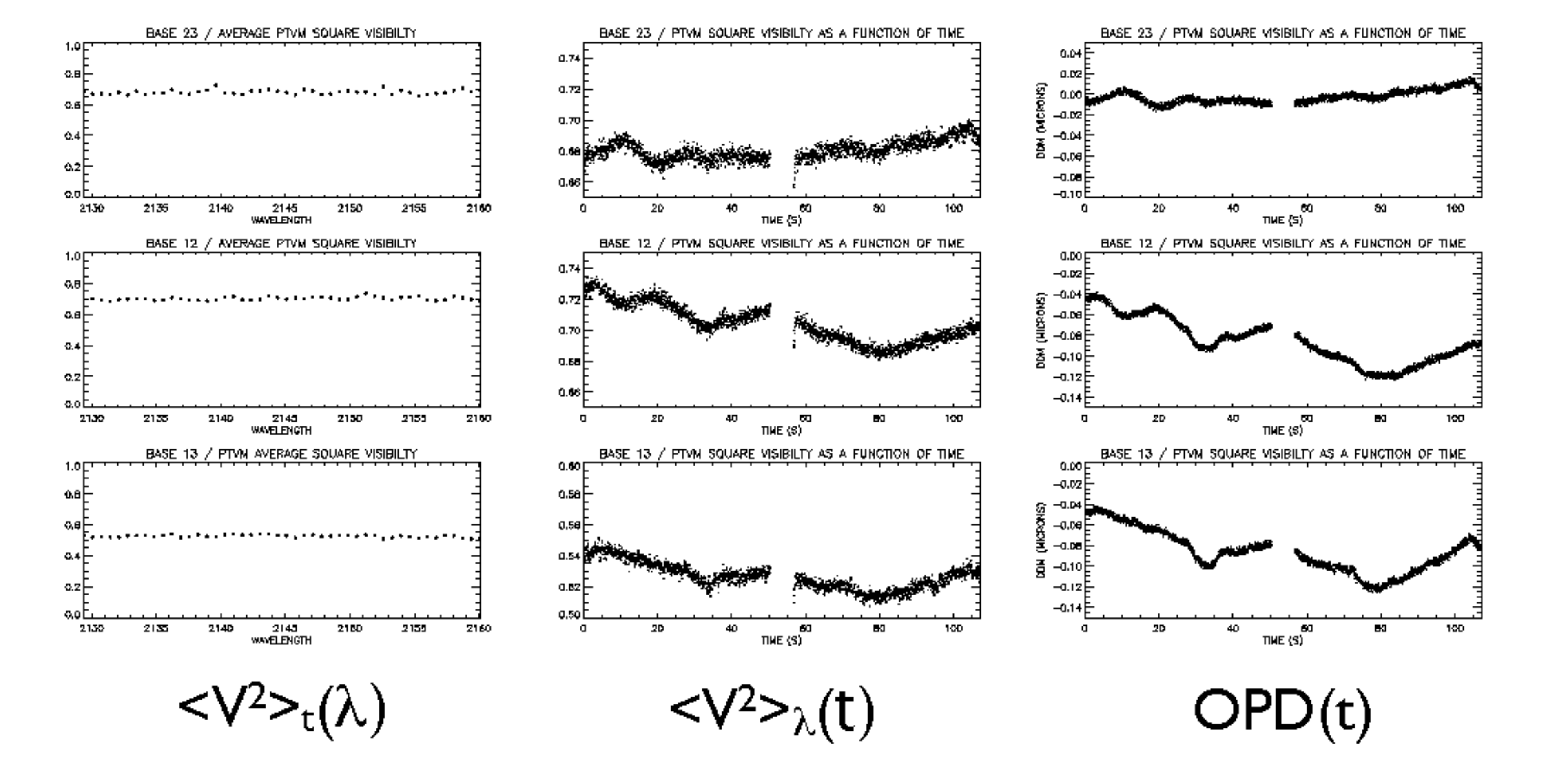}
  \caption{Square visibilities observed on the AMBER internal source with the
    \textbf{piezos activated} averaged in time plotted in function of wavelength
    (left), averaged in wavelength and plotted in function of time (middle) and
    the piston in function of time for the three baselines (from top to
    bottom).}
  \label{fig:piezos-on}
\end{figure}

\begin{figure}[hbp]
  \centering
  \includegraphics[width=0.8\hsize]{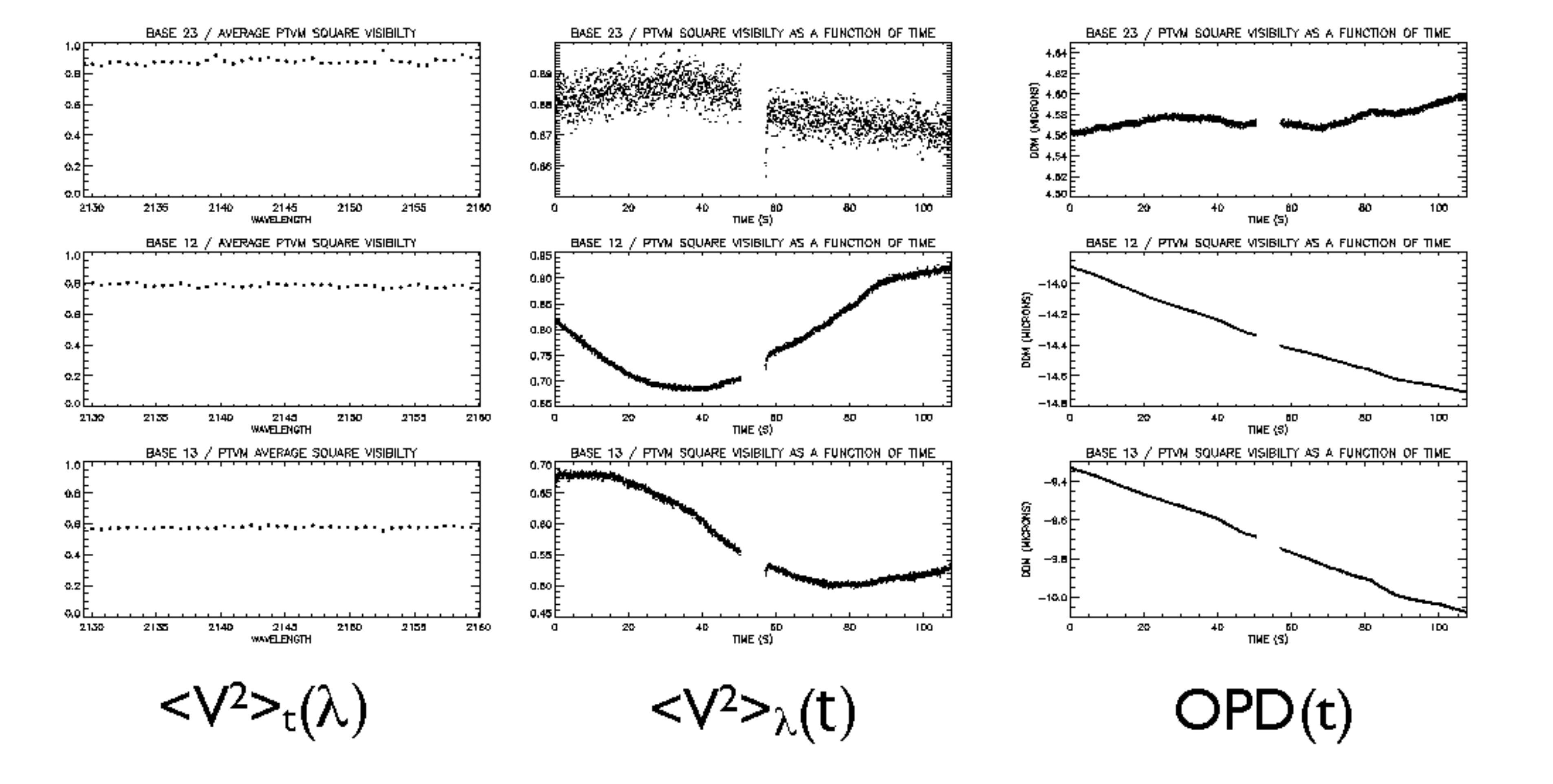}
  \caption{Square visibilities observed on the AMBER internal source with the
    \textbf{piezos desactivated}. The lay-out is the same as the one of
    Fig.~\ref{fig:piezos-on}.}
  \label{fig:piezos-off}
\end{figure}

During the ATF investigation, we were interested to know if the piezos which
drives the fiber injection to control the differential OPDs cold be at the
origin of vibrations that could lead to visibility fluctuations. We have
therefore taken exposures with the piezos on and off (see
Figs.~\ref{fig:piezos-on} and \ref{fig:piezos-off}). This experoence was made
with the polarizers still on the table.

We have not detected any changes in the behavior of the visibility.  When the
piezos are off then all piezos move by 40 microns at the beginning of the
stroke. We had to compensate these
OPDs manually with the translation stage of the input dichroics which are known
to unstable. Therefore the increase of the OPD changes seen on the right part of
Fig.~\ref{fig:piezos-off} are probably relaxation of these translation stages.


\clearpage
\section{Influence of injection into the fibers}
\label{app:fibers}

We took the opportunity of the new piezo-controlled 'IRIS Fast
Guiding' (IFG) positioning of the AMBER entrance beams and
the availability of the AT beacons to test whether the injection of
the beam creates or augments the ``phase beating'' effect (see
sect.~\ref{sec:phase-beating}).

\begin{figure}[hp]
  \centering  
  \includegraphics[width=0.6\hsize]{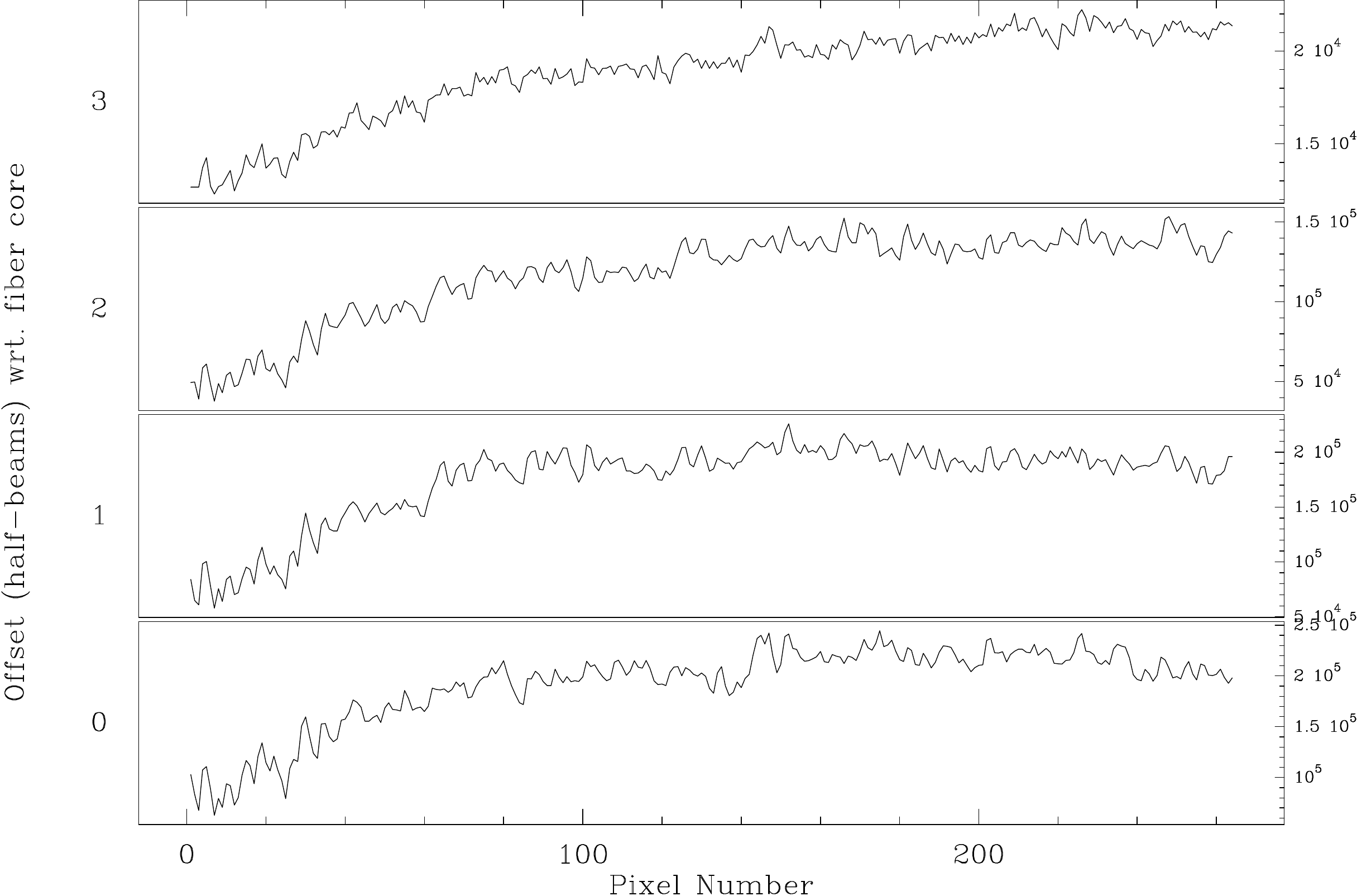}  
  \includegraphics[width=0.6\hsize]{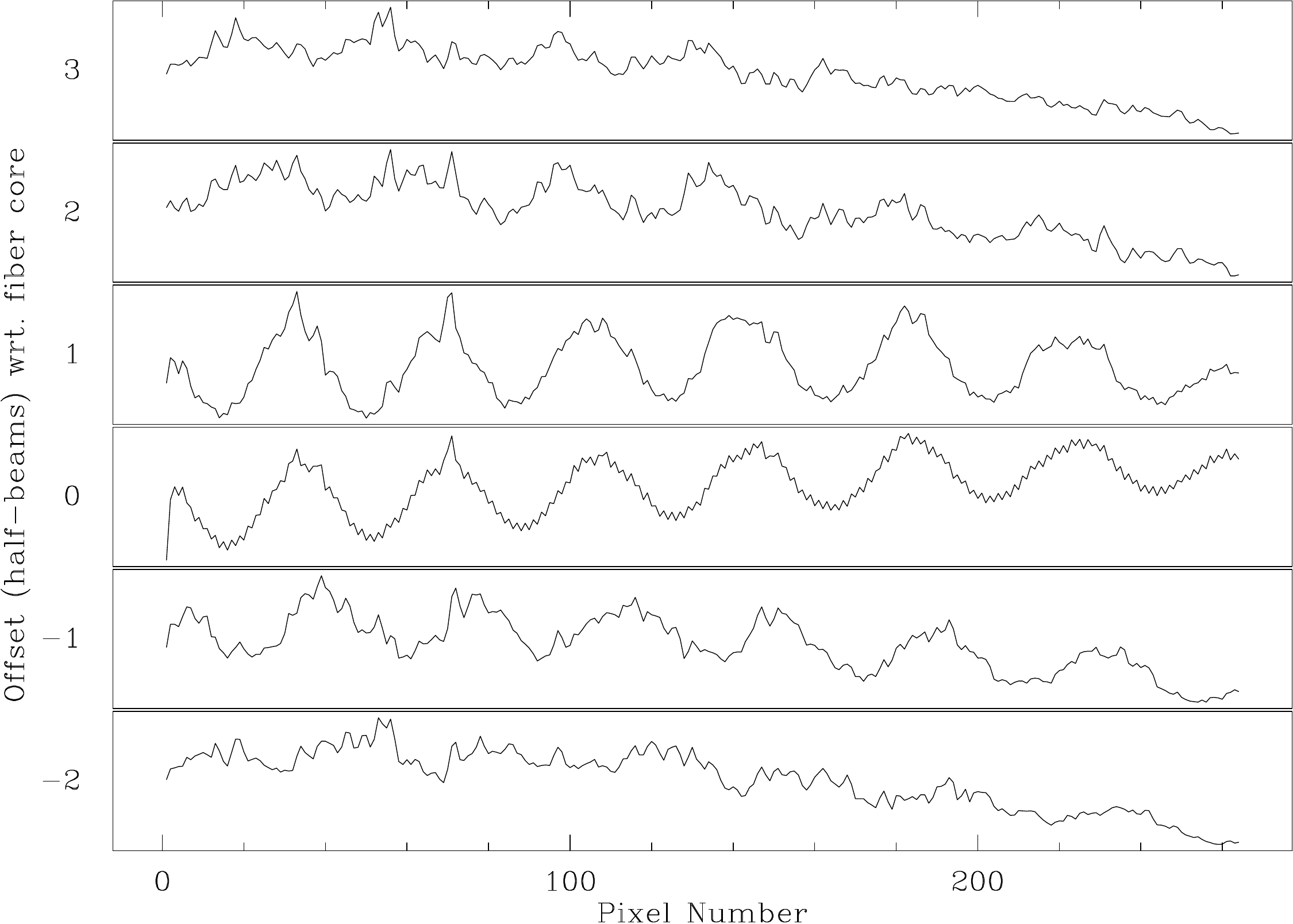}
  \caption{Flux recorded in one fiber with one AT beacon in function
    of the injection with (top) and without the polarizers
    (bottom). Injection condition were changed without touching AMBER
    but using the AMBER fast guiding piezos and moving by half an Airy
    disk.} 
  \label{fig:fibers-ifg}
\end{figure}

We found (see Fig.~\ref{fig:fibers-ifg}) that there is no effect
detectable of the changes of injection (up to five core diameters)
when the polarizer (Appendix~\ref{app:POL}) is removed. The dispersed
spectra of the beacons are devoid of amplitude structures (phase was
not mesurable since beacons are not one coherent source), and the flux
injected simply diminishes as the injection worsens.  In the presence
of the polarisers, however, amplitude effects are present and vary in
shape with the injection. That the small changes in the positioning of
the beams induced by the IFG would produce this effect demonstrates
how sensitive the experiment was to the differential Fabry-P\'erot
effect induced by the polarizer.

\clearpage
\section{Observing calibrators without polarizers}
\label{app:observations}

\begin{figure}[p]
  \centering
  \includegraphics[width=0.85\hsize]{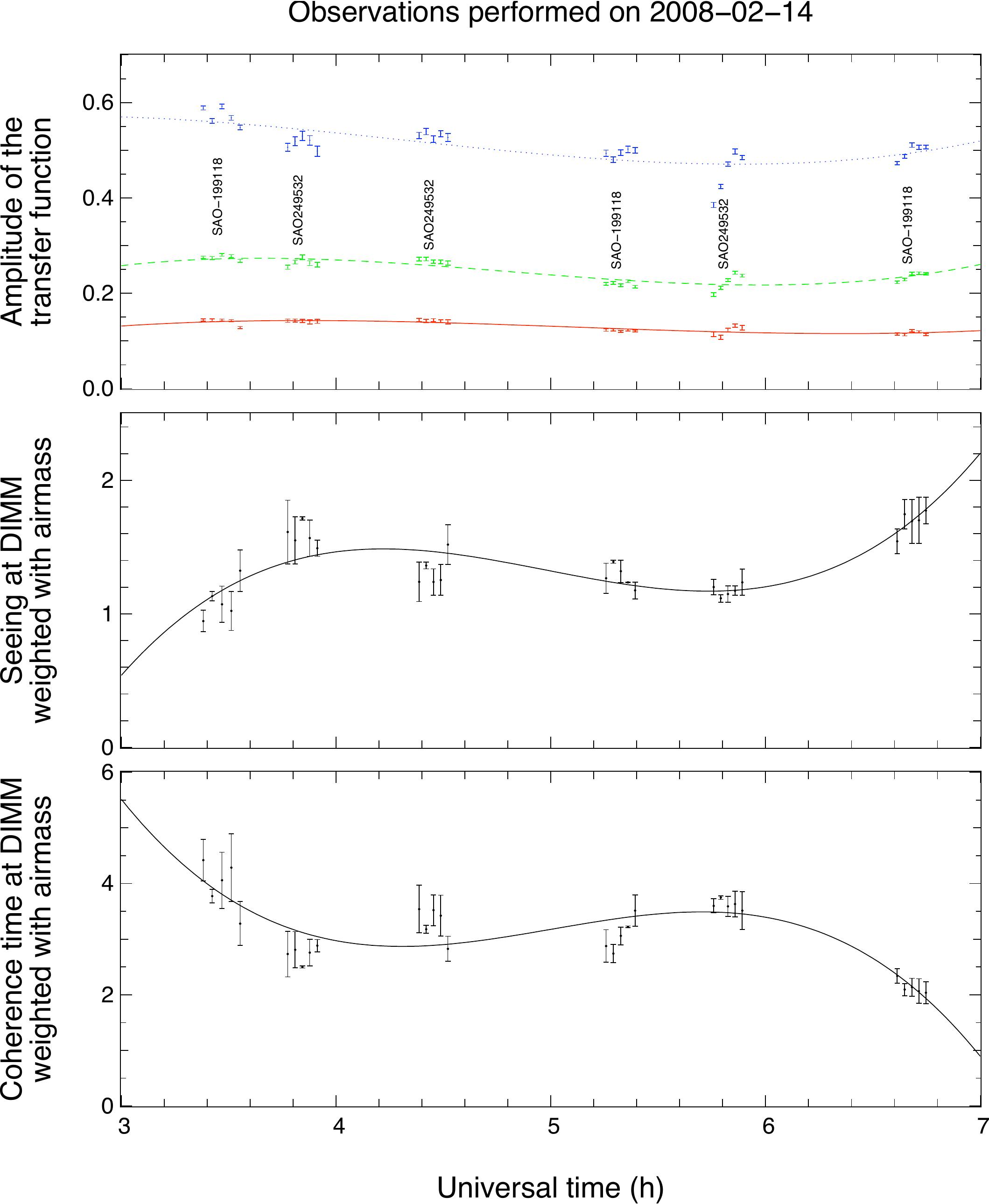}
  \caption{Variations of the transfer function averaged over the
    wavelength for the two calibrators SAO\,199118 and SAO\,249932
    observed over 4 hours in MR, without FINITO and whithout the
    polarizer. }
  \label{fig:tf-stability-sky}
\end{figure}

The last night of ATF run was focused on the observations of 2
calibrators without FINITO in order to
characterize AMBER without the polarizers:
\begin{itemize}
\item SAO\,199118: diameter of $2.54\pm0.037\,\mbox{mas}$ (Bordé), spectral
  type K2.5 II-III and
\item SAO\,249932: diameter of $2.19\pm0.053\,\mbox{mas}$ (Bordé),
  spectral type K5 III,
\end{itemize}
and a larger star SAO\,218755. We have reduced these data in order to
display the curve of the transfer function (see
Fig.~\ref{fig:tf-stability-sky}). We have also computed the
differential phase and the closure phase. 

\begin{figure}[p]
  \centering
  \includegraphics[width=0.85\hsize]{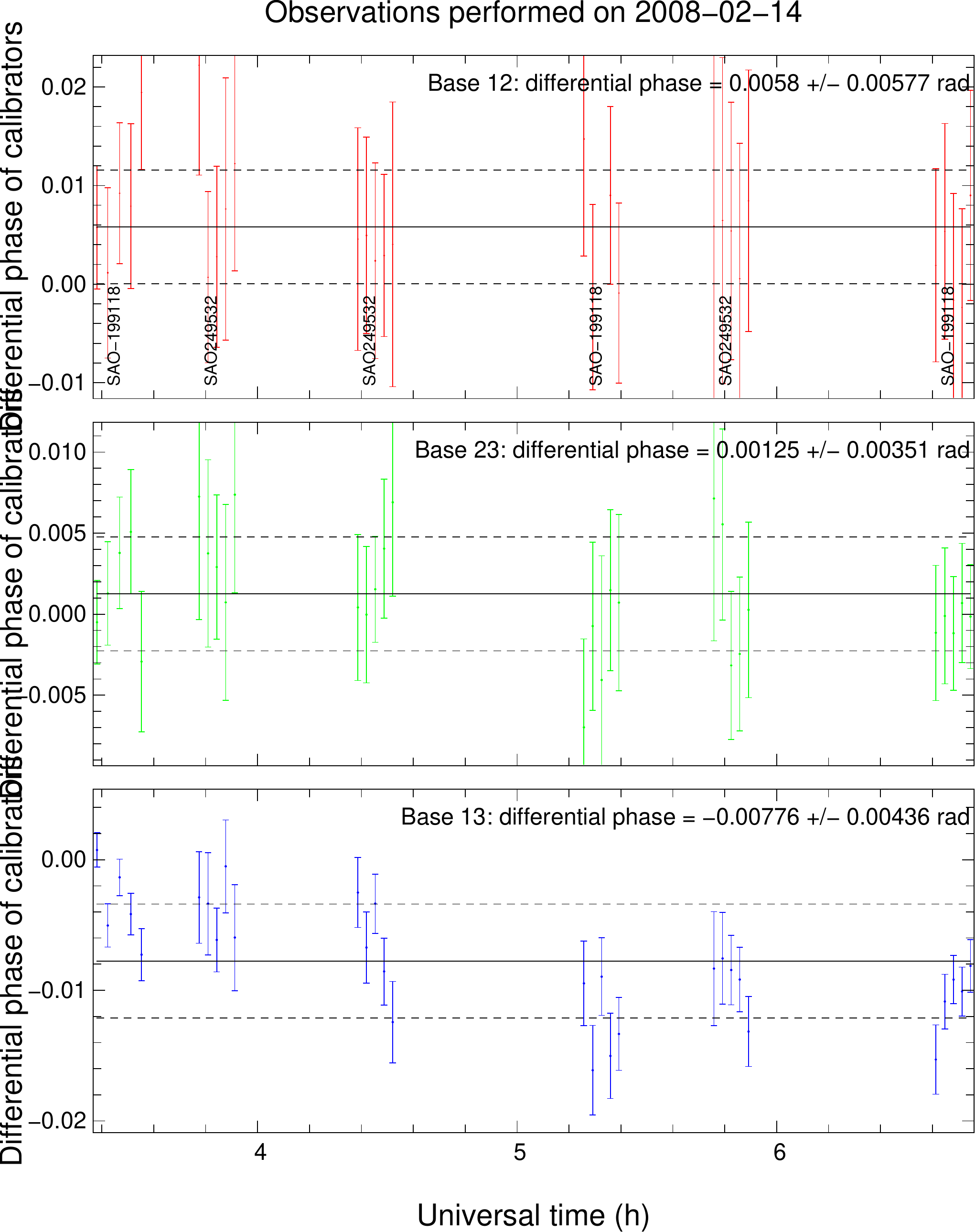}
  \caption{Variations of the differential phases averaged over the
    wavelength for the two calibrators SAO\,199118 and SAO\,249932
    observed over 4 hours in MR, without FINITO and whithout the
    polarizer. }
  \label{fig:diffpha-stability-sky}
\end{figure}

\begin{figure}[p]
  \centering
  \includegraphics[width=0.85\hsize]{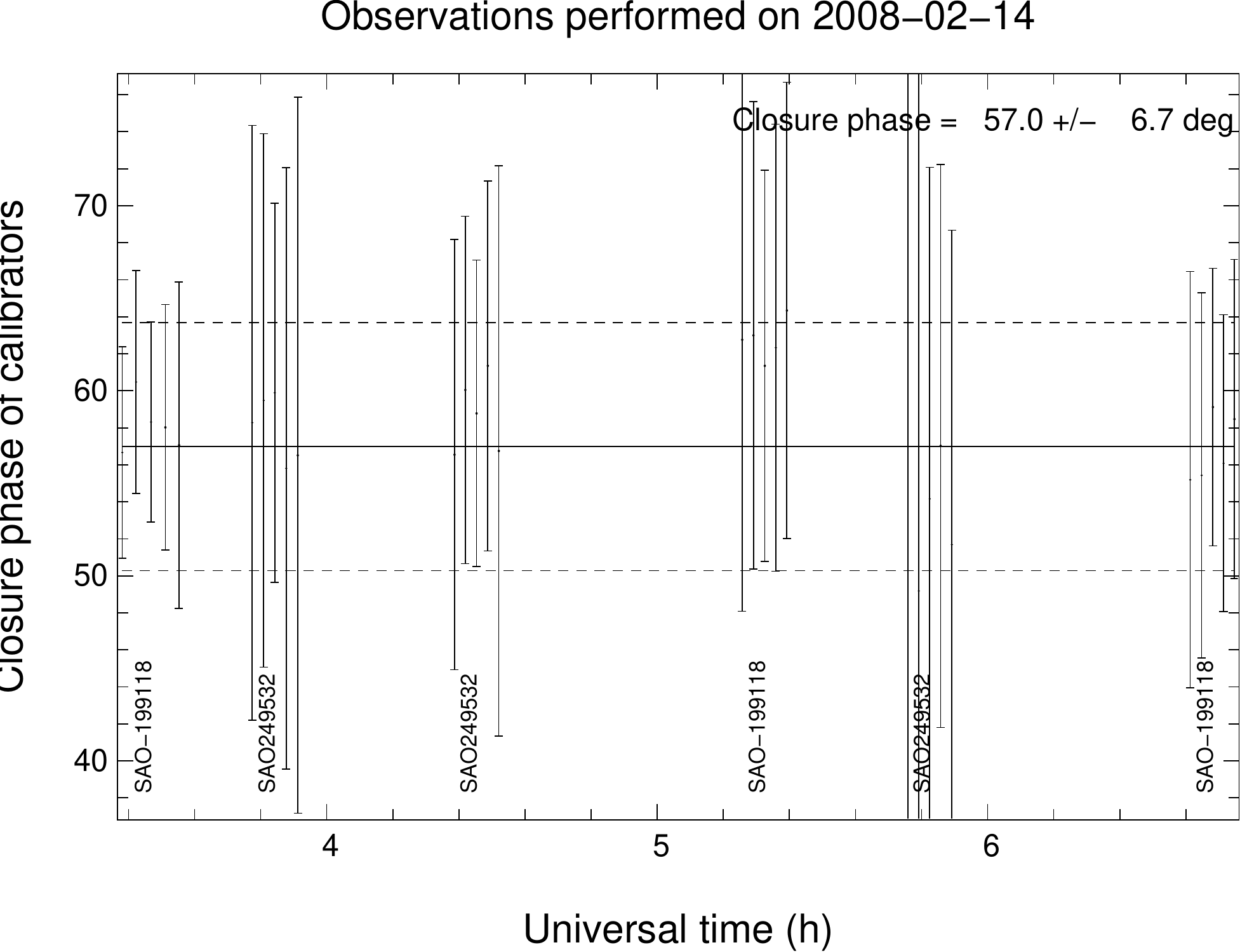}
  \caption{Variations of the closure phase averaged over the
    wavelength for the two calibrators SAO\,199118 and SAO\,249932
    observed over 4 hours in MR, without FINITO and whithout the
    polarizer.}
  \label{fig:clopha-stability-sky}
\end{figure}

The result of this night is:
\begin{itemize}
\item transfer function (see Fig.~\ref{fig:tf-stability-sky}): accuracy of the order of the
  percent. However it changes with seeing conditions and coherence
  time and therefore calibrations should be done at close as possible
  from the target science.
\item Differential phase (see Fig.~\ref{fig:diffpha-stability-sky}) are zero at veru good accuracy (less than
  $10^{-2}$\,rad, i.e. within specifications).
\item Closure phases (see Fig.~\ref{fig:clopha-stability-sky}) are not very accurate on this sample night. A
  reason might be that the SNR depends on the value of the
  visibilities which were small for two baselines due to the absence
  of polarizers (see values in Fig.~\ref{fig:tf-stability-sky}).
\end{itemize}

\begin{figure}[hbp]
  \centering
  \includegraphics[width=0.45\hsize]{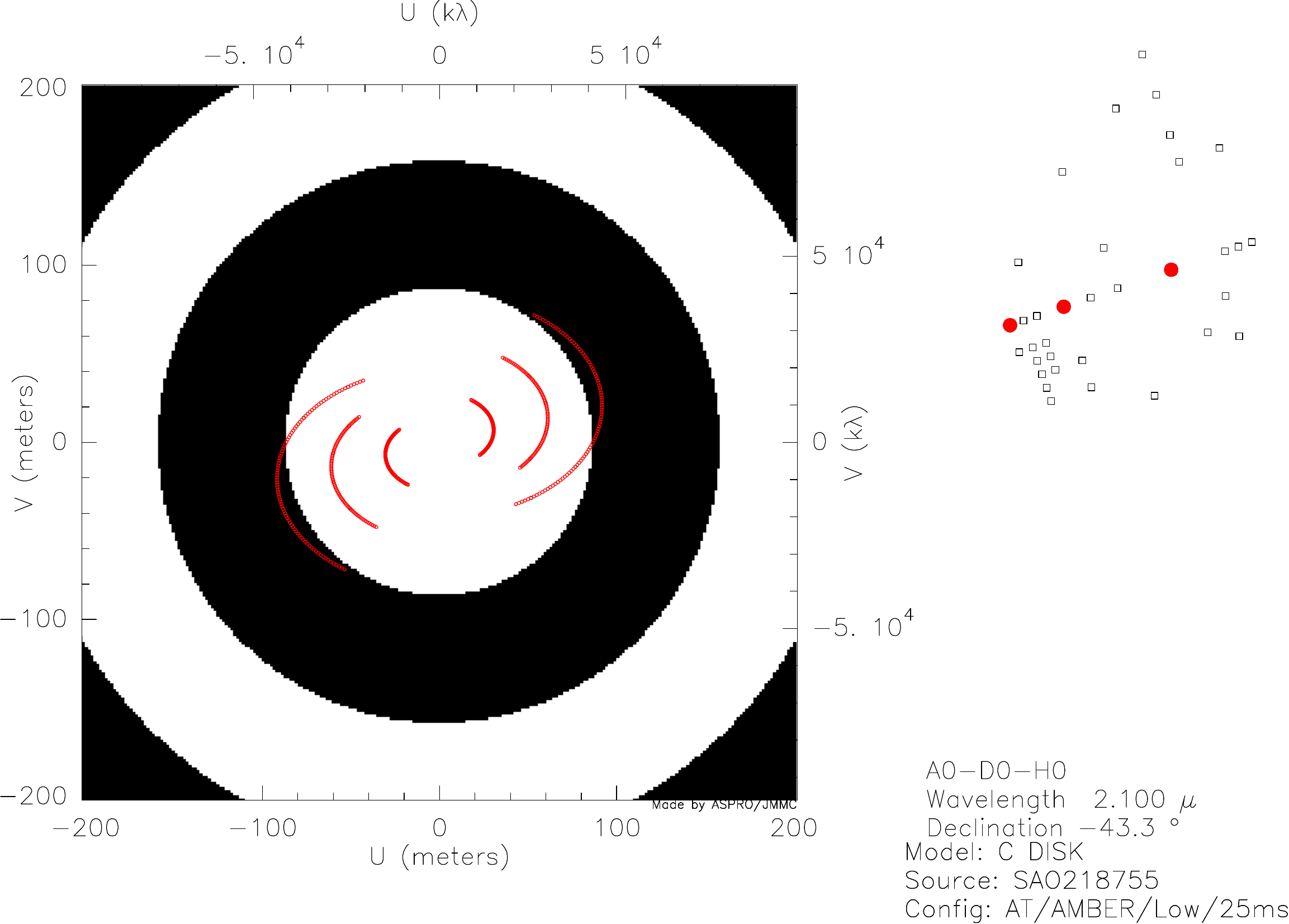}
  \includegraphics[width=0.45\hsize]{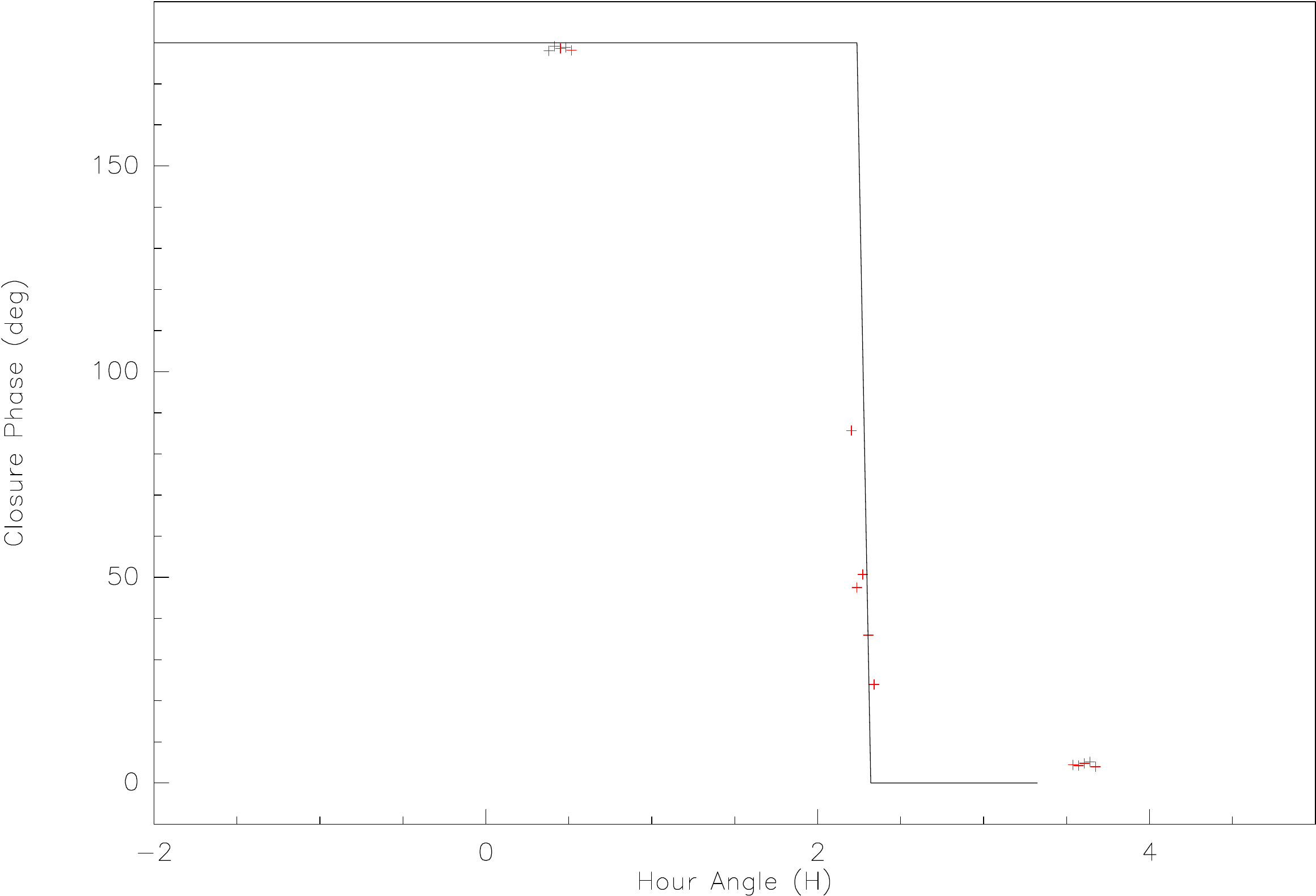}
  \caption{Closure phase measured on the star SAO\,218755. Left: uv
    coverage overplot on the closure phase change due to the
    diameter. Right: measured closure phase (red crosses) with the
    expected closure phase change (solid black line).}
  \label{fig:clopha-crossing}
\end{figure}

During this night, we have interlaced the observation of the star
SAO\,218755 which has an epxected diameter of 6.4\,mas. Interestingly,
the longest baseline crosses the first lobe and therefore we were able
to see a $\pi$ shift in the closure phase validating the closure phase
measurement (see Fig.~\ref{fig:clopha-crossing}) 

\vfill

\begin{sidewaystable}[p]
  \centering
  \caption{AMBER performance on calibrators without the polarizers.}
  \medskip
  \begin{tabular}{|l|r|rr|rr|rr|rrr|r|}
\hline
SAO& Time& $\mathrm{V}^2_{12}$ & $\sigma(\mathrm{V}^2_{23})$ & $\mathrm{V}^2_{13}$ &
 $\sigma(\mathrm{V}^2_{12})$ 
& $\mathrm{V}^2_{23}$ & $\sigma(\mathrm{V}^2_{13})$ & $\phi_{12}$ &$\phi_{23}$ &$\phi_{13}$    & $\psi_{123}$  \\
Number& \multicolumn{1}{c|}{(hours)}& &&&&&&\multicolumn{3}{c|}{(degrees)}&  \multicolumn{1}{c|}{(degrees)}\\
\hline                                                                                 
199118 &0.00 &0.156& 0.003& 0.067 &0.005& 0.012 &0.004&0.00& -0.002& -0.01 & 55.74\\
199118 &0.04 &0.141& 0.004& 0.065 &0.005& 0.011 &0.005&0.01& -0.004& -0.02 & 59.51\\
199118 &0.09 &0.158& 0.003& 0.069 &0.005& 0.012 &0.004&0.00& -0.018&  0.00 & 57.16\\
199118 &0.13 &0.146& 0.004& 0.066 &0.006& 0.011 &0.004&0.01& -0.007& -0.02 & 56.83\\
199118 &0.17 &0.135& 0.004& 0.060 &0.007& 0.008 &0.006&0.02& -0.002& -0.07 & 55.90\\
249532 &0.39 &0.160& 0.012& 0.062 &0.009& 0.017 &0.007&0.01& -0.040& -0.01 & 55.64\\
249532 &0.43 &0.168& 0.012& 0.067 &0.009& 0.017 &0.007&0.01& -0.008& -0.11 & 58.07\\
249532 &0.46 &0.176& 0.012& 0.072 &0.009& 0.016 &0.006&0.02& -0.007&  0.04 & 57.67\\
249532 &0.50 &0.170& 0.013& 0.066 &0.011& 0.016 &0.008&0.01& -0.022& -0.06 & 54.96\\
249532 &0.53 &0.155& 0.013& 0.064 &0.011& 0.016 &0.009&0.02& -0.019& -0.09 & 56.95\\
249532 &1.01 &0.180& 0.009& 0.070 &0.007& 0.017 &0.008&0.01&  0.010&  0.10 & 57.18\\
249532 &1.04 &0.186& 0.008& 0.070 &0.006& 0.017 &0.007&0.02&  0.004&  0.10 & 59.50\\
249532 &1.07 &0.175& 0.008& 0.068 &0.006& 0.017 &0.006&0.01&  0.007&  0.07 & 58.33\\
249532 &1.11 &0.183& 0.008& 0.068 &0.006& 0.016 &0.006&0.03& -0.009&  0.04 & 60.60\\
199118 &1.88 &0.127& 0.008& 0.045 &0.006& 0.012 &0.005&0.05&  0.021& -0.04 & 61.20\\
199118 &1.91 &0.122& 0.008& 0.045 &0.006& 0.012 &0.004&0.08& -0.002&  0.01 & 60.82\\
199118 &1.94 &0.157& 0.008& 0.082 &0.006& 0.038 &0.004&0.05& -0.140& -5.08 & 57.71\\
199118 &1.98 &0.134& 0.008& 0.047 &0.006& 0.012 &0.004&0.06& -0.007&  0.01 & 61.99\\
199118 &2.01 &0.133& 0.007& 0.042 &0.005& 0.012 &0.004&0.06&  0.006& -0.02 & 62.30\\
249532 &1.14 &0.178& 0.011& 0.066 &0.009& 0.016 &0.010&0.04&  0.027&  0.09 & 55.29\\
249532 &2.38 &0.100& 0.008& 0.037 &0.009& 0.011 &0.012&0.04&  0.009& -0.07 & 26.20\\
249532 &2.41 &0.121& 0.007& 0.043 &0.007& 0.010 &0.010&0.05& -0.059&  0.19 & 46.11\\
249532 &2.44 &0.149& 0.006& 0.050 &0.006& 0.013 &0.008&0.04& -0.004&  0.09 & 52.37\\
249532 &2.48 &0.167& 0.007& 0.057 &0.006& 0.015 &0.008&0.03& -0.001& -0.05 & 54.33\\
249532 &2.51 &0.158& 0.007& 0.054 &0.006& 0.014 &0.010&0.06& -0.015&  0.01 & 51.49\\
199118 &3.23 &0.146& 0.005& 0.047 &0.005& 0.011 &0.005&0.05&  0.010& -0.06 & 54.59\\
199118 &3.27 &0.156& 0.005& 0.050 &0.005& 0.011 &0.005&0.05& -0.012&  0.02 & 55.35\\
199118 &3.30 &0.173& 0.005& 0.055 &0.004& 0.013 &0.005&0.04&  0.020&  0.09 & 57.73\\
199118 &3.33 &0.170& 0.005& 0.055 &0.004& 0.012 &0.005&0.04& -0.007& -0.02 & 56.69\\
199118 &3.37 &0.171& 0.004& 0.055 &0.004& 0.011 &0.005&0.03& -0.014& -0.01 & 57.66\\
\hline
\end{tabular}
\end{sidewaystable}

\clearpage
\section{Log book of the ATF run}
\label{app:logbook}

\begin{verbatim}
*  31-Jan to 5-Feb:
    o pupil alignment
    o OPD de-saturation 

* 6 Feb:
    o H-band pupil alignment + OPD de-saturation
    o ghost characterization
    o overall stability control 

* 7-Feb:
    o find the fringe beating, characterize it at LR, MR and HR
    o Tests: crossing fibers. No improvement. Crossing does not change 
      injection significantly, but adds a 7.5mm piston. Fringes still visible i
      HR. No significant improvement on phase beating. 
    o remove the polarizers and test at LR, MR and HR: no more fringe beating
      (but lower --by half-- fringe contrast). 
          + flux in K much higher and spectrum very different
          + still a phase closure on the CAU of 6°
          + We realize that by shutting down the RAS JH fiber, the K spectrum
            becomes very different --> strong K pollution through the JH fiber. 
          + Additionnally, now, the RAS light gives a closure of 0° ! 
    o tests on sky on Sirius in MR and HR with 2 telescopes 

* 8-Feb:
    o discussion with RPe on status
    o discussion with Pedro on alignment procedures
    o test of a single fibered source in RAS: strongly reduced flux beating
    o put POL3 after output f fibers close to the periscope in // beams: higher
      contrast: 0.9 0.9 0.85 
    o discussion with Andres on software procedures.
    o try some Sirius 3T obs with this setup 

* 9-Feb
    o put back the POL3 at its 'nominal' position
    o AMBER is fully realigned
    o test of an Ocean Optics source in entrance of the fiber K: flux in J and H
      is higher than with normal RAS lamp but there is absorption due to dichroics. 
    o tests of piezos: 10mV noise corresponding to 0.1 microns ampltitude
    o Observations on CAU show OPD drifts and visibility fluctuations.
    o laser for spectral calibration: PKe explains to FMa how it works.
    o test on switched-off piezo: it seems that visibilities are higher with
      piezo off than on, but we still see changes of several % in minutes.  

* 10-Feb
    o test of the stability of the output beams in 0th-order and also in MR-K:
      conclusion, this is not the reason for the visibility fluctuations. 
    o fibers to be inverted: not possible in LR and MR because the OPD
      difference is about 7mm. 
    o test with the BCD to identify if the source of fluctuations is before or
      after the BCD: BCD in & out with piezos off in MR. 
    o OPD on or off to see differences
    o bang tests in H and K.
    o foreseen: test switching off all motors. Piezos have been shut down. 

* 11-Feb
    o different DIT in MR to see the origin of vibrations if any...: no dependance
    o different DIT in H & K LR to see the origin of vibrations if any: no
      dependance -> is there really vibrations? 
    o check the CAU source: PKe's source enlight K, H & J but presence of
      absorption line due to dichroics 
    o display the K-LR fringes on RTD and locate sensitive optical parts: the
      large parabola of CAU is very sensitive 
    o P2VM? + obs. in LR with no piezo to check that the variation of V^2 come
      or not from piezzos: impossible to do because of OS software
      limitations. P2VM? requires piezo on for l/4 dephasing and the software
      forces them to be at 40um. No piezos means an OPD of 40um.... We stopped. 
    o P2VM? + obs. in MR: OK but we have still fluctuations: does it mean that
      piezos are out of questions? 
    o Observation with the beacons for bias estimation
    o Scan of the fiber inputs with the beacon to see influence of incidence
      (polarizer on) 
    o Polarizer is taken out: no more photometry socks
    o Data reduction of previous data: the fluctuations are large, changing in
      wavelength, in amplitude,...: This is probably teh origin of V^2 fluctuations. 
    o Record of data without polarizers for comparison: V12 and V23 are about
      0.4 when V23 is about 0.8 without polarizers. Still fluctuations.  

* 12-Feb
    o New measurements in MR-K: the data is much more stable
    o meeting and visit of the lab with P. Haguenauer
    o discussion with JBLB
    o experiment with perturbation of the large parabola in the CAU
    o telecon with R. Petrov
    o On sky measurements in MR-K: bright, medium and faint stars. Measurements
      with and w/o FINITO with DIT from 25ms to 200ms 
    o level of fluxes in K (H and J not realigned after the POL removed): gain
      between 500% and 700% without the POL. 
    o adjustment of LR-H for observations in LR-JHK (J needs to be adjusted)
    o observations in LR-JHK with FINITO. 

* 13-Feb
    o Discussion on the objective of the night: focused on MR-K
    o readjustment of H and J + OPD
    o beampos (careful with RAS)
    o LR- spectral calibration: script edition
    o start preset the telescopes at 22:30 
    o observe all night in MR-K without polarizers on calibrator stars

* 14-Feb
    o put back the polarizers
    o check the optical alignment in all bands
    o perform the spectral calibration with piezo scan (give also loss of
      coherence in LR)  
\end{verbatim}

\end{document}